\UseRawInputEncoding
\documentclass[12pt,a4paper]{article}
\usepackage{geometry}
\usepackage{amsmath}
\usepackage{amsmath}
\usepackage{amssymb}
\usepackage[dvips]{graphicx}
\usepackage{cite}
\usepackage{pstricks}
\usepackage{bm}
\usepackage{pbox}
\usepackage{placeins}
\usepackage{graphicx}
\usepackage{caption}
\usepackage{subcaption}
\usepackage[T1]{fontenc}
\usepackage{footnote}
\usepackage{pdfpages}
\usepackage{hhline}
\usepackage{multirow}
\usepackage{feynmf}
\usepackage{multicol}
\usepackage{enumitem}
\usepackage{multirow}
\usepackage{amsmath}
\usepackage{relsize}
\usepackage{slashed}
\usepackage{graphicx}
\usepackage[toc,page]{appendix}
\usepackage[compat=1.0.0]{tikz-feynman}
\allowdisplaybreaks

\usepackage{slashed}
\usepackage{cases}

\usepackage{empheq}

\definecolor{MyDarkBlue}{rgb}{0.1, 0.1, 0.8} %defining the color 'MyDarkBlue'
\definecolor{SBlue}{rgb}{0.2, 0.4, 0.7} %defining the color 'MyDarkBlue'
\definecolor{MyLightBlue}{rgb}{0.22,0.51,0.9}
\definecolor{MyGreen}{rgb}{0.0, 0.5, 0.0}
\definecolor{BrickRed}{rgb}{0.8, 0.25, 0.33}
\RequirePackage{hyperref}
\hypersetup{colorlinks, citecolor=SBlue,linkcolor=MyDarkBlue, urlcolor=BrickRed}

\begin{document}
\vspace*{-0.2in}
\begin{flushright}
\end{flushright}
\vspace{0.1cm}
\begin{center}
{\Large \bf
Investigating New Physics Models with  \\ \vspace{0.05in} Signature of Same-Sign Diboson$+\slashed{E}_{T}$ 
}
\end{center}
\renewcommand{\thefootnote}{\fnsymbol{footnote}}
\begin{center}
{
{}~\textbf{Cheng-Wei Chiang$^{1,2}$}\footnote{ E-mail: \textcolor{MyDarkBlue}{chengwei@phys.ntu.edu.tw }},
{}~\textbf{Sudip Jana$^3$}\footnote{ E-mail: \textcolor{MyDarkBlue}{sudip.jana@mpi-hd.mpg.de}},
{}~\textbf{Dibyashree Sengupta$^1$}\footnote{ E-mail: \textcolor{MyDarkBlue}{dsengupta@phys.ntu.edu.tw}}
}
\vspace{0.5cm}
{
\\
\em $^1$Department of Physics, National Taiwan University, Taipei, Taiwan 10617, R.O.C.
\\
$^2$Physics Division, National Center for Theoretical Sciences, Taipei, Taiwan 10617, R.O.C.
\\
$^3$Max-Planck-Institut f{\"u}r Kernphysik, Saupfercheckweg 1, 69117 Heidelberg, Germany
} 
\end{center}

%\vspace{0.6cm}
\renewcommand{\thefootnote}{\arabic{footnote}}
\setcounter{footnote}{0}
\thispagestyle{empty}

%%%%%%%%%%%%%%%%%%%%%%%%%%%%%%%%%%%%%%%%%%%%%%%
%%%%%%%%%%%%%%%%%%%%%%%%%%%%%%%%%%%%%%%%%%%%%%%
\begin{abstract}
We investigate the prospect of searching for new physics via the novel signature of same-sign diboson + $\slashed{E}_{T}$ at current and future LHC.  We study three new physics models: (i) natural SUSY models, (ii) type-III seesaw model and (iii) type-II seesaw/Georgi-Machacek model. In the first two class of models, this signature arises due to the presence of a singly-charged particle which has lifetime long enough to escape detection, while in the third model this signature originates resonantly from a doubly-charged particle produced along with two forward jets that, most likely, would escape detection.  We analyze in great detail the discovery prospects of the signal in these three classes of models in the current as well as the upcoming runs of the LHC (such as HL-LHC, HE-LHC and FCC-hh) by showing a distinction among these scenarios. 
\end{abstract}
%%%%%%%%%%%%%%%%%%%%%%%%%%%%%%%%%%%%%%%%%%%%%%%
%%%%%%%%%%%%%%%%%%%%%%%%%%%%%%%%%%%%%%%%%%%%%%%

%\newpage
\setcounter{footnote}{0}

%{
%  \hypersetup{linkcolor=black}
%  \tableofcontents
%}
\newpage

%%%%%%%%%%%%%%%%%%%%%%%%%%%%%%%%%%%%%%%%%%%%%%%
%%%%%%%%%%%%%%%%%%%%%%%%%%%%%%%%%%%%%%%%%%%%%%%

%%%%%%%%%%%%%%%%%%%%%%%%%%%%%%%%%%%%%%%%%%%%%%%
\section{Introduction}
\label{sec:intro}
%%%%%%%%%%%%%%%%%%%%%%%%%%%%%%%%%%%%%%%%%%%%%%%

In the past few decades, there have been several major discoveries in particle physics, culminating in the observation of the Higgs boson in 2012~\cite{Aad:2012tfa,Chatrchyan:2012ufa}. Despite this tremendous success of the Standard Model (SM), it is incomplete in its current form.  There is strong theoretical as well as experimental evidence (such as the hierarchical pattern seen in the fermion masses and mixings, the origin of neutrino masses, an understanding of dark matter, and the origin of the matter-antimatter asymmetry in the Universe) which calls for new physics beyond the Standard Model (BSM).

At the LHC, several searches have been performed to look for clues of these BSM models.  However, we have not seen any clear new physics signals so far.  In this work, we investigate the novel signal of same-sign diboson (SSdB) + $\slashed{E}_{T}$ which has been less studied and deserves more attention.  This signal is of interest because it has negligibly small background in the SM.  Hence, an observation of this signal will give a clear sign of BSM physics.  After a careful study, we find that it is possible to observe such a unique signature in three well-motivated BSM scenarios, namely: (i) natural supersymmetry models~\cite{Matalliotakis:1994ft, Baer:2005bu, Baer:2016hfa, Randall:1998uk, Baer:2018hwa, Baer:2020kwz}, (ii) type-III seesaw model~\cite{Foot:1988aq}, and (iii) type-II seesaw~\cite{Magg:1980ut, Schechter:1980gr, Mohapatra:1979ia, Lazarides:1980nt}/Georgi-Machacek model~\cite{Georgi:1985nv}~, while still being consistent with the existing theoretical and experimental limits.

\begin{figure} [h]
\begin{center}
\includegraphics [width=0.35\textwidth] {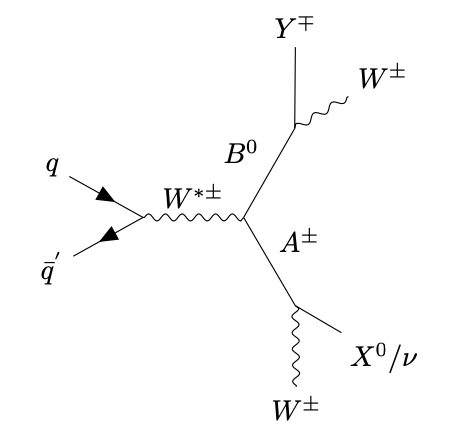}
\end{center}
\vspace*{-0.3in}
\caption{A generic Feynman diagram for SSdB + $\slashed{E}_{T}$ production at the LHC in BSM models, where $B^0$, $A^{\pm}$, $X^0$ and $Y^{\pm}$ are new particles.}
\label{fig:sample}
\end{figure}

Being a well-motivated BSM framework, supersymmetry (SUSY) provides an elegant solution to the Higgs mass hierarchy problem, accommodates a valid cold dark matter candidate, explains electroweak symmetry breaking, and features gauge coupling unification~\cite{Baer:2006rs}.  Although LHC searches for SUSY particles have pushed the masses of squarks and gluino high enough to expose weak scale SUSY to the risk of being unnatural/highly fine-tuned, there exist a class of SUSY models which can be natural as well as accommodate such highly massive sparticles well beyond the reach of the current LHC~\cite{Baer:2018hpb}. Since experiments do not put such high mass bounds on the masses of wino, bino, higgsino (and also the singlino which appears in some extended SUSY models~\cite{Nilles:1982dy, Frere:1983ag, Derendinger:1983bz, Maniatis:2009re, Ellwanger:2009dp, ATLAS:2019xhj, Barger:2010aq}), these natural SUSY models can have a wino-like or a bino-like lightest supersymmetric particle (LSP), provided gaugino mass unification is not considered~\cite{Baer:2015tva}. However, natural SUSY models can have a higgsino-like LSP irrespective of whether it assumes gaugino mass unification or not~\cite{Baer:2015tva}. Ref.~\cite{Barman:2017swy} and Ref.~\cite{Barman:2020vzm} study SUSY models with bino-like and singlino-like LSP, respectively, the latter being an extension of the minimal supersymmetric standard model (MSSM). Here, we consider a specific class of natural SUSY models that have a higgsino-like LSPs  which, under R-parity conservation, cannot decay to lighter SM particles, and hence can give rise to the novel SSdB + $\slashed{E}_{T}$ signature via the generic process shown in Fig.~\ref{fig:sample}. Earlier analyses have been done in this regard in Refs.~\cite{Baer:2013yha, Baer:2017gzf}.  Our current SUSY analysis differs from these earlier analyses in several aspects to be discussed in detail in Sec.~\ref{ssec:susy}.

We consider another interesting theoretical framework, type-III seesaw model, which has been proposed~\cite{Foot:1988aq} to explain the tiny neutrino masses and mixings.  In the type-III seesaw model, the SM particle spectrum is extended by three generations of $SU(2)_L$ triplet fermions with hypercharge $Y=0$, the lightest of which has a lifetime long enough to escape detection~\cite{Jana:2019tdm}, provided they have mass around a few hundred~GeV.  Hence, this model can also give rise to the novel signature of SSdB + $\slashed{E}_{T}$ via the generic process shown in Fig.~\ref{fig:sample}.

Another framework that can generate Majorana neutrino mass at tree level is type-II seesaw model~\cite{Magg:1980ut, Schechter:1980gr, Mohapatra:1979ia, Lazarides:1980nt}.  In addition to the SM particles, the model is extended by at least one $SU(2)_L$ triplet scalar $\Delta$ with hypercharge $Y=1$. Another model, called Georgi-Machacek (GM) model~\cite{Georgi:1985nv}, further has a real $SU(2)_L$ triplet scalar.  The doubly-charged scalar from the complex $SU(2)_L$ triplet scalar can be produced via vector boson fusion (VBF) process and decays into two $W$ bosons with same electric charge along with two forward jets coming from the initial state.  The forward jets may not be caught by the detector and hence the resultant final state will mimic our signature of interest.  However, later in Sec.~\ref{ssec:type2}, we will show that due to a stringent $T$-parameter constraint, the type-II seesaw model cannot give a sizeable cross section for this signature, whereas the GM model can.

There are several other channels that have potential to lead to the discovery of the three BSM scenarios considered here. Note that this work does not intend to make a comparative study between the SSdB+$\slashed{E}_{T}$ channel of interest here and the other more promising discovery channels.  The gist of this paper is to point out possible BSM models that can be a potential source of such a novel signature, if seen in experiments, since it is very unlikely for this signature process to appear within the SM.  Since more than one BSM scenario qualify, thus a need to distinguish among them is called for and such a distinction can be accomplished by the use of different sets of cuts.

In this article, we analyze the prospects of discriminating the above-mentioned BSM models through the SSdB + $\slashed{E}_{T}$ signature at the LHC. We find that the HL-LHC ($\sqrt{s}=14$~TeV and an integrated luminosity of 3 ab$^{-1}$) is insufficient to probe any of these three models through the signature process.  The natural SUSY models can be probed at the HE-LHC ($\sqrt{s}=27$~TeV) with an integrated luminosity of 3 ab$^{-1}$ while the type-II seesaw/GM model and the type-III seesaw model can be probed at the HE-LHC with an integrated luminosity of 15 ab$^{-1}$.  For completeness, we also extend our analysis to the Future Circular Collider of hadrons (FCC-hh with $\sqrt{s}=100$~TeV) with an integrated luminosity of 3 ab$^{-1}$ and 15 ab$^{-1}$. The rest of the paper is organized as follows. In Sec.~\ref{sec:models}, we review the three BSM models and how they give rise to the SSdB + $\slashed{E}_{T}$ signature. In Sec.~\ref{sec:eval}, the signals from all the three BSM models are optimized against the SM background. We show how each BSM model stands out for a particular set of cuts and discuss the discovery prospect for each scenario. Finally, we conclude in Sec.~\ref{sec:conclude}.

%%%%%%%%%%%%%%%%%%%%%%%%%%%%%%%%%%%%%%%%%%%%%%%
\section{SSdB + $\slashed{E}_{T}$ Signature from BSM Models}
\label{sec:models}
%%%%%%%%%%%%%%%%%%%%%%%%%%%%%%%%%%%%%%%%%%%%%%%

In this section, we briefly review the three classes of new physics models considered in this work, and how each of them leads to the SSdB + $\slashed{E}_{T}$ signature at the LHC.

%%%%%%%%%%%%%%%%%%%%%%%%%%%%%%%%%%%%%%%%%%%%%%%
\subsection{Supersymmetry}
\label{ssec:susy}
%%%%%%%%%%%%%%%%%%%%%%%%%%%%%%%%%%%%%%%%%%%%%%%

Although weak scale SUSY is a well-motivated BSM framework, experimental searches for sparticles have pushed the masses for squarks and gluino above 1 TeV. For example, current LHC data indicate that $m_{\tilde{g}}$ > 2.2~TeV~\cite{Aaboud:2017vwy, Vami:2019slp} and $m_{\tilde{t}_1}$ > 1.2~TeV~\cite{Vami:2019slp, ATLAS:2019oho, CMS:2019ysk}. Such large lower bounds on the masses of sparticles question the naturalness of weak scale SUSY~\cite{Craig:2013cxa}.  According to older notions of naturalness, SUSY models with such heavy sparticles are highly fine-tuned or unnatural~\cite{Barbieri:1987fn, Papucci:2011wy, Kitano:2006gv}. However, these earlier notions of naturalness can be updated to a more conservative electroweak naturalness measure, denoted by $\Delta_{EW}$~\cite{Baer:2013gva, Mustafayev:2014lqa, Baer:2014ica}.  A numerical expression for $\Delta_{EW}$ is obtained from minimizing the MSSM scalar potential that equates the $Z$ boson mass to weak scale SUSY parameters as  
\begin{equation}
m_Z^2/2  =\frac{m_{H_d}^2+\Sigma_d^d-(m_{H_u}^2+\Sigma_u^u)\tan^2\beta}
{\tan^2\beta -1}-\mu^2\\ 
\simeq -m_{H_u}^2-\mu^2-\Sigma_u^u(\tilde{t}_{1,2}) 
~,
\label{eq:mzs}
\end{equation}
where $\mu$ is the superpotential higgsino mass parameter, $m_{H_u}^2$ and $m_{H_d}^2$ are the soft SUSY breaking up-type and down-type Higgs mass parameters, respectively, $\tan\beta$ is the ratio of up-type Higgs vacuum expectation value (VEV) to the down-type Higgs VEV, and $\Sigma_u^u$ and $\Sigma_d^d$ denote radiative corrections as given in the Appendix of Ref.~\cite{Baer:2012cf}).

The electroweak naturalness measure, denoted by $\Delta_{EW}$, is defined as 
\begin{equation}
\Delta_{EW}=|(\mbox{max RHS contribution in Eq.~\eqref{eq:mzs}})|/(m_Z^2/2)
\label{eq:DEW}
\end{equation}
 It is suggested that a conservative choice for natural SUSY models is $\Delta_{EW} < 30$. Therefore, every point in the parameter space of a SUSY model that yields $\Delta_{EW}$ < 30 is considered to be natural.  As can be derived from Eq.~\eqref{eq:mzs} and Eq.~\eqref{eq:DEW}, $\Delta_{EW}$ < 30 demands: 
\begin{itemize}
    \item $\mu$ $\sim$ $100-300 \rm~GeV$. \footnote{The space of $\mu$ < 100~GeV has been ruled out by LEP2 experiment~\cite{LEP:2003aa}.}
    \item $m_{H_u}^2$ should acquire a small negative value $\sim$ $-(100-300)^2$ $ \rm~GeV^2$ at the weak scale. This occurs when $m_{H_u}^2$ is driven radiatively from high-energy scales to the weak scale augmented by a combination of statistical and anthropic pull such that electroweak symmetry is only barely broken~\cite{Baer:2016lpj}.
    \item $\Sigma_u^u$ should also be below $(300)^2$ $ \rm~GeV^2$. This is attainable with $m_{\tilde{t}_1}$ > 1.2~TeV and $m_{\tilde{g}}$ > 2.2~TeV. 
\end{itemize}

\begin{figure} [h!]
\begin{center}
\includegraphics [width=0.35\textwidth] {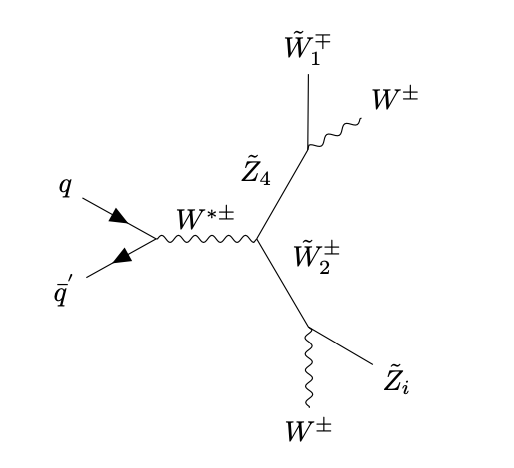}
\end{center}
\vspace*{-0.3in}
\caption{Feynman diagram for SSdB production at the LHC in SUSY models with light higgsinos ($\tilde{W}_{1}^{\mp}$ and $\tilde{Z}_{i}$ with $i = 1,2$).  Here $\tilde{Z}_{4}$ and $\tilde{W}_{2}^{\pm}$ in the intermediate step are winos.}
\label{fig:susy}
\end{figure}

Ref.~\cite{Baer:2018hpb} shows several natural SUSY models that satisfy all the above criteria, with huge parameter space still left to be probed experimentally.  As seen from the above conditions, natural SUSY models, assuming gaugino mass unification, have a unique property that $\mu$ $\ll$ $M_{1,2}$ $<$ $M_3$ where $M_1$, $M_2$ and $M_3$ refer to the masses of bino, wino and gluino, respectively, at the weak scale.  Thus, in these natural SUSY models, the LSP is almost purely higgsino-like. Under assumed R-parity conservation, the LSP becomes a good dark matter candidate in the model and manifests as $\slashed{E}_{T}$ in collider experiments. However, such a higgsino-like LSP of mass $100-300$~GeV are thermally under-produced. Hence, such an LSP only partially contributes to the total dark matter content of the Universe. In order to account for the entire dark matter content of the Universe, either other particles suitable to form dark matter must be present in the model or the rest of the dark matter must be non-thermally produced. It has been shown in Ref.~\cite{Baer:2018rhs} that the latter case is excluded by experiments. Therefore, the rest of the dark matter must be formed by some other particles.  Axion, which arises in a completely different context of solving the strong CP problem via the Peccei-Quinn solution~\cite{Peccei:1977hh, Peccei:1977ur,Weinberg:1977ma, Wilczek:1977pj, Kim:1979if, Shifman:1979if, Dine:1981rt, Zhitnitsky:1980tq}, turns out to be an excellent candidate for serving this purpose.  Out of various natural SUSY models listed in Ref.~\cite{Baer:2018hpb}, we choose the two extra parameter non-universal Higgs (NUHM2) model~\cite{Matalliotakis:1994ft, Baer:2005bu} with $\mu \ll M_{1} < M_{2} < M_3$ which can give rise to a clean SSdB + $\slashed{E}_{T}$ signature via wino-pair production, as pointed out in Ref.~\cite{Baer:2013yha, Baer:2017gzf}.  The corresponding Feynman diagram is shown in Fig.~\ref{fig:susy}.

In this work, we have analyzed this signal in detail using the most up-to-date constraints on $m_{\tilde{g}}$ and $\sigma^{SI}(\tilde{z}_1, p)$ obtained from the LHC data with an integrated luminosity ($\mathcal{L}$) of 139~fb$^{-1}$~\cite{Aaboud:2017vwy, Vami:2019slp} and the XENON1T experiment~\cite{Aprile:2017iyp}, respectively.
The relevant benchmark point is given in Table~\ref{tab:susybm}. 

\begin{table}[h!]
\centering
\begin{tabular}[h]{|p{3.9cm}|c|}
\hline
parameter & NUHM2 \\
\hline
$m_0$      & 5000~GeV \\
$A_0$      & $-8000$~GeV \\
$\tan\beta$    & 12  \\
\hline
$M_1 ({\rm GUT})$   & 1250~GeV \\
$M_2 ({\rm GUT})$   & 895~GeV \\
$M_3 ({\rm GUT})$   & 1250~GeV \\
\hline
$\mu$          & 150~GeV  \\
$m_A$          & 2500~GeV \\
\hline
$m_{\tilde{g}}$   & 2938.2~GeV \\
$m_{\tilde{u}_L}$ & 5458.3~GeV \\
$m_{\tilde{u}_R}$ & 5591.1~GeV \\
$m_{\tilde{e}_R}$ & 4840.1~GeV \\
$m_{\tilde{t}_1}$ & 1820.5~GeV \\
$m_{\tilde{t}_2}$ & 3925.7~GeV \\
$m_{\tilde{b}_1}$ & 3959.9~GeV \\
$m_{\tilde{b}_2}$ & 5301.7~GeV \\
\hline
\end{tabular}
\hfill
\begin{tabular}[h]{|p{4.2cm}|c|}
\hline
parameter & NUHM2 \\
\hline
$m_{\tilde{\tau}_1}$ & 4728.9~GeV \\
$m_{\tilde{\tau}_2}$ & 5061.8~GeV \\
$m_{\tilde{\nu}_{\tau}}$ & 5067.2~GeV \\
$m_{\tilde{w}_1}$ & 156.6~GeV \\
$m_{\tilde{w}_2}$ & 762.9~GeV \\
$m_{\tilde{z}_1}$ & 146.0~GeV \\ 
$m_{\tilde{z}_2}$ & 157.9~GeV \\ 
$m_{\tilde{z}_3}$ & 559.8~GeV \\ 
$m_{\tilde{z}_4}$ & 775.4~GeV \\ 
$m_h$       & 125.1~GeV \\ 
\hline
$\Omega_{\tilde{z}_1}^{\rm std}h^2$ & 0.007 \\
$BF(b\to s\gamma)\times 10^4$ & $3.06$ \\
$BF(B_s\to \mu^+\mu^-)\times 10^9$ & $3.8$ \\
$\sigma^{SI}(\tilde{z}_1, p)$ (pb) & $2.08\times 10^{-9}$ \\
$\sigma^{SD}(\tilde{z}_1, p)$ (pb)  & $8.4\times 10^{-5}$ \\
$\langle\sigma v\rangle |_{v\to 0}$  (cm$^3$/sec)  & $2.99\times 10^{-25}$ \\
$\Delta_{EW}$ & 29.6 \\
\hline
\end{tabular}
\caption{Input parameters and masses
for a SUSY benchmark point from the NUHM2 model
with $m_t=173.2$~GeV using {\tt Isajet}~7.88~\cite{Paige:2003mg}.
}
\label{tab:susybm}
\end{table}

Here we have assumed a more general scenario without gaugino mass unification~\cite{Baer:2015tva}. This also has the advantage of having the wino mass $\sim 770$~GeV while satisfying the LHC constraints on gluino mass~\cite{Aaboud:2017vwy, Vami:2019slp}.  Note that the above choice of wino mass is for a comparison of this signal with a similar one obtained from the type-III seesaw model, as discussed in the next subsection. Also note that even though gaugino mass unification is not assumed, the benchmark point given in Table~\ref{tab:susybm} satisfies $\mu \ll M_{1} < M_{2} < M_3$ which is an essential criteria for obtaining the signal of our interest via the process shown in Fig.~\ref{fig:susy}.

%%%%%%%%%%%%%%%%%%%%%%%%%%%%%%%%%%%%%%%%%%%%%%%
\subsection{The type-III Seesaw Model}
\label{ssec:type3}
%%%%%%%%%%%%%%%%%%%%%%%%%%%%%%%%%%%%%%%%%%%%%%%
In the type-III seesaw model, the SM particle spectrum is extended by multiple $SU(2)_L$ triplet fermions ($\Sigma$s) which have hypercharge $Y=0$. In order to generate tiny neutrino mass and proper flavor structure in the neutrino sector, one needs to introduce at least two generations of $SU(2)_L$ triplet fermions.  The tiny neutrino masses are generated at the tree level and can be expressed as $m_\nu \simeq Y^2_\nu v^2/M_\Sigma$, where $Y_\nu$ is the Yukawa coupling, $v$ is the SM Higgs VEV, and $M_\Sigma$ is the triplet fermion mass~\cite{Foot:1988aq}.  In general, the type-III seesaw scenario with $M_\Sigma \simeq {\cal O} (1)$~TeV is technically natural and opens up a plethora of implications in collider experiments~\cite{Franceschini:2008pz, Arhrib:2009mz, Jana:2019tdm, Bandyopadhyay:2011aa, Goswami:2017jqs, Das:2020gnt, Ashanujjaman:2021jhi, Das:2020uer, Biggio:2019eeo, delAguila:2008cj}.

In our analysis, we consider three generations of $SU(2)_{L}$ triplet fermions, $\Sigma_i$ ($i = 1, 2, 3$), with a non-degenerate mass spectrum.  The relevant Lagrangian is given by 
\begin{equation}
\mathcal{L}_{\Sigma}=
{\rm Tr}\left[{\overline \Sigma}_i {\slashed D} \Sigma_i\right]
- \left( \frac{1}{2} M_{\Sigma}^{ij} {\rm Tr}
\left[\overline{\Sigma^c}_{i}  \Sigma_j\right] + {\rm h.c}\right) 
-\left({\sqrt 2} Y_{\Sigma }^{ij} {\overline L}_i  \Sigma_j {H}+ {\rm h.c}\right)
~, 
\label{lag}
\end{equation}
where $D_\mu$ is the covariant derivative for $\Sigma_i$, $M_\Sigma$ denotes the triplet fermions mass matrix, and $Y_{\Sigma}$ is the Yukawa coupling matrix.  For the rest of our paper, we refer to the lightest generation of heavy fermions as $\tilde{\Sigma}$ and their masses as $m_{\tilde{\Sigma}}$.  Depending on a normal or inverted hierarchy, the lightest fermion triplet will be $\Sigma_{1}$ or $\Sigma_{3}$, respectively.  For simplicity, we set the other two generations of heavy fermions to be almost degenerate.

\begin{figure} [h!]
\begin{center}
\includegraphics [width=0.35\textwidth] {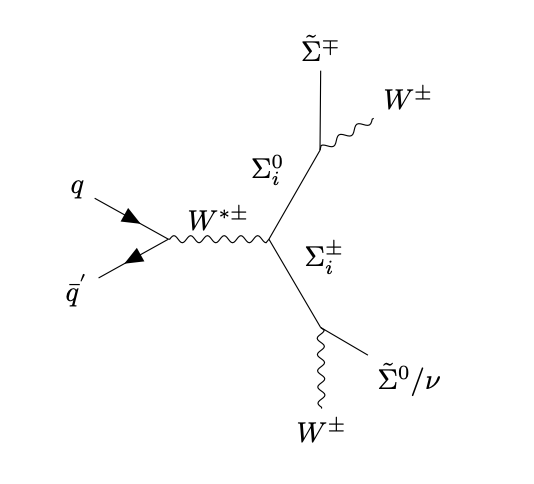}
\end{center}
\vspace*{-0.3in}
\caption{Feynman diagram for the SSdB + $\slashed{E}_{T}$ signature at the LHC in the type-III seesaw model, where $\tilde{\Sigma}^0$ and $\tilde{\Sigma}^{\pm}$ are members of the lightest fermionic triplets.
}
\label{fig:type3seesaw}
\end{figure}
\begin{table}[h!]
\centering
\begin{tabular}{lc}
\hline
parameter & Type III Seesaw \\
\hline
$m(\Sigma_i^0)$      & 770.39 GeV \\
$m(\Sigma_i^\pm)$      & 770.57 GeV \\
$m(\tilde{\Sigma}^0)$      & 670.39 GeV \\
$m(\tilde{\Sigma}^\pm)$      & 670.57 GeV \\
\hline
%$\Gamma(\Sigma_2^0)$      & $3.70\times 10^{-2}$ GeV \\
%$\Gamma(\Sigma_2^\pm)$      & $1.91\times 10^{-2}$ GeV \\
%$\Gamma(\Sigma_1^0)$      & $1.70\times 10^{-19}$ GeV \\
%$\Gamma(\Sigma_1^\pm)$      & $4.45\times 10^{-16}$ GeV \\
%\hline
$c\tau(\Sigma_i^0)$      &  $5.34\times 10^{-13}$ cm \\
$c\tau(\Sigma_i^\pm)$      &  $1.034\times 10^{-12}$ cm \\
$c\tau(\tilde{\Sigma}^0)$      &  $1.2\times 10^{5}$ cm \\
$c\tau(\tilde{\Sigma}^\pm)$      &  44.37 cm\\
\hline
\end{tabular}
\caption{Masses and decay widths
for a Type III Seesaw benchmark point.
}
\label{tab:type3bm}
\end{table}

As can be inferred from a detailed calculation of partial decay widths of these $SU(2)_L$ triplet fermions in Ref.~\cite{Jana:2019tdm}, depending on the neutrino parameters, the electically neutral member of the lightest generation of fermionic triplets $\tilde{\Sigma}^{0}$ of mass around a few hundred~GeV can have lifetime long enough to escape detection and hence shows up as large $\slashed{E}_{T}$  in collider experiments. $\tilde{\Sigma}^{\pm}$, being only a few MeV heavier than its neutral partner $\tilde{\Sigma}^{0}$, travels a short distance before primarily decaying into $\tilde{\Sigma}^{0}$ and a charged pion of momentum low enough to be reconstructed as a track. This results in a disappearing track signature from $\tilde{\Sigma}^{\pm}$ as can also be seen in  Ref.~\cite{Jana:2019tdm}. There are several dedicated searches for the disappearing track signature at the LHC~\cite{Sirunyan:2020pjd}.
%{ATLAS:2021ttq, Aaboud:2017mpt} : similar atlas references. 
We recast a recent LHC limit in Ref.~\cite{Sirunyan:2020pjd} to derive a bound on the charged heavy fermion in the type-III seesaw model.  We find the lower bound on mass of $\tilde{\Sigma}^{\pm}$ to be around 670~GeV in order to be consistent with the collider data. In our analysis, we set the other two pairs of heavy fermions to have mass at 770~GeV, so that they primarily decay to a $W^{\pm}$ boson and a $\tilde{\Sigma}^{\pm,0}$ particle through a tiny mixing.  This leads to a clean SSdB + $\slashed{E}_{T}$ signature from pair production of $\Sigma_i^{\pm,0}$ at the LHC via the Feynman diagram shown in Fig.~\ref{fig:type3seesaw}. This particular benchmark point is shown in Table.~\ref{tab:type3bm}.

%%%%%%%%%%%%%%%%%%%%%%%%%%%%%%%%%%%%%%%%%%%%%%%
\subsection{Type-II seesaw/Georgi-Machacek model}
\label{ssec:type2}
%%%%%%%%%%%%%%%%%%%%%%%%%%%%%%%%%%%%%%%%%%%%%%%

In this subsection, we focus on the scenario where the SSdB signature originates from the decay of a doubly-charged scalar.  Generally, these doubly-charged scalars appear in several BSM frameworks~\cite{Magg:1980ut, Schechter:1980gr, Lazarides:1980nt, Mohapatra:1980yp, Pati:1974yy, Mohapatra:1974hk, Senjanovic:1975rk, Kuchimanchi:1993jg, Babu:2008ep, Babu:2014vba, Basso:2015pka, Zee:1985id, Babu:1988ki, ArkaniHamed:2002qx, Babu:2020hun, Georgi:1985nv, Gunion:1989ci, Babu:2009aq, Bonnet:2009ej, Bhattacharya:2016qsg, Kumericki:2012bh}. One such framework is the simplest type-II seesaw model~\cite{Magg:1980ut, Schechter:1980gr, Mohapatra:1979ia, Lazarides:1980nt} which introduces an $SU(2)_L$ triplet scalar $\Delta = (\Delta^{++}, \Delta^{+}, \Delta^{0})$ with hypercharge $Y=1$. Tiny neutrino masses are generated while the neutral component of the $SU(2)_L$ triplet, $\Delta^0$, acquires a small VEV, $v_{\Delta}$.  This type of $SU(2)_L$ triplet scalar which contains a doubly-charged scalar $\Delta^{++}$ also appears in Minimal Left-Right Symmetric Model~\cite{Pati:1974yy, Mohapatra:1974hk, Senjanovic:1975rk} as well as GM model~\cite{Georgi:1985nv}.

\begin{figure} [h]
\begin{center}
\includegraphics [width=0.5\textwidth] {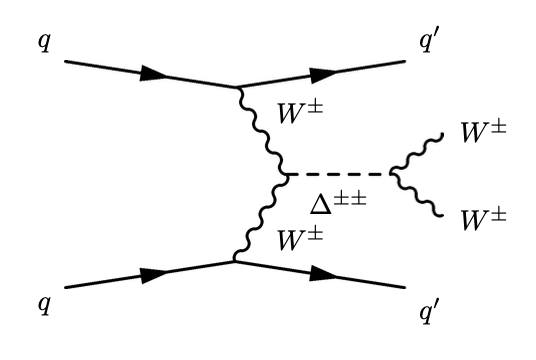}
\end{center}
\vspace*{-0.3in}
\caption{Feynman diagram for SSdB + forward jets production at LHC in the type-II seesaw models.}
\label{fig:type2seesaw}
\end{figure}

At the LHC, these doubly-charged scalars ($\Delta^{\pm\pm}$) can be pair-produced via the Drell-Yan process (s-channel $Z/\gamma$ exchanges). There are several extensive phenomenological studies as well as experimental studies on pair-production of doubly-charged scalars which also look at the $W^{+}W^{+}W^{-}W^{-}/ \ell^{+} \ell^{+} \ell^{-} \ell^{-}$ final state signatures (for a review, see Ref.~\cite{Cai:2017mow}). However, we are focusing on the resonant production of the doubly-charged scalar through the VBF process here, as shown in Fig.~\ref{fig:type2seesaw}. This production rate is proportional to $v_{\Delta}^2$ and becomes more dominant than the Drell-Yan process for $v_\Delta \sim {\cal O}(10)$~GeV and $m_{\Delta^{\pm\pm}} \sim {\cal O}(100)$~GeV~\cite{Chiang:2012dk, Chiang:2015amq}.  Note that $v_{\Delta}$ in the simplest type-II seesaw model~\cite{Magg:1980ut, Schechter:1980gr, Mohapatra:1979ia, Lazarides:1980nt} is tightly bounded by the electroweak $T$ parameter, giving $v_{\Delta} \lesssim 3$~GeV~\cite{Zyla:2020zbs}.  In the GM model that also contains $SU(2)_L$ scalar triplet fields, $\xi$ with hypercharge $Y=0$ and $\chi$ with hypercharge $Y=1$, to preserve the custodial symmetry at tree level, $v_{\Delta}$ can be as high as $\sim 50$~GeV~\cite{Blasi:2017xmc, Chiang:2018xpl}.  As a consequence, the resonant production rate could be much larger than in the simplest type-II seesaw scenario.

\begin{table}[h!]
\centering
\begin{tabular}{lc}
\hline
parameter & Type II Seesaw \\
\hline
$m(\Delta^{\pm\pm})$      & 300.0 GeV \\
$m(\Delta^\pm)$      & 302.0 GeV \\
\hline
$\Gamma(\Delta^{\pm\pm})$      & $3.83\times 10^{-4}$ GeV \\
\hline
$v_{\Delta}$              & 1 GeV \\
\hline
\end{tabular}
\caption{Masses and decay widths
for a Type II Seesaw benchmark point.
}
\label{tab:type2bm}
\end{table}

\begin{figure}[h!]
\centering
\includegraphics [width=0.55\textwidth] {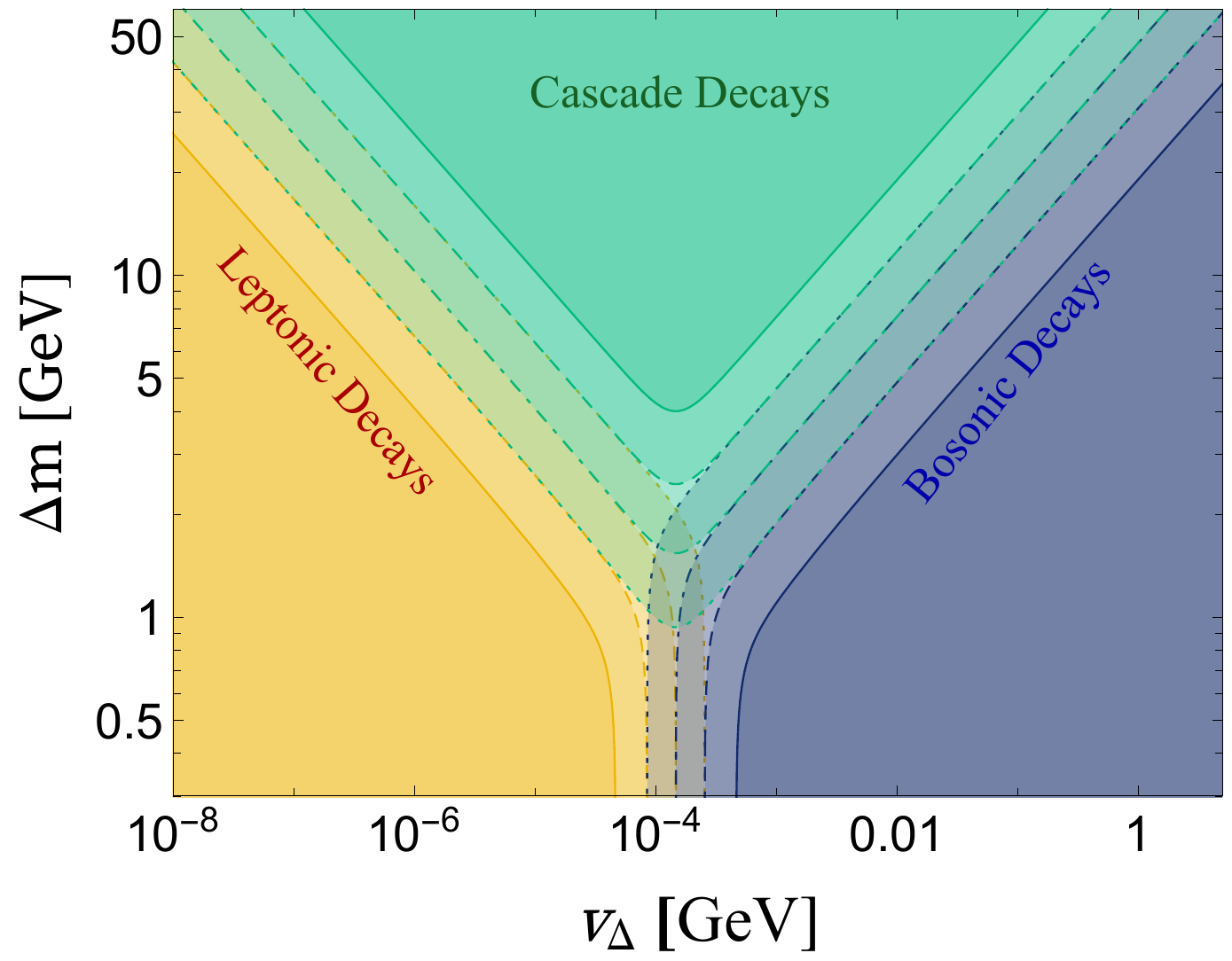}
\vspace*{-0.1in}
\caption{Decay phase diagram of doubly-charged scalar ($\Delta^{\pm \pm}$) with mass = 300~GeV. The solid, dashed, dot-dashed and dotted contours indicate 99$\%$, 90$\%$, 50$\%$ and 10$\%$ branching ratios respectively, for the bosonic, leptonic or cascade decays.  The mass splitting $\Delta m$ is defined in the main text.}
\label{fig:phaset2}
\end{figure}

In general the doubly-charged scalar can be either lightest or heaviest depending on the sign of the quartic coupling in the potential.  In the rest of our analysis, we consider the scenario where $\Delta^{++}$ is the lightest among all members in the triplet fields.  In this scenario, $\Delta^{++}$ dominantly decays into same-sign dilepton (SSdL) ($\Delta^{\pm\pm} \to \ell^{\pm} \ell^{\pm}$) or SSdB ($\Delta^{\pm\pm} \to W^{\pm} W^{\pm}$), depending on the value of $v_{\Delta}$~\cite{Melfo:2011nx,Aoki:2011pz,Chiang:2012cn}. In Fig.~\ref{fig:phaset2}, we show a complete decay phase diagram of the doubly-charged scalar of mass 300~GeV.  As shown in the plot, $\Delta^{\pm\pm}$ dominantly decays to two same-sign $W$ bosons for $v_{\Delta}$ $\gtrsim$ 1~MeV, provided the mass splitting $\Delta m \equiv m_{\Delta^{\pm\pm}} - m_{\Delta^{\pm}}$ $\lesssim$ 5~GeV.  For our analysis, we set the mass splitting $\Delta m$ = 2~GeV and $v_{\Delta} \sim$ 1~GeV so that this benchmark point (shown in Table~\ref{tab:type2bm}) lies in the blue shaded region of Fig.~\ref{fig:phaset2} that is of our interest.  Thus, after being produced at the LHC along with two forward jets, $\Delta^{\pm\pm}$ decays primarily to two same-sign $W$ bosons.  These jets may escape detection, especially in the forward region with lower detector efficiency. Assuming leptonic decay of the $W$ bosons, we obtain SSdL + $\slashed{E}_{T}$ in the final state. In this case, the final state mimics the signature of our interest.

Being proportional to $v_\Delta^2$, the cross section obtained for this signature in the type-II seesaw models, even before any cuts, is negligibly small even for $v_{\Delta}$ = 1~GeV.  As stated above, the GM model can accommodate $v_{\Delta}$ as high as 50~GeV.  Hence, from now on, we will be considering the GM model, instead of the simplest type-II seesaw model, assuming $v_{\Delta}$ = 10~GeV, which is an arbitrary choice. The cross section for the stated signal is now sufficiently large to be detectable at the LHC and can scale easily with $v_{\Delta}^2$.

%%%%%%%%%%%%%%%%%%%%%%%%%%%%%%%%%%%%%%%%%%%%%%%
\section{Signal and Background Evaluations}
\label{sec:eval}
%%%%%%%%%%%%%%%%%%%%%%%%%%%%%%%%%%%%%%%%%%%%%%%

Here in this section we systematically investigate the signal and the background for the aforementioned models. Considering leptonic decays of the $W$ bosons, the signal of interest here has a final state of SSdL + $\slashed{E}_{T}$, where the leptons include the electron and muon.  As stated in Sec.~\ref{sec:models}, we can obtain such a signal from wino pair production in NUHM2 model in SUSY, pair production of heavy $SU(2)_L$ triplet in type-III seesaw model, and the resonant production of the doubly-charged scalar in the GM model when the forward jets go undetected. Note that, a final state of SSdL + $\slashed{E}_{T}$ can also be obtained from gluino/squark pair production in SUSY models~\cite{Barnett:1988mx, Baer:1989hr, Baer:1991xs, Barnett:1993ea}. However, this signature can be distinguished from the signal studied here because the SSdL + $\slashed{E}_{T}$ from gluino/squark pair production appears along with large number of hard central jets.  We evaluate the signal from all the three models and the background from the SM, and optimize cuts to efficiently reduce the background. As compared to an earlier analysis in Ref.~\cite{Baer:2017gzf}, three new backgrounds, namely, $ZZ$, $W^{\pm}W^{\mp}Z$ and $W^{\pm}ZZ$ have been included here. Therefore, the relevant SM background processes are: $t\bar{t}$, $t\bar{t}t\bar{t}$, $t\bar{t}W^{\pm}$, $t\bar{t}Z$, $W^{\pm}W^{\pm}jj$, $W^{\pm}W^{\pm}W^{\mp}$, $W^{\pm}Z$, $ZZ$, $W^{\pm}W^{\mp}Z$ and $W^{\pm}ZZ$.

As discussed in Sec.~\ref{ssec:type3}, for our analysis the mass of $\tilde{\Sigma}^{\pm,0}$ is taken to be around 670~GeV so as to satisfy the mass constraint in Ref.~\cite{Sirunyan:2020pjd} and we take ${\Sigma}_i^{\pm,0}$ of mass 770~GeV to ensure that ${\Sigma}_i^{\pm,0}$ primarily decay to $W^{\pm}$ and $\tilde{\Sigma}^{\pm,0}$.  Hence, we take a suitable benchmark point in the NUHM2 model as well, with the wino-like particles ($\tilde{Z}_4$ and $\tilde{W}_2^{\pm}$) also attaining a mass of around 770~GeV, so that the signals from the type-III seesaw model and the NUHM2 model are at par with each other, as stated earlier in Sec.~\ref{ssec:susy}.  However, for the GM model, we have considered $m(\Delta^{\pm\pm})$ = $300 \rm~GeV$ since the limit is less stringent on the mass of $\Delta^{\pm\pm}$ ($> 200 \rm~GeV$~\cite{Chiang:2012dk}) while looking for the bosonic final state signatures at the LHC.  Using the same argument, we could have taken lower mass for the wino-like particles ($\tilde{Z}_4$ and $\tilde{W}_2^{\pm}$) in the NUHM2 model but then the NUHM2 benchmark point would not satisfy various constraints such as the mass limit on gluino from the LHC, dark matter constraints from direct detection experiments, etc.

For  simulations, we have used {\tt MadGraph5$\_$aMC@NLO}~\cite{Alwall:2011uj,Alwall:2014hca} for event generation, interfaced with {\tt Pythia}~8.2 ~\cite{Sjostrand:2014zea} for parton showering and hadronization, followed by {\tt Delphes}~3.4.2~\cite{deFavereau:2013fsa} for detector simulation where the default Delphes card is employed. Therefore, in the detector level simulation:
\begin{itemize}
    \item The anti-$k_T$ jet algorithm~\cite{Cacciari:2008gp} has been used with $R = 0.5$ and $p_T({\rm jet}) > 20$~GeV. 
    \item The default jet flavor association module has been used that identifies a jet containing a $b$-hadron as a potential $b$-jet if the $b$-hadron lies within $\Delta R = 0.5$ of the jet axis.  Such $b$-jets are tagged with $85 \%$ efficiency with a mistagging probability of $10 \%$ for other lighter jets following CMS $b$-tagging algorithm~\cite{CMS:2012feb}.\footnote{The $b$-tagging and misidentification rates considered here are from those used for $\sqrt{s}=7$~TeV collisions. We have also used the $b$-tagging and misidentification rates from those used for $\sqrt{s}=13$~TeV collisions, where $b$-jets are tagged with an efficiency of $68 \%$ and a mistagging probability of $1 \%$ for other lighter jets following a more recent CMS $b$-tagging algorithm~\cite{CMS:2017wtu} for the $t\bar{t}$ process and found similar results after the same cuts.}
    \item The $\tau$ lepton is identified with an efficiency of $60 \%$ (fixed in {\tt Delphes}~3.4.2~\cite{deFavereau:2013fsa}) if it lies within $\Delta R = 0.5$ of the jet axis~\cite{CMS-DP-2018-026,CMS:2015pac}. 
    \item The $e$ and $\mu$ leptons are isolated following the criterion that the ratio of the sum of transverse momenta higher than 0.5~GeV of all particles that lie within a cone of $\Delta R < 0.5$ around the lepton to the transverse momentum of that lepton is less than 0.25.
\end{itemize}

We have used {\tt Isajet}~7.88~\cite{Paige:2003mg} to generate the Les Houches Accord (LHA) file for the NUHM2 signal and pass it through the above-mentioned simulation chain.

We have used {\tt Prospino}~\cite{Beenakker:1996ed} to calculate the leading-order (LO) and next-to-leading-order (NLO) cross sections for the NUHM2 signal process and type-III seesaw signal process for 14~TeV LHC.  Since {\tt Prospino} is designed specifically for calculating NLO cross sections of SUSY processes, using it to calculate the same for the type-III seesaw model is made possible by utilizing the analogy between the type-III seesaw model and the minimal anomaly-mediated SUSY breaking (mAMSB) model~\cite{Randall:1998uk, Gherghetta:1999sw, Feng:1999hg}.  The mAMSB model has a wino-like LSP ($\tilde{Z_1}$) and a wino-like next-to-LSP (NLSP) ($\tilde{W_1^{\pm}}$), analogous to the type-III seesaw model with its lightest and next-to-lightest particles being $\tilde{\Sigma}^0$ and $\tilde{\Sigma}^{\pm}$, respectively.  Thus, a suitable mAMSB parameter space point has been used to calculate the LO and NLO cross sections for the type-III seesaw model. 

We have used the K-factor for the type-II seesaw models (and hence the GM model), as done in Ref.~\cite{ATLAS:2016pbt}.  The K-factors for the SM background processes are used as in Ref.~\cite{Baer:2017gzf}. We obtain the K-factor for the $ZZ$ process from Ref.~\cite{CMS:2017dzg}. The K-factor for the $W^{\pm}W^{\mp}Z$ process has been used as in Ref.~\cite{Nhung:2013tfu, Nhung:2013jta, Hankele:2007sb}. For the $W^{\pm}ZZ$ process, we use the same K-factor as for $W^{\pm}W^{\mp}Z$, as suggested in Ref.~\cite{NLOMultilegWorkingGroup:2008bxd}. We generate $10^7$ events for the $t\bar{t}$ process and $10^6$ events for all the other background and signal processes for $\sqrt{s}=27$~TeV. While for $\sqrt{s}=100$~TeV, we generate $10^7$ events for the $t\bar{t}$ and $W^{\pm}Z$ processes and $10^6$ events for all the other background and signal processes.

%Taking into account the NLO cross sections for each signal and background process, cross sections before any cut at single event level are as follows: 

%\begin{table}[h!]
%\begin{center}
%\begin{tabular}{|c|c|c|c|c|c|c|c|c|c|c|}
%\hline
%\hline
%Process  & NUHM2 & type-III & GM & $t\bar{t}$ & $t\bar{t}t\bar{t}$ & $t\bar{t}W^{\pm}$  & $t\bar{t}Z$ &  $W^{\pm}W^{\pm}jj$ & $W^{\pm}W^{\pm}W^{\mp}$ & $W^{\pm}Z$\\ 
%\hline
%K-factor & 1.17 & 1.16 & 1.26 & 1.72 & 1.27 & 1.24 & 1.39 & 1.04& 2.45 & 1.88 \\
%\hline
%$\sigma$/events [ab]  & 0.042 & 0.043 & 0.056 & 410 & 0.11 & 1.5 & 4.4 & 1.1 & 0.8 & 120\\
%\hline\hline
%\end{tabular}
%\end{center}
%\caption{Cross sections for each signal and background process before any cut at single event level at $\sqrt{s} = 27$~TeV.}
%\label{single event 27tev}
%\end{table}

Motivated by the earlier analyses~\cite{Baer:2013yha, Baer:2017gzf}, we put a set of basic cuts, dubbed the S1-cuts, to reduce the SM backgrounds.  Explicitly, the S1-cuts include: 
\begin{itemize}
    \item Require exactly two same-sign isolated leptons, where the isolated leptons are defined as those with $p_T(\ell) > 10$~GeV and $\eta(\ell) < 2.5$.
    \item Veto events with any identified $b$-jet.
    \item Require $p_T(\ell_1) > 20$~GeV, where $\ell_1$ denotes the leading lepton. 
\end{itemize}

In the following subsections, we show how each BSM model stands out by further imposing a particular set of additional cuts.

%%%%%%%%%%%%%%%%%%%%%%%%%%%%%%%%%%%%%%%%%%%%%%%
\subsection{Supersymmetry Analysis}
\label{ssec:susyanalysis}
%%%%%%%%%%%%%%%%%%%%%%%%%%%%%%%%%%%%%%%%%%%%%%%

\begin{figure}[h]
\centering
\begin{subfigure}[h]{0.5\textwidth}
  \centering
  \includegraphics[width=1\linewidth]{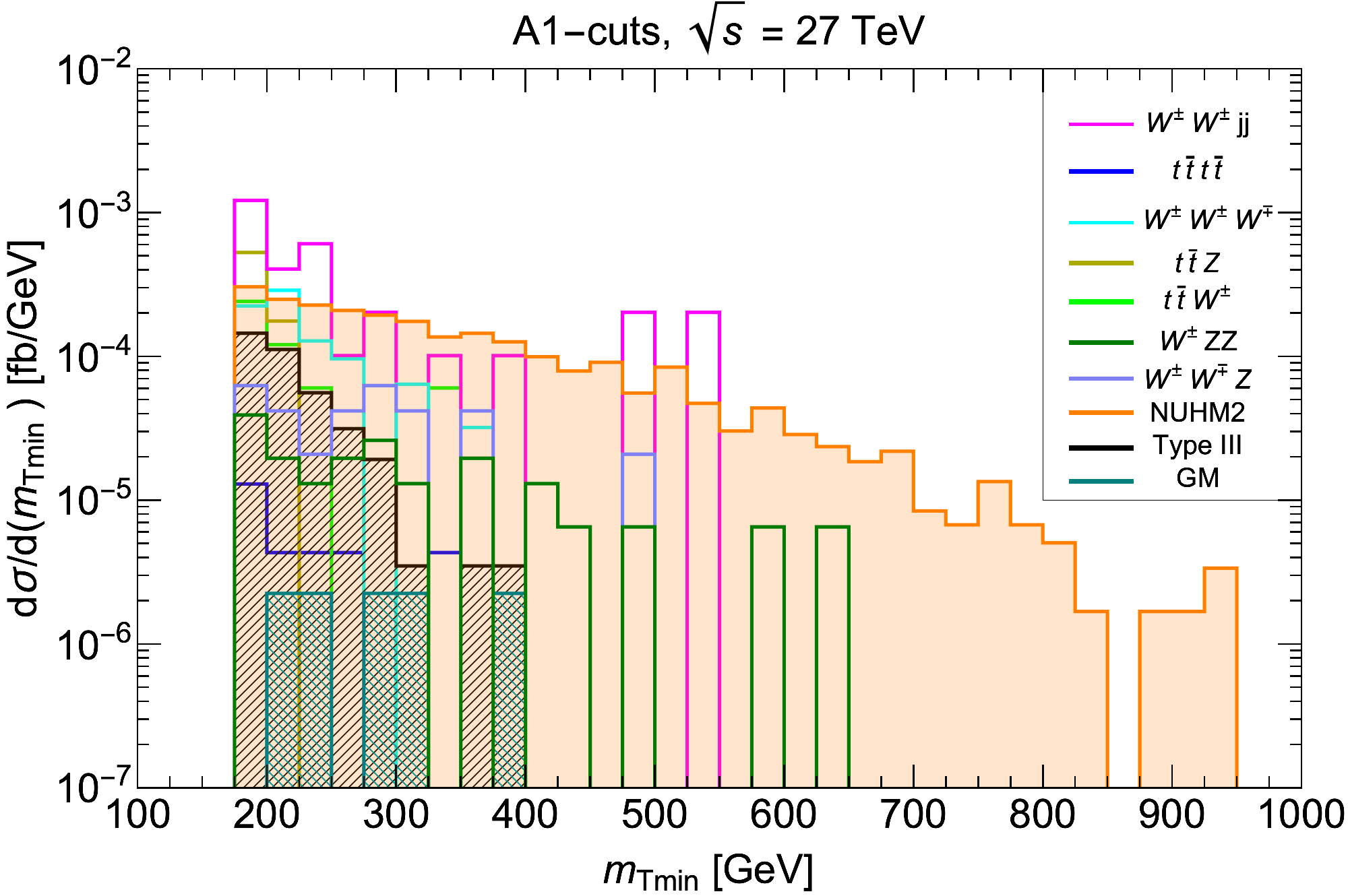}
  \caption{}
  \label{fig:mtmina1susy27}
\end{subfigure}%
\begin{subfigure}[h]{0.5\textwidth}
  \centering
  \includegraphics[width=1\linewidth]{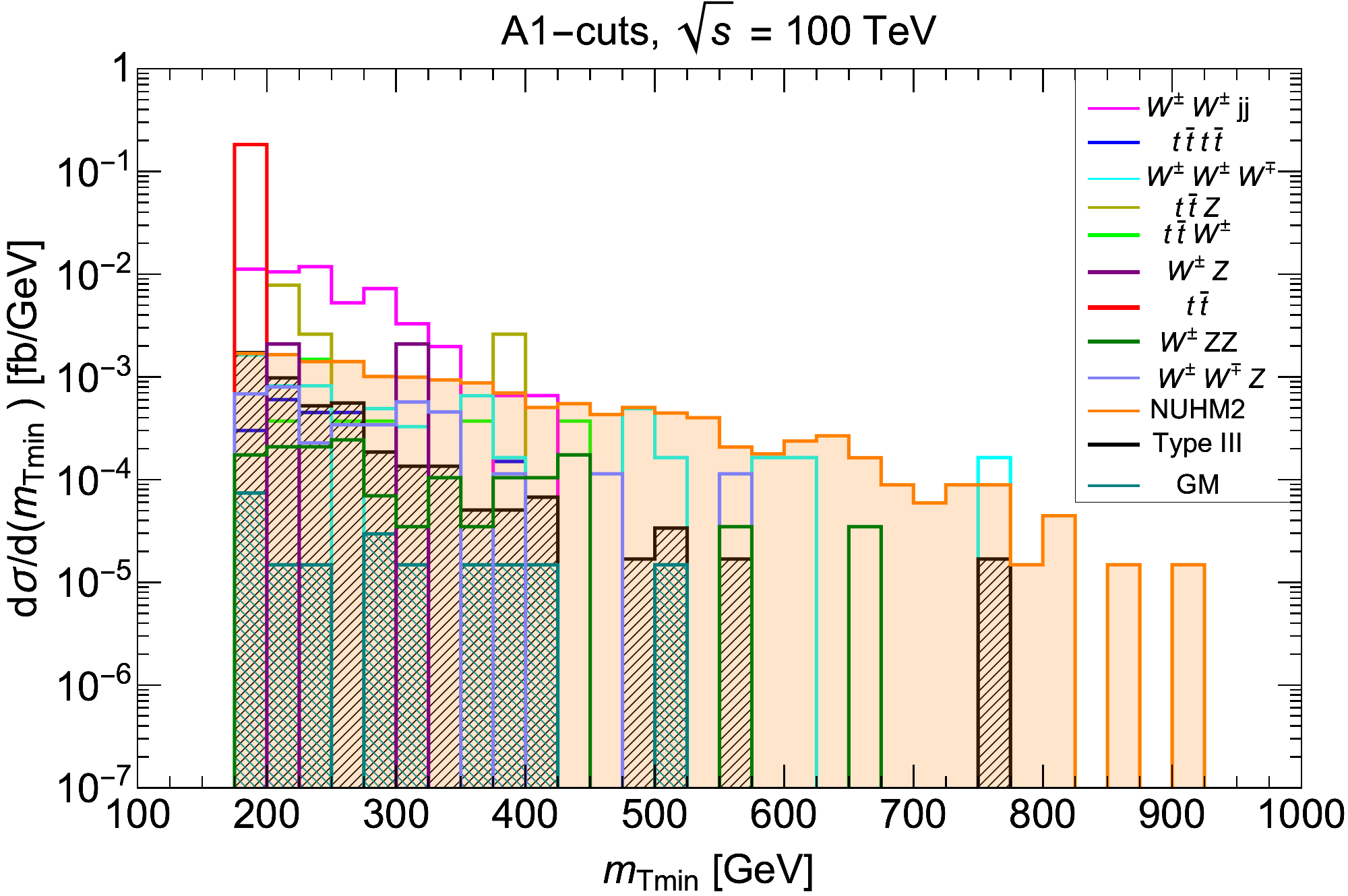}
  \caption{}
  \label{fig:mtmina1susy100}
\end{subfigure}
\vspace*{-0.1in}
\caption{$m_{T_{\rm min}}$ distribution after A1-cuts for (a) $\sqrt{s}=27$~TeV and (b) $\sqrt{s}=100$~TeV.}
\end{figure}

In the NUHM2 signal, the LSP $\tilde{Z}_1$, due to R-parity conservation, is stable and shows up as $\slashed{E}_{T}$ at the LHC.  The particles $\tilde{Z}_2$ and $\tilde{W}_1^{\pm}$, being not much heavier than the LSP, promptly decay into very soft leptons and the LSP.  These leptons are so soft that they pass the detector undetected.  Hence, $\tilde{Z}_2$ and $\tilde{W}_1^{\pm}$ also show up as $\slashed{E}_{T}$ at the LHC.  Hence the NUHM2 signal has large missing transverse energy ($\slashed{E}_{T}$). Similarly, the NUHM2 signal also has large minimum transverse mass ($m_{T_{\rm min}}$) which is defined as:
\begin{equation}
    m_{T_{\rm min}} = {\rm min}  (m_T(\ell_1, \slashed{E}_{T}), m_T(\ell_2, \slashed{E}_{T}))
\end{equation}
It turns out that we cannot gain a sufficient cross section for $\sqrt{s}=14$~TeV.  Therefore, we extend the analysis to $\sqrt{s}=27$~TeV. Although $\sqrt{s}=27$~TeV is sufficient to render a significance above 5$\sigma$, we also show an analysis for $\sqrt{s}=100$~TeV.  Inspired by the earlier analyses done in this context~\cite{Baer:2013yha, Baer:2017gzf}, we further impose the following set of cuts: 
\begin{itemize}
    \item Require $\slashed{E}_{T} > 200$~GeV.
    \item Require $m_{T_{\rm min}} > 175$~GeV.  
\end{itemize}
Together with the S1-cuts, we call the entire set of cuts as the A1-cuts.  After the A1-cuts, we plot the $m_{T_{\rm min}}$ distribution in Fig.~\ref{fig:mtmina1susy27} for $\sqrt{s}=27$~TeV  and in Fig.~\ref{fig:mtmina1susy100} for $\sqrt{s}=100$~TeV.

%\begin{figure}[h!]
%\centering
%\includegraphics [height=0.35\textheight] {mtm_a1.png}
%\vspace*{-0.1in}
%\caption{$m_{T_{\rm min}}$ distribution after A1-cuts.}
%\label{fig:mtmina1susy27}
%\end{figure}

As suggested by Fig.~\ref{fig:mtmina1susy27}, a cut of $m_{T_{\rm min}} > 200$~GeV, if employed after the A1-cuts, would reduce the SM background to some extent. Therefore, we have the A2-cuts:
\begin{itemize}
    \item A1-cuts + $m_{T_{\rm min}} > 200$~GeV.
\end{itemize}

Similarly, Fig.~\ref{fig:mtmina1susy100} suggests that a cut of $m_{T_{\rm min}} > 325$~GeV would be highly beneficial in significantly reducing the SM background at $\sqrt{s}=100$~TeV. Therefore, we have the A2$^{\prime}$-cuts:
\begin{itemize}
    \item A1-cuts + $m_{T_{\rm min}} > 325$~GeV.
\end{itemize}

\begin{figure}[h!]
\centering
\begin{subfigure}[h]{0.5\textwidth}
  \centering
  \includegraphics[width=1\linewidth]{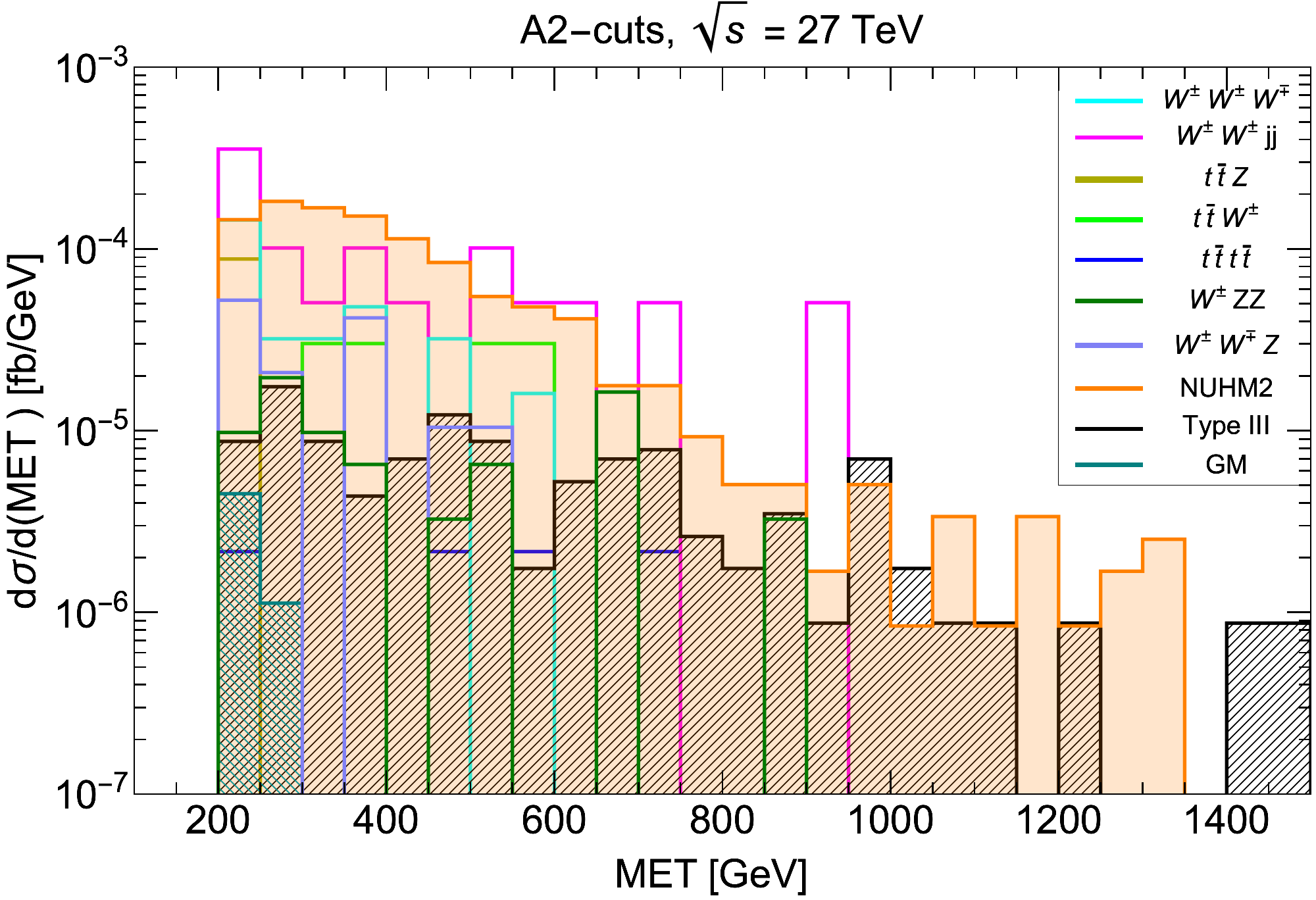}
  \caption{}
  \label{fig:meta2susy27}
\end{subfigure}%
\begin{subfigure}[h]{0.5\textwidth}
  \centering
  \includegraphics[width=1\linewidth]{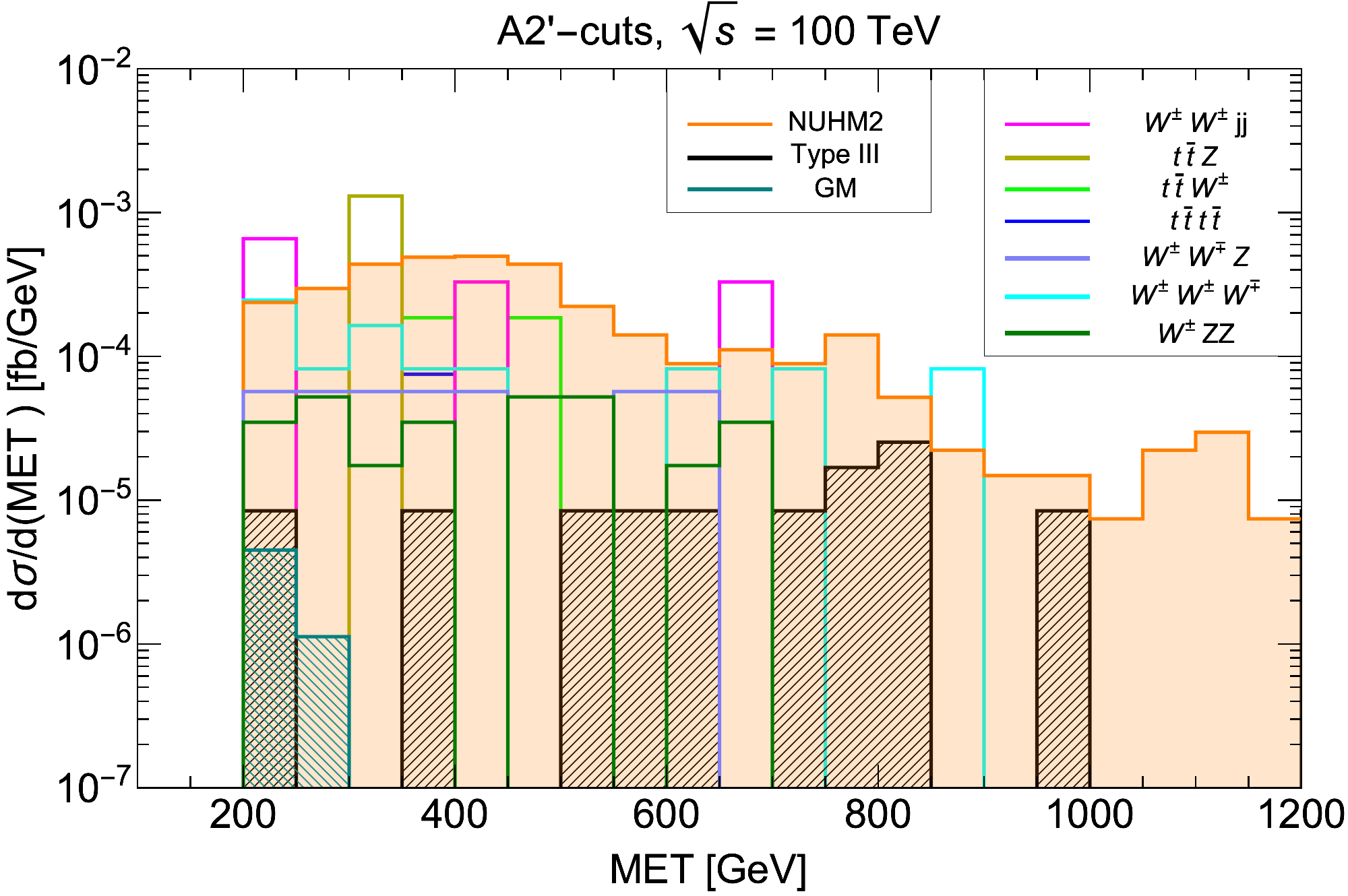}
  \caption{}
  \label{fig:meta2susy100}
\end{subfigure}
\vspace*{-0.1in}
\caption{$\slashed{E}_{T}$ distribution after (a)  A2-cuts at  $\sqrt{s}=27$~TeV and (b) A2$^{\prime}$-cuts at $\sqrt{s}=100$~TeV.}
\end{figure}

After the A2-cuts and A2$^{\prime}$-cuts, we plot the $\slashed{E}_{T}$ distribution for $\sqrt{s}=27$~TeV and $\sqrt{s}=100$~TeV in Fig.~\ref{fig:meta2susy27} and Fig.~\ref{fig:meta2susy100}, respectively.

As shown in Fig.~\ref{fig:meta2susy27} (\ref{fig:meta2susy100}), a cut of $\slashed{E}_{T} > 250$~GeV ($\slashed{E}_{T} > 350$~GeV), applied after the A2 (A2$^{\prime}$)-cuts, would result in a cleaner NUHM2 signal with efficiently reduced SM background as well as heavily reduced signal cross sections for the other two BSM models at $\sqrt{s}=27$~TeV ($100$~TeV).  Therefore, we finally have the A3-cuts:
\begin{itemize}
    \item A2-cuts + $\slashed{E}_{T} > 250$~GeV.
\end{itemize}
and the A3$^{\prime}$-cuts:
\begin{itemize}
    \item A2$^{\prime}$-cuts + $\slashed{E}_{T} > 350$~GeV.
\end{itemize}

The cut flow is summarized in Table~\ref{susy 27tev}.

\begin{table}[h!]
\centering
\resizebox{\columnwidth}{!}{%
\begin{tabular}{|l|l|llllll|}
\hline
\hline
                        &          & \multicolumn{6}{c|}{$\sqrt{s}=27$~TeV}                                                                                    \\ \hline 
\multirow{2}{*}{Process}                & \multirow{2}{*}{K-factor} & \multicolumn{1}{l|}{\multirow{2}{*}{$\sigma$ (NLO) {[}ab{]}}} & \multicolumn{1}{l|}{\multirow{2}{*}{A1 {[}ab{]}}} & \multicolumn{1}{l|}{\multirow{2}{*}{A2 {[}ab{]}}} & \multicolumn{1}{l|}{\multirow{2}{*}{A3 {[}ab{]}}} & \multicolumn{2}{c|}{Significance} \\ \cline{7-8} 
\multicolumn{1}{|l|}{} & \multicolumn{1}{|l|}{} & \multicolumn{1}{|l|}{} & \multicolumn{1}{|l|}{} & \multicolumn{1}{|l|}{} & \multicolumn{1}{|l|}{} & \multicolumn{1}{c|}{$\mathcal{L} = $ 3 ab$^{-1}$} & \multicolumn{1}{c|}{$\mathcal{L} = $ 15 ab$^{-1}$} \\ \hline
NUHM2                   & 1.17     & \multicolumn{1}{l|}{$4.2 \cdot 10^4$}        & \multicolumn{1}{l|}{60.9}        & \multicolumn{1}{l|}{53.3}        & \multicolumn{1}{l|}{46.1}        & \multicolumn{1}{l|}{$8.06$ ($7.8$) }                                       &      $18.01$ ($15.4$)                                     \\ 
type-III                & 1.16     & \multicolumn{1}{l|}{$4.36 \cdot 10^4$}       & \multicolumn{1}{l|}{9.33}        & \multicolumn{1}{l|}{5.7}         & \multicolumn{1}{l|}{5.3}         & \multicolumn{1}{l|}{$1.21$ ($1.14$)}                                         &          $2.71$ ($2.12$)                      \\ 
GM                      & 1.26     & \multicolumn{1}{l|}{$5.6 \cdot 10^4$}        & \multicolumn{1}{l|}{0.28}        & \multicolumn{1}{l|}{0.28}        & \multicolumn{1}{l|}{0.0562}      & \multicolumn{1}{l|}{$0.0135$ ($0.013$)}                                       &     $0.03$ ($0.023$)                            \\\hline
$t\bar{t}$              & 1.72     & \multicolumn{1}{l|}{$4.1 \cdot 10^9$}        & \multicolumn{1}{l|}{0}           & \multicolumn{1}{l|}{0}           & \multicolumn{1}{l|}{0}           & \multicolumn{1}{l|}{-}                                           & -        \\ 
$t\bar{t}t\bar{t}$      & 1.27     & \multicolumn{1}{l|}{$1.1 \cdot 10^5$}        & \multicolumn{1}{l|}{0.8}         & \multicolumn{1}{l|}{0.4}         & \multicolumn{1}{l|}{0.3}         & \multicolumn{1}{l|}{-}                                           & -    \\ 
$t\bar{t}W^{\pm}$       & 1.24     & \multicolumn{1}{l|}{$1.5 \cdot 10^6$}        & \multicolumn{1}{l|}{12.1}        & \multicolumn{1}{l|}{6.03}        & \multicolumn{1}{l|}{6.03}        & \multicolumn{1}{l|}{-}                                           & -     \\ 
$t\bar{t}Z$             & 1.39     & \multicolumn{1}{l|}{$4.4 \cdot 10^6$}        & \multicolumn{1}{l|}{17.6}        & \multicolumn{1}{l|}{4.4}         & \multicolumn{1}{l|}{0}           & \multicolumn{1}{l|}{-}                                           & -    \\ 
$W^{\pm}W^{\pm}jj$      & 1.04     & \multicolumn{1}{l|}{$2.5 \cdot 10^6$}        & \multicolumn{1}{l|}{78.4}        & \multicolumn{1}{l|}{48.04}       & \multicolumn{1}{l|}{30.3}        & \multicolumn{1}{l|}{-}                                           & -   \\ 
$W^{\pm}W^{\pm}W^{\mp}$ & 2.45     & \multicolumn{1}{l|}{$8.0 \cdot 10^5$}        & \multicolumn{1}{l|}{20.8}        & \multicolumn{1}{l|}{15.2}        & \multicolumn{1}{l|}{8.0}         & \multicolumn{1}{l|}{-}                                           & -   \\ 
$W^{\pm}Z$              & 1.88     & \multicolumn{1}{l|}{$1.2 \cdot 10^8$}        & \multicolumn{1}{l|}{0}           & \multicolumn{1}{l|}{0}           & \multicolumn{1}{l|}{0}           & \multicolumn{1}{l|}{-}                                           & -    \\ 
$ZZ$              & 1.7     & \multicolumn{1}{l|}{$4.1 \cdot 10^7$}        & \multicolumn{1}{l|}{0}           & \multicolumn{1}{l|}{0}           & \multicolumn{1}{l|}{0}           & \multicolumn{1}{l|}{-}                                           & -  \\ 
$W^{\pm}W^{\mp}Z$              & 2.0    & \multicolumn{1}{l|}{$5.2 \cdot 10^5$}        & \multicolumn{1}{l|}{8}           & \multicolumn{1}{l|}{7}           & \multicolumn{1}{l|}{4.2}           & \multicolumn{1}{l|}{-}                                           & -   \\
$W^{\pm}ZZ$              & 2.0     & \multicolumn{1}{l|}{$1.6 \cdot 10^5$}        & \multicolumn{1}{l|}{4.7}           & \multicolumn{1}{l|}{3.7}           & \multicolumn{1}{l|}{3.2}           & \multicolumn{1}{l|}{-}                                           & -  \\\hline
Total BG                & $-$      & \multicolumn{1}{l|}{$4.3 \cdot 10^9$}        & \multicolumn{1}{l|}{142.4}       & \multicolumn{1}{l|}{84.77}       & \multicolumn{1}{l|}{52.1}        & \multicolumn{1}{l|}{-}                                           & -  \\ \hline \hline
                        &          & \multicolumn{6}{c|}{$\sqrt{s}=100$~TeV}                                                                                    \\ \hline 
\multirow{2}{*}{Process}                 & \multirow{2}{*}{K-factor}  & \multicolumn{1}{l|}{\multirow{2}{*}{$\sigma$ (NLO) {[}ab{]}}} & \multicolumn{1}{l|}{\multirow{2}{*}{A1 {[}ab{]}}} & \multicolumn{1}{l|}{\multirow{2}{*}{A2$^{\prime}$ {[}ab{]}}} & \multicolumn{1}{l|}{\multirow{2}{*}{A3$^{\prime}$ {[}ab{]}}} & \multicolumn{2}{c|}{Significance} \\ \cline{7-8}
\multicolumn{1}{|l|}{} & \multicolumn{1}{|l|}{} & \multicolumn{1}{|l|}{} & \multicolumn{1}{|l|}{} & \multicolumn{1}{|l|}{} & \multicolumn{1}{|l|}{} & \multicolumn{1}{c|}{$\mathcal{L} = $ 3 ab$^{-1}$} & \multicolumn{1}{c|}{$\mathcal{L} = $ 15 ab$^{-1}$} \\ \hline
NUHM2                   & 1.17     & \multicolumn{1}{l|}{$3.71\cdot 10^5$}        & \multicolumn{1}{l|}{374.4}       & \multicolumn{1}{l|}{170.5}                  & \multicolumn{1}{l|}{121.9}                  & \multicolumn{1}{l|}{$13.6$ ($12.7$)}           &                $30.5$ ($22.9$)                        \\ 
type-III                & 1.16     & \multicolumn{1}{l|}{$4.2\cdot 10^5$}         & \multicolumn{1}{l|}{112.55}      & \multicolumn{1}{l|}{10.1}                   & \multicolumn{1}{l|}{9.7}                    & \multicolumn{1}{l|}{$1.5$ ($1.3$)}     &               $3.3$ ($2.1$)                     \\ 
GM                      & 1.26     & \multicolumn{1}{l|}{$3.7\cdot 10^5$}         & \multicolumn{1}{l|}{5.2}         & \multicolumn{1}{l|}{1.5}                    & \multicolumn{1}{l|}{0.4}                    & \multicolumn{1}{l|}{$0.064$ ($0.056$)}     &          $0.14$ ($0.1$)                           \\ \hline
$t\bar{t}$              & 1.72   & \multicolumn{1}{l|}{$4.6\cdot 10^{10}$}        & \multicolumn{1}{l|}{4570.0}      & \multicolumn{1}{l|}{0}                      & \multicolumn{1}{l|}{0}                      & \multicolumn{1}{l|}{-}                                           & -                                            \\ 
$t\bar{t}t\bar{t}$      & 1.27     & \multicolumn{1}{l|}{$3.75\cdot 10^6$}        & \multicolumn{1}{l|}{48.8}        & \multicolumn{1}{l|}{4.0}                    & \multicolumn{1}{l|}{4.0}                    & \multicolumn{1}{l|}{-}                                           & -                                            \\ 
$t\bar{t}W^{\pm}$       & 1.24      & \multicolumn{1}{l|}{$9.3\cdot 10^6$}         & \multicolumn{1}{l|}{83.5}        & \multicolumn{1}{l|}{18.5}                   & \multicolumn{1}{l|}{18.5}                   & \multicolumn{1}{l|}{-}                                           & -                                            \\ 
$t\bar{t}Z$             & 1.39     & \multicolumn{1}{l|}{$6.51\cdot 10^7$}        & \multicolumn{1}{l|}{325.5}       & \multicolumn{1}{l|}{65.1}                   & \multicolumn{1}{l|}{0}                      & \multicolumn{1}{l|}{-}                                           & -                                            \\ 
$W^{\pm}W^{\pm}jj$      & 1.04     & \multicolumn{1}{l|}{$1.6\cdot 10^7$}         & \multicolumn{1}{l|}{1315.2}      & \multicolumn{1}{l|}{82.0}                   & \multicolumn{1}{l|}{49.3}                   & \multicolumn{1}{l|}{-}                                           & -                                            \\ 
$W^{\pm}W^{\pm}W^{\mp}$ & 2.45     & \multicolumn{1}{l|}{$4.1\cdot 10^6$}         & \multicolumn{1}{l|}{151.5}       & \multicolumn{1}{l|}{49.0}                   & \multicolumn{1}{l|}{24.6}                   & \multicolumn{1}{l|}{-}                                           & -                                            \\ 
$W^{\pm}Z$              & 1.88      & \multicolumn{1}{l|}{$5.2\cdot 10^8$}         & \multicolumn{1}{l|}{104.5}       & \multicolumn{1}{l|}{0}                      & \multicolumn{1}{l|}{0}                      & \multicolumn{1}{l|}{-}                                           & -                                            \\ 
$ZZ$              & 1.7  & \multicolumn{1}{l|}{$1.8 \cdot 10^8$}         & \multicolumn{1}{l|}{0}       & \multicolumn{1}{l|}{0}                      & \multicolumn{1}{l|}{0}                      & \multicolumn{1}{l|}{-}                                           & -                                            \\ 
$W^{\pm}W^{\mp}Z$              & 2.0   & \multicolumn{1}{l|}{$2.8 \cdot 10^6$}         & \multicolumn{1}{l|}{94}       & \multicolumn{1}{l|}{20}                      & \multicolumn{1}{l|}{11.4}                      & \multicolumn{1}{l|}{-}                                           & -                                            \\
$W^{\pm}ZZ$              & 2.0   & \multicolumn{1}{l|}{$8.7 \cdot 10^5$}         & \multicolumn{1}{l|}{38.3}       & \multicolumn{1}{l|}{14.8}                      & \multicolumn{1}{l|}{9.6}                      & \multicolumn{1}{l|}{-}                                           & -                                            \\\hline
Total BG                & $-$      & \multicolumn{1}{l|}{$4.72 \cdot 10^{10}$}        & \multicolumn{1}{l|}{6731.3}      & \multicolumn{1}{l|}{253.4}                  & \multicolumn{1}{l|}{117.4}                   & \multicolumn{1}{l|}{-}                                           & -                                            \\ \hline \hline
\end{tabular}%
}
\caption{Cut flow table for cleaner NUHM2 signal.}
\label{susy 27tev}
\end{table}

After the A3 and A3$^{\prime}$-cuts for $\sqrt{s}=27$~TeV and $100$~TeV, respectively, a sufficient NUHM2 signal cross section is retained, while the SM background and signals from the other two BSM models are greatly reduced.

The statistical significance of the signal has been computed using the relation $S/\sqrt{S+B}$.  We have shown the significance for center of mass energy ($\sqrt{s}$) at $27$~TeV and $100$~TeV with an integrated luminosity $\mathcal{L}$ = 3~ab$^{-1}$ and 15~ab$^{-1}$ in Table~\ref{susy 27tev}. Note that, on calculating the significance at $\sqrt{s}=14$~TeV and $\mathcal{L}$ = 3~ab$^{-1}$, the NUHM2 BM point chosen here yields $S/\sqrt{S+B} = 4.7$. This result is different from that obtained in an earlier analysis in Ref.~\cite{Baer:2017gzf} mainly due to differences in the detector-level simulations in these two cases. %\footnote{The reader might observe a difference between our current SUSY analysis and an earlier analysis in Ref.~\cite{Baer:2017gzf}. The main cause of difference is that the detector-level simulations are very different in these two cases. On calculating the significance at $\sqrt{s}=14~$ TeV and $\mathcal{L}$ = 3~ab$^{-1}$, the NUHM2 BM point chosen here yields $S/\sqrt{S+B} = 4.7$.} 
 Moreover, in order to show the impact of systematic uncertainties, we compute the significance  considering 3$\%$ systematic uncertainties using the relation $S/\sqrt{S+B+\Delta^2B^2}$ with $\Delta = 0.03$. We find that the significance drops down while we consider systematic uncertainties, as shown in parenthesis in Table~\ref{susy 27tev}. From Table~\ref{susy 27tev}, we can see that after the A3 and A3$^{\prime}$-cuts for $\sqrt{s}=27$~TeV and $100$~TeV, respectively, the NUHM2 signal yields a significance above 5$\sigma$ while the type-III and GM models don't, with or without considering the 3$\%$ systematic uncertainties. 

\begin{figure}[t]
\centering
\begin{subfigure}[t]{0.5\textwidth}
  \centering
  \includegraphics[width=1\linewidth]{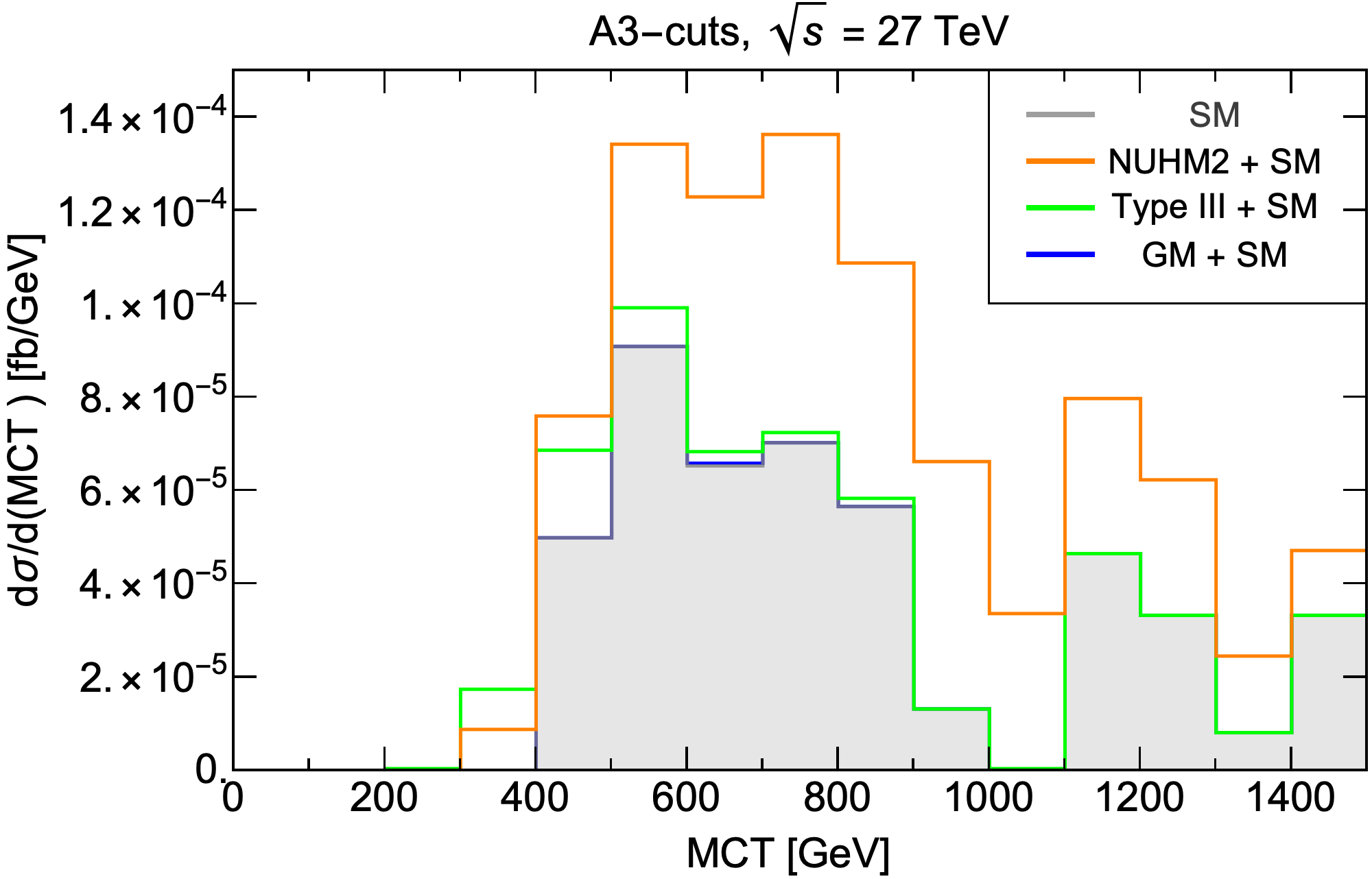}
  \caption{}
  \label{fig:mcta3susy}
\end{subfigure}%
\begin{subfigure}[t]{0.5\textwidth}
  \centering
  \includegraphics[width=1\linewidth]{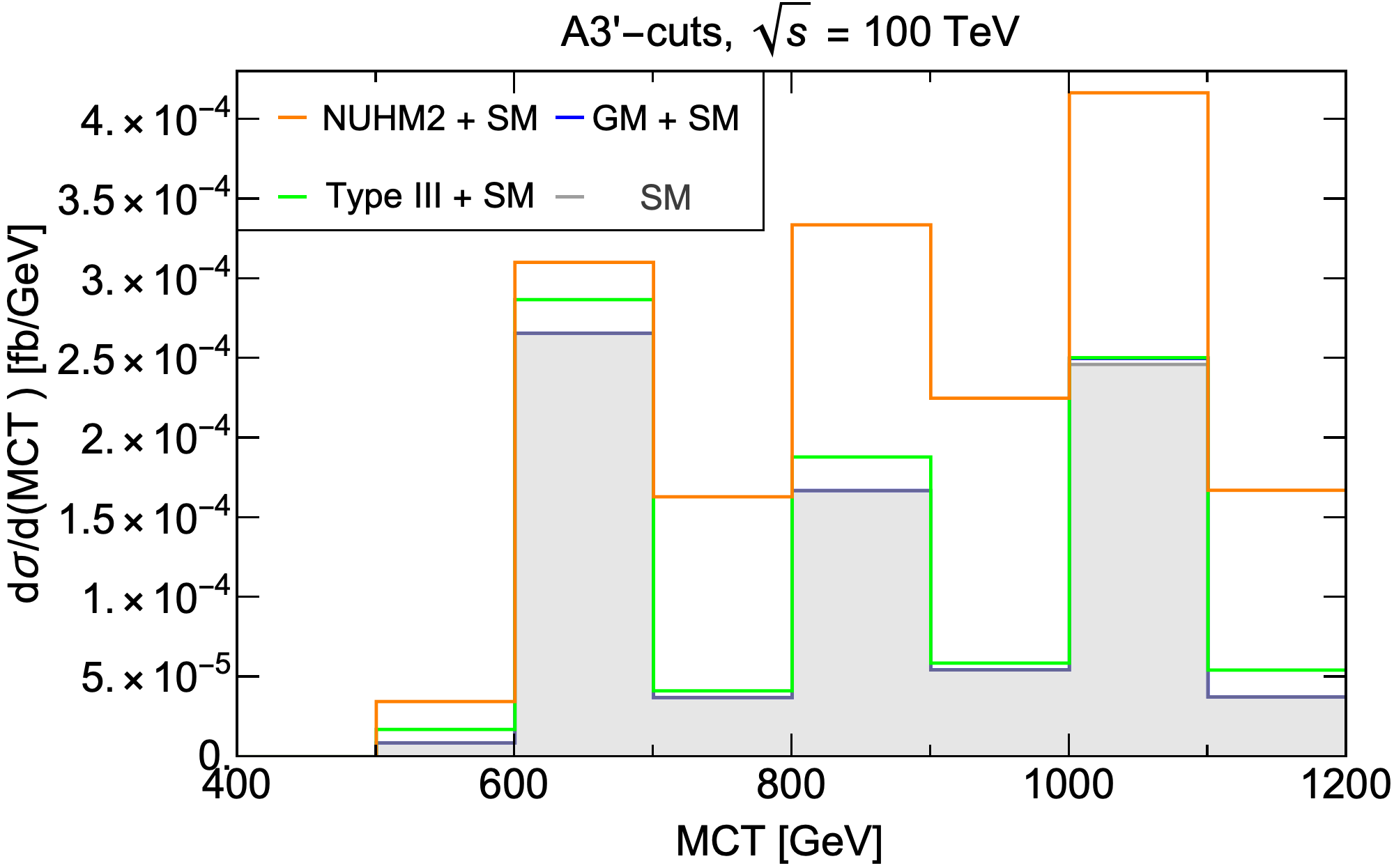}
  \caption{}
   \label{fig:mcta3susy100}
\end{subfigure}
\begin{subfigure}[t]{0.5\textwidth}
  \centering
  \includegraphics[width=1\linewidth]{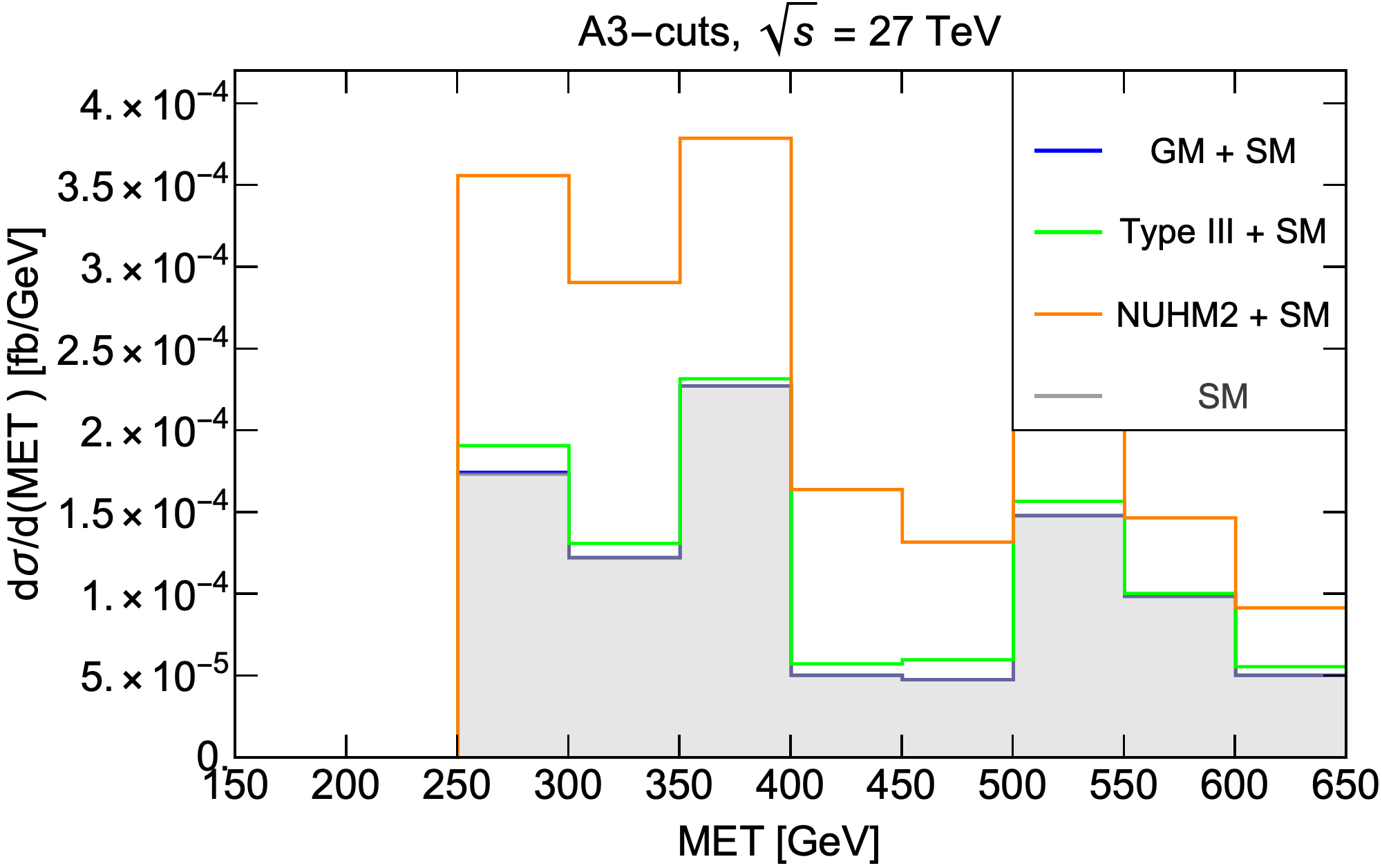}
  \caption{}
  \label{fig:meta3susy}
\end{subfigure}%
\begin{subfigure}[t]{0.5\textwidth}
  \centering
  \includegraphics[width=1\linewidth]{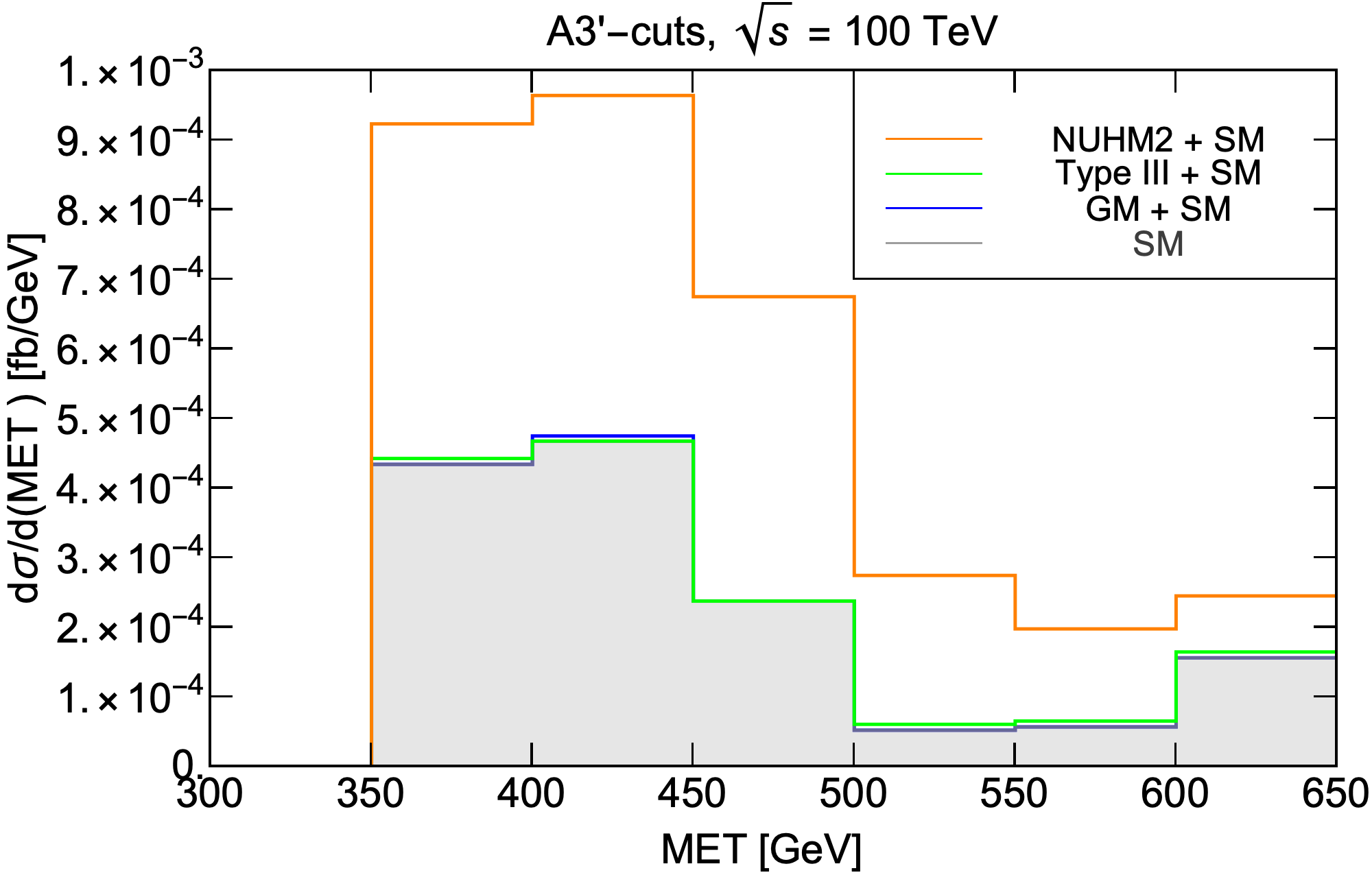}
  \caption{}
  \label{fig:meta3susy100}
\end{subfigure}
\vspace*{-0.1in}
\caption{MCT distribution after (a) A3-cuts at  $\sqrt{s}=27$~TeV and (b) A3$^{\prime}$-cuts at $\sqrt{s}=100$~TeV; and $\slashed{E}_{T}$ distribution after (c) A3-cuts at  $\sqrt{s}=27$~TeV and (d) A3$^{\prime}$-cuts at $\sqrt{s}=100$~TeV.}
\label{fig:a3}
\end{figure}

We have plotted the cluster transverse mass (MCT) distribution and the $\slashed{E}_{T}$ distribution after A3 and A3$^{\prime}$-cuts at the respective energies for the total SM background and various signals on top of it in Fig.~\ref{fig:a3}.

\clearpage
%%%%%%%%%%%%%%%%%%%%%%%%%%%%%%%%%%%%%%%%%%%%%%%
\subsection{Type-III Seesaw Analysis}
\label{ssec:type3analysis}
%%%%%%%%%%%%%%%%%%%%%%%%%%%%%%%%%%%%%%%%%%%%%%%
We cannot gain a sufficient cross section for $\sqrt{s}=14$~TeV for the type-III seesaw signal either, and therefore consider $\sqrt{s}=27$~TeV and $\sqrt{s}=100$~TeV.  We start with S1-cuts as defined earlier.

It is expected that after the S1-cuts all the SM backgrounds, considered here, should have numerous jets while the type-III seesaw signal can have jets only from initial state QCD radiation. Therefore, requiring only those events that have less than two jets would significantly reduce the SM background while retaining enough type-III seesaw signal. 

Similar to the NUHM2 signal, the type-III signal will also show high $\slashed{E}_{T}$. However, since in type-III signal, the mass difference between the intermediate state and the final state is not as high as that in the NUHM2 signal, hence the $\slashed{E}_{T}$ distribution does not tail out as high as the NUHM2 signal. Therefore, for $\sqrt{s}=27$~TeV ($100$~TeV), we apply a cut of $\slashed{E}_{T} > 100$~GeV ($> 120$~GeV), a less stringent cut on $\slashed{E}_{T}$ as compared to that applied in case of NUHM2 signal. 

For the same reason as mentioned above, we employ a less stringent cut on  $m_{T_{\rm min}}$ as well: a cut of 105~GeV~$< m_{T_{\rm min}} < 195$~GeV. 

A cut on the upper limit of $m_{T_{\rm min}}$ has been applied in order to reduce the NUHM2 signal and yet retain most of the type-III signal as for NUHM2 signal the $m_{T_{\rm min}}$ distribution tails out to a much higher value than that in the type-III signal. The cut on the upper limit of $m_{T_{\rm min}}$ is thus necessary to differentiate between the NUHM2 and the type-III signal. Therefore after S1-cuts we apply three additional cuts, namely, njet $\le 1$, $\slashed{E}_{T} > 100$~GeV and  105~GeV~$< m_{T_{\rm min}} < 195$~GeV and name this entire set of cut as B1-cuts:

\begin{itemize}
    \item S1-cuts + njet $\le 1$ + $\slashed{E}_{T} > 100$~GeV + 105~GeV~$< m_{T_{\rm min}} < 195$~GeV.
\end{itemize}

At $\sqrt{s}=100$~TeV, we use a slightly tougher cut on $\slashed{E}_{T}$ and apply B1$^{\prime}$-cuts, which include: 

\begin{itemize}
    \item S1-cuts + njet $\le 1$ + $\slashed{E}_{T} > 120$~GeV + 105~GeV~$< m_{T_{\rm min}} < 195$~GeV.
\end{itemize}

\begin{figure}[h]
\centering
\begin{subfigure}[h]{0.5\textwidth}
  \centering
  \includegraphics[width=1\linewidth]{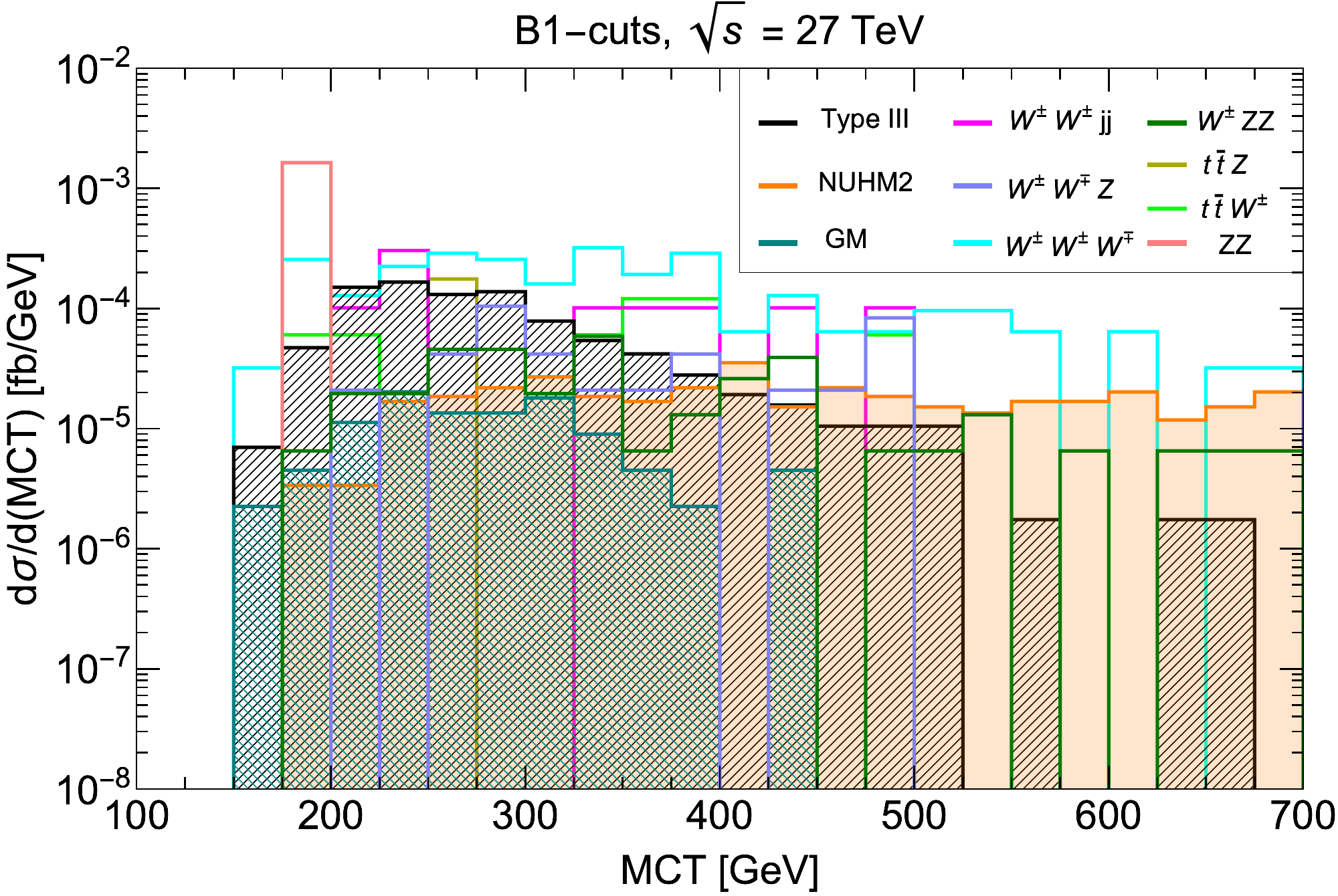}
  \caption{}
  \label{fig:mctb1type3}
\end{subfigure}%
\begin{subfigure}[h]{0.5\textwidth}
  \centering
  \includegraphics[width=1\linewidth]{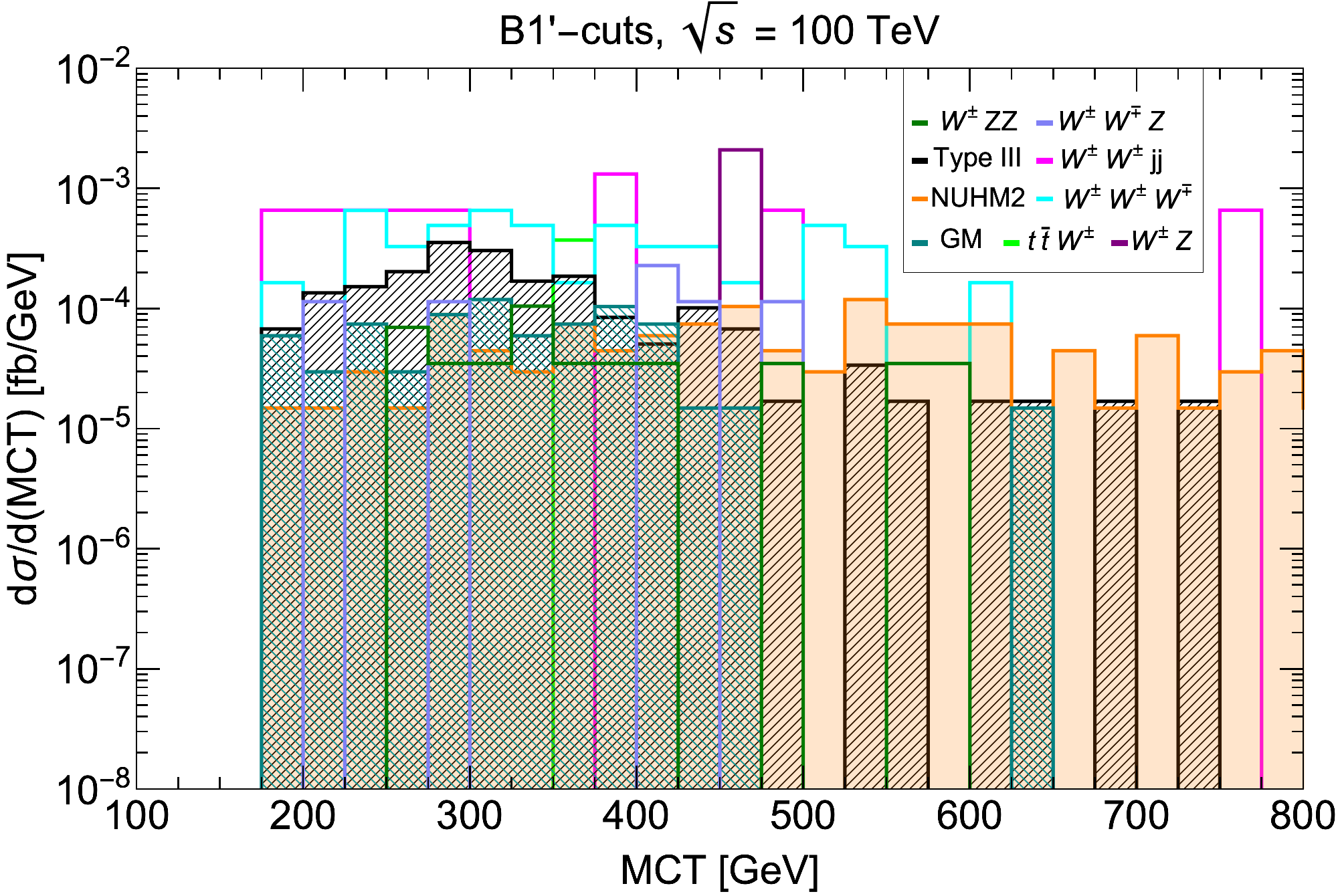}
  \caption{}
  \label{fig:mctb1prime}
\end{subfigure}
\vspace*{-0.1in}
\caption{MCT distribution after (a) B1-cuts for $\sqrt{s}=27$~TeV (b) B1$^{\prime}$-cuts for $\sqrt{s}=100$~TeV .}
\end{figure}

After the B1-cuts and B1$^{\prime}$-cuts for $\sqrt{s}=27$~TeV and $100$~TeV, respectively, the MCT distribution is plotted in Fig.~\ref{fig:mctb1type3} and in Fig.~\ref{fig:mctb1prime}. From Fig.~\ref{fig:mctb1type3} we see that a cut of 200~GeV < MCT < 325~GeV can further  reduce the SM background and NUHM2 signal as well while retaining enough type-III signals to be visible. Therefore, next we apply the B2-cuts defined as:
\begin{itemize}
    \item B1-cuts + 200~GeV < MCT < 325~GeV.
\end{itemize}

Similarly, for $\sqrt{s}=100$~TeV, Fig.~\ref{fig:mctb1prime} shows that a suitable cut would be 200~GeV < MCT < 350~GeV. Therefore, next we apply the B2$^{\prime}$-cuts defined as:
\begin{itemize}
    \item B1$^{\prime}$-cuts + 200~GeV < MCT < 350~GeV.
\end{itemize}

The cut flow for this scenario is summarized in Table~\ref{type3 27tev}.

\begin{table}[h!]
\centering
\resizebox{\columnwidth}{!}{%
\begin{tabular}{|l|l|lllll|}
\hline
\hline
                   &         & \multicolumn{5}{c|}{$\sqrt{s}=27$~TeV}                                                                                                                                    \\ \hline 
\multirow{2}{*}{Process}                & \multirow{2}{*}{K-factor} & \multicolumn{1}{l|}{\multirow{2}{*}{$\sigma$ (NLO) {[}ab{]}}} & \multicolumn{1}{l|}{\multirow{2}{*}{B1 {[}ab{]}}} & \multicolumn{1}{l|}{\multirow{2}{*}{B2 {[}ab{]}}}  & \multicolumn{2}{c|}{Significance} \\ \cline{6-7} 
\multicolumn{1}{|l|}{} & \multicolumn{1}{|l|}{} & \multicolumn{1}{|l|}{} & \multicolumn{1}{|l|}{} & \multicolumn{1}{|l|}{} &  \multicolumn{1}{c|}{$\mathcal{L} = $ 3 ab$^{-1}$} & \multicolumn{1}{c|}{$\mathcal{L} = $ 15 ab$^{-1}$} \\ \hline
NUHM2                   & 1.17     & \multicolumn{1}{l|}{$4.2 \cdot 10^4$}        & \multicolumn{1}{l|}{13.2}        & \multicolumn{1}{l|}{2.2}              & \multicolumn{1}{l|}{$0.52$ ($0.5$)}                                        & $1.2$ ($0.9$)                                       \\ 
type-III                & 1.16     & \multicolumn{1}{l|}{$4.36 \cdot 10^4$}       & \multicolumn{1}{l|}{22.8}        & \multicolumn{1}{l|}{16.6}               & \multicolumn{1}{l|}{$3.5$ ($3.3$)}                                         & $7.8$ ($6.3$)                                        \\ 
GM                      & 1.26     & \multicolumn{1}{l|}{$5.6 \cdot 10^4$}        & \multicolumn{1}{l|}{2.6}        & \multicolumn{1}{l|}{1.9}            & \multicolumn{1}{l|}{$0.45$ ($0.42$)}                                       & $1.0$ ($0.8$)                                      \\ \hline
$t\bar{t}$              & 1.72     & \multicolumn{1}{l|}{$4.1 \cdot 10^9$}        & \multicolumn{1}{l|}{0}           & \multicolumn{1}{l|}{0}                     & \multicolumn{1}{l|}{-}                                           & -                                            \\ 
$t\bar{t}t\bar{t}$      & 1.27     & \multicolumn{1}{l|}{$1.1 \cdot 10^5$}        & \multicolumn{1}{l|}{0}         & \multicolumn{1}{l|}{0}                 & \multicolumn{1}{l|}{-}                                           & -                                             \\ 
$t\bar{t}W^{\pm}$       & 1.24     & \multicolumn{1}{l|}{$1.5 \cdot 10^6$}        & \multicolumn{1}{l|}{12.1}        & \multicolumn{1}{l|}{1.5}               & \multicolumn{1}{l|}{-}                                           & -                                            \\ 
$t\bar{t}Z$             & 1.39     & \multicolumn{1}{l|}{$4.4 \cdot 10^6$}               & \multicolumn{1}{l|}{4.4}         & \multicolumn{1}{l|}{4.4 }           & \multicolumn{1}{l|}{-}                                           & -                                             \\ 
$W^{\pm}W^{\pm}jj$      & 1.04     & \multicolumn{1}{l|}{$2.5 \cdot 10^6$}        & \multicolumn{1}{l|}{25.3}        & \multicolumn{1}{l|}{10.1}              & \multicolumn{1}{l|}{-}                                           & -                                             \\ 
$W^{\pm}W^{\pm}W^{\mp}$ & 2.45     & \multicolumn{1}{l|}{$8.0 \cdot 10^5$}        & \multicolumn{1}{l|}{76.0 }        & \multicolumn{1}{l|}{26.4}                 & \multicolumn{1}{l|}{-}                                           & -                                             \\ 
$W^{\pm}Z$              & 1.88     & \multicolumn{1}{l|}{$1.2 \cdot 10^8$}        & \multicolumn{1}{l|}{0}           & \multicolumn{1}{l|}{0}                      & \multicolumn{1}{l|}{-}                                           & -                                             \\ 
$ZZ$              & 1.7     & \multicolumn{1}{l|}{$4.1 \cdot 10^7$}        & \multicolumn{1}{l|}{41}           & \multicolumn{1}{l|}{0}                  & \multicolumn{1}{l|}{-}                                           & -                                            \\ 
$W^{\pm}W^{\mp}Z$              & 2.0    & \multicolumn{1}{l|}{$5.2 \cdot 10^5$}        & \multicolumn{1}{l|}{10.9}           & \multicolumn{1}{l|}{5.2}                     & \multicolumn{1}{l|}{-}                                           & -                                             \\
$W^{\pm}ZZ$              & 2.0     & \multicolumn{1}{l|}{$1.6 \cdot 10^5$}        & \multicolumn{1}{l|}{9.8}           & \multicolumn{1}{l|}{4}                      & \multicolumn{1}{l|}{-}                                           & -                                            \\\hline
Total BG                & $-$      & \multicolumn{1}{l|}{$4.3 \cdot 10^9$}        & \multicolumn{1}{l|}{179.5}       & \multicolumn{1}{l|}{51.6}             & \multicolumn{1}{l|}{-}                                           & -                                             \\ \hline \hline
                   &         & \multicolumn{5}{c|}{$\sqrt{s}=100$~TeV}                                                                                                                                    \\ \hline 
\multirow{2}{*}{Process}                 & \multirow{2}{*}{K-factor} & \multicolumn{1}{l|}{\multirow{2}{*}{$\sigma$ (NLO) {[}ab{]}}} &  \multicolumn{1}{l|}{\multirow{2}{*}{B1$^{\prime}$ {[}ab{]}}} & \multicolumn{1}{l|}{\multirow{2}{*}{B2$^{\prime}$ {[}ab{]}}} & \multicolumn{2}{c|}{Significance} \\ \cline{6-7} 
\multicolumn{1}{|l|}{} & \multicolumn{1}{|l|}{} & \multicolumn{1}{|l|}{} & \multicolumn{1}{|l|}{} & \multicolumn{1}{|l|}{} &  \multicolumn{1}{c|}{$\mathcal{L} = $ 3 ab$^{-1}$} & \multicolumn{1}{c|}{$\mathcal{L} = $ 15 ab$^{-1}$} \\ \hline
NUHM2                   & 1.17     & \multicolumn{1}{l|}{$3.71\cdot 10^5$}        & \multicolumn{1}{l|}{36.3}       & \multicolumn{1}{l|}{5.6}                                 & \multicolumn{1}{l|}{$0.8$ ($0.7$)}                                        & $1.8$ ($1.1$)                                       \\ 
type-III                & 1.16      & \multicolumn{1}{l|}{$4.2\cdot 10^5$}         & \multicolumn{1}{l|}{50.2}      & \multicolumn{1}{l|}{33}                                    & \multicolumn{1}{l|}{$4.3$ ($3.8$)}                                        & $9.6$ ($6.0$)                                    \\ 
GM                      & 1.26      & \multicolumn{1}{l|}{$3.7\cdot 10^5$}         & \multicolumn{1}{l|}{18.9}         & \multicolumn{1}{l|}{10}                                 & \multicolumn{1}{l|}{$1.4$ ($1.2$)}                                        & $3.1$ ($1.9$)                                       \\ \hline
$t\bar{t}$              & 1.72      & \multicolumn{1}{l|}{$4.6\cdot 10^{10}$}             & \multicolumn{1}{l|}{0}                      & \multicolumn{1}{l|}{0}                      & \multicolumn{1}{l|}{-}                                           & -                                            \\ 
$t\bar{t}t\bar{t}$      & 1.27        & \multicolumn{1}{l|}{$3.75\cdot 10^6$}        & \multicolumn{1}{l|}{0}        & \multicolumn{1}{l|}{0}                                      & \multicolumn{1}{l|}{-}                                           & -                                            \\ 
$t\bar{t}W^{\pm}$       & 1.24        & \multicolumn{1}{l|}{$9.3\cdot 10^6$}         & \multicolumn{1}{l|}{9.3}        & \multicolumn{1}{l|}{0}                                      & \multicolumn{1}{l|}{-}                                           & -                                            \\ 
$t\bar{t}Z$             & 1.39       & \multicolumn{1}{l|}{$6.51\cdot 10^7$}        & \multicolumn{1}{l|}{0}                         & \multicolumn{1}{l|}{0}                      & \multicolumn{1}{l|}{-}                                           & -                                            \\ 
$W^{\pm}W^{\pm}jj$      & 1.04      & \multicolumn{1}{l|}{$1.6\cdot 10^7$}         & \multicolumn{1}{l|}{213.7}      & \multicolumn{1}{l|}{65.8}                              & \multicolumn{1}{l|}{-}                                           & -                                            \\ 
$W^{\pm}W^{\pm}W^{\mp}$ & 2.45      & \multicolumn{1}{l|}{$4.1\cdot 10^6$}         & \multicolumn{1}{l|}{135.1}       & \multicolumn{1}{l|}{65.5}                                     & \multicolumn{1}{l|}{-}                                           & -                                            \\ 
$W^{\pm}Z$              & 1.88      & \multicolumn{1}{l|}{$5.2\cdot 10^8$}         & \multicolumn{1}{l|}{156.7}       & \multicolumn{1}{l|}{0}                                            & \multicolumn{1}{l|}{-}                                           & -                                            \\ 
$ZZ$              & 1.7      & \multicolumn{1}{l|}{$1.8 \cdot 10^8$}         & \multicolumn{1}{l|}{0}       & \multicolumn{1}{l|}{0}                                           & \multicolumn{1}{l|}{-}                                           & -                                            \\ 
$W^{\pm}W^{\mp}Z$              & 2.0     & \multicolumn{1}{l|}{$2.8 \cdot 10^6$}         & \multicolumn{1}{l|}{17.1}       & \multicolumn{1}{l|}{5.7}                     & \multicolumn{1}{l|}{-}                                           & -                                            \\
$W^{\pm}ZZ$              & 2.0      & \multicolumn{1}{l|}{$8.7 \cdot 10^5$}         & \multicolumn{1}{l|}{14.0}       & \multicolumn{1}{l|}{6.1}                                            & \multicolumn{1}{l|}{-}                                           & -                                            \\\hline
Total BG                & $-$      & \multicolumn{1}{l|}{$4.72\cdot 10^{10}$}        & \multicolumn{1}{l|}{545.9}                        & \multicolumn{1}{l|}{143.1}                   & \multicolumn{1}{l|}{-}                                           & -                                            \\ \hline \hline
\end{tabular}%
}
\caption{Cut flow table for cleaner type-III signal.}
\label{type3 27tev}
\end{table}

\begin{figure}[h!]
\centering
\begin{subfigure}[t]{0.5\textwidth}
  \centering
  \includegraphics[width=1\linewidth]{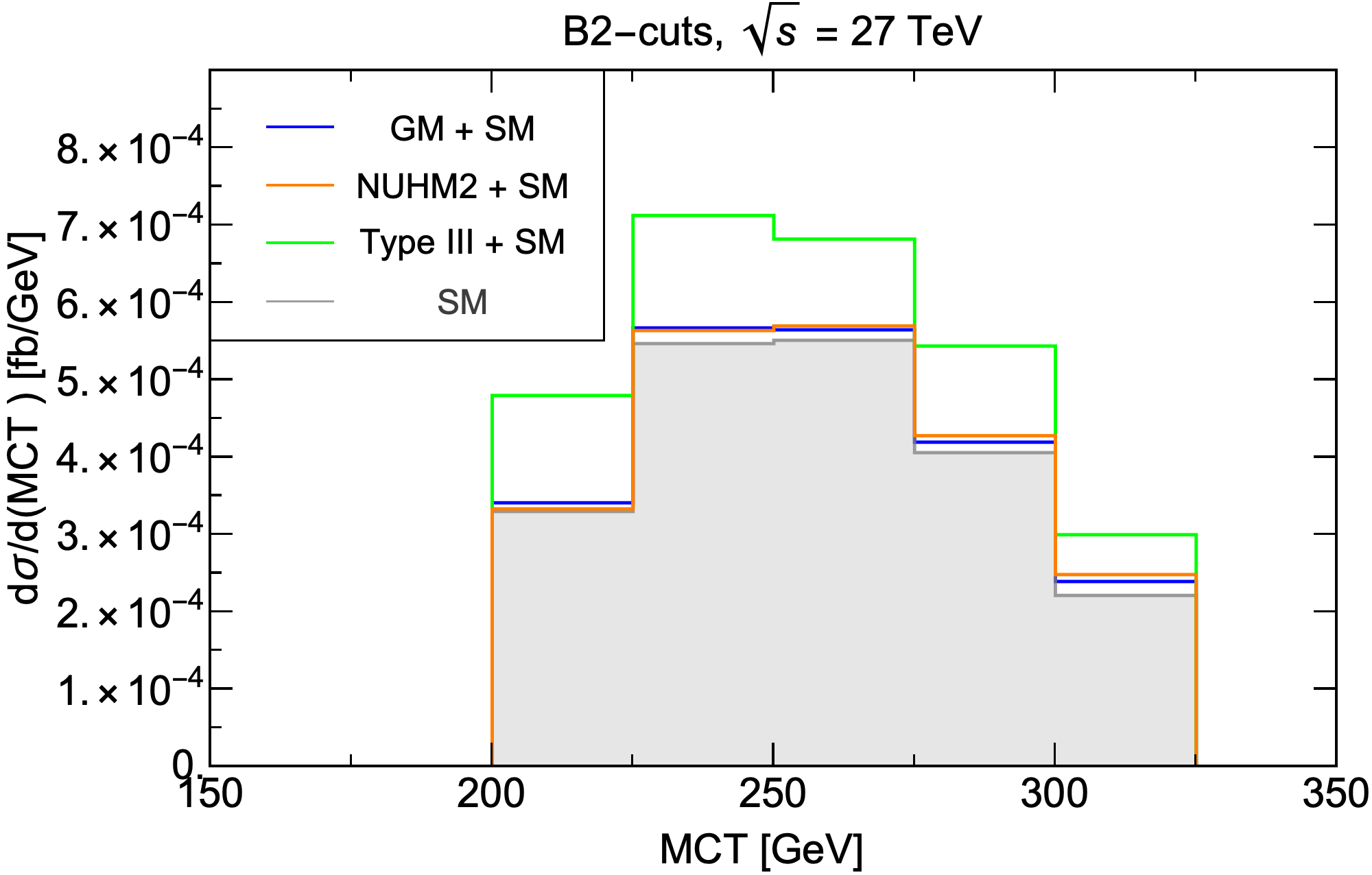}
  \caption{}
\end{subfigure}%
\begin{subfigure}[t]{0.5\textwidth}
  \centering
  \includegraphics[width=1\linewidth]{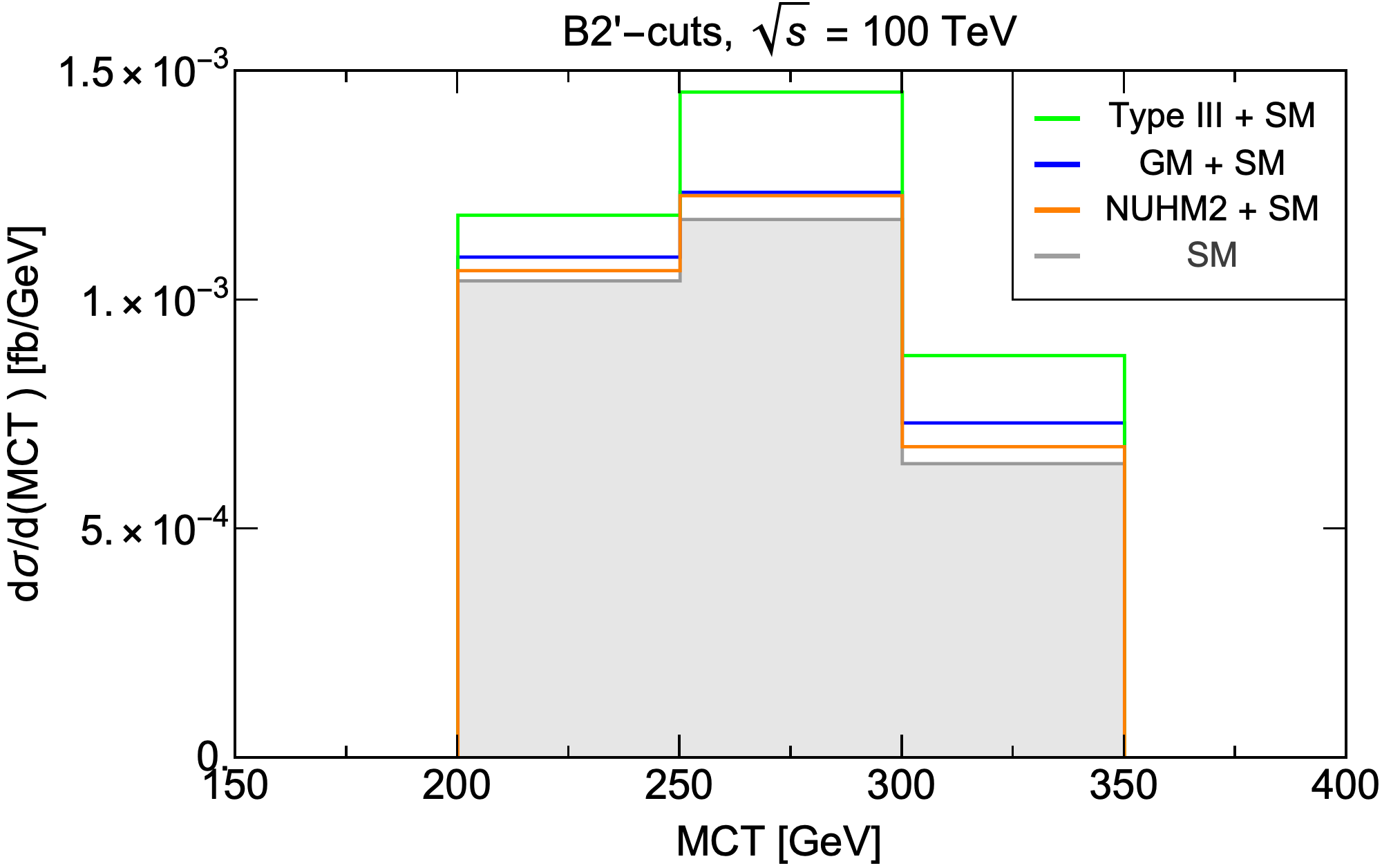}
  \caption{}
\end{subfigure}
\begin{subfigure}[t]{0.5\textwidth}
  \centering
  \includegraphics[width=1\linewidth]{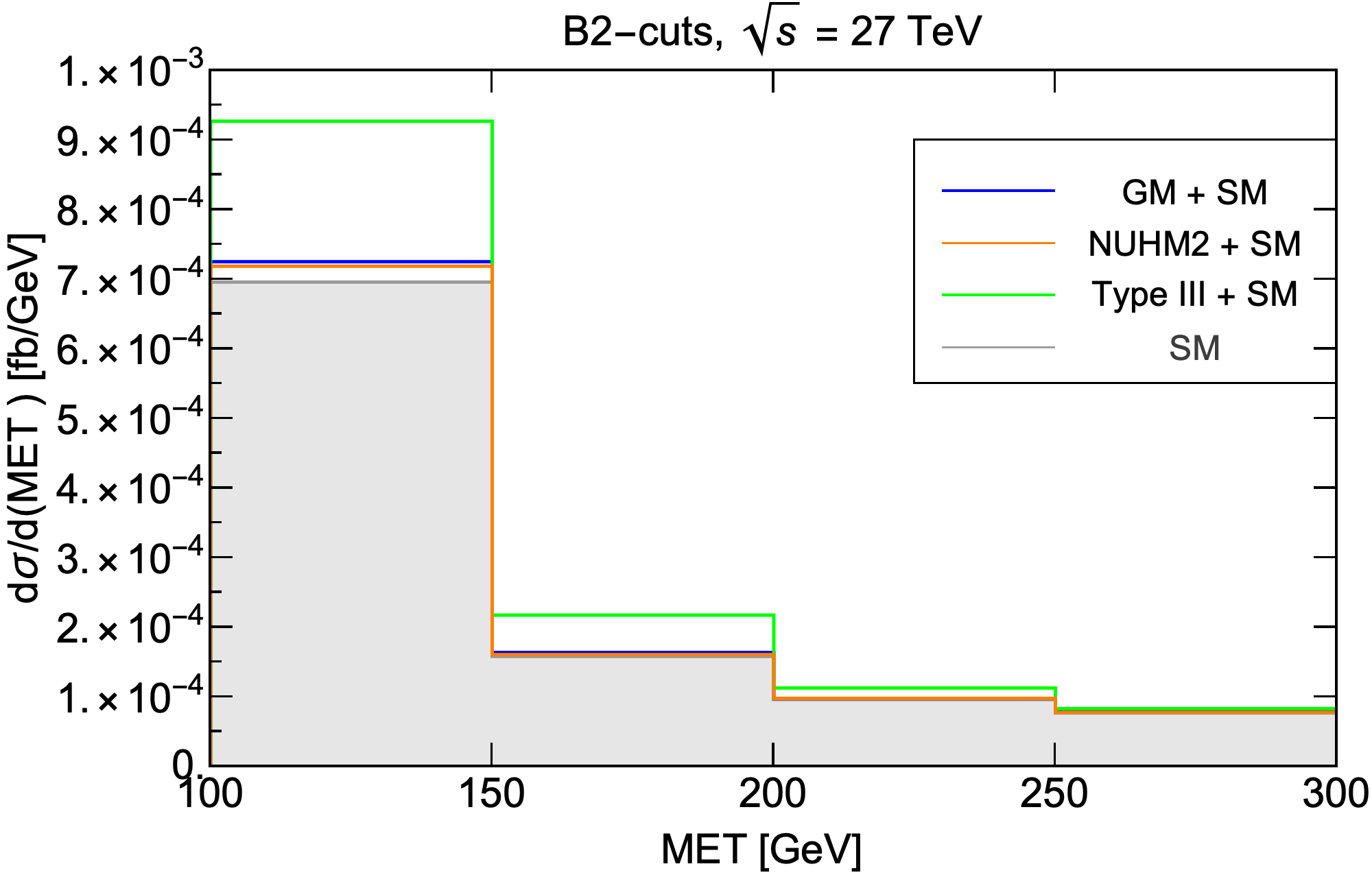}
  \caption{}
\end{subfigure}%
\begin{subfigure}[t]{0.5\textwidth}
  \centering
  \includegraphics[width=1\linewidth]{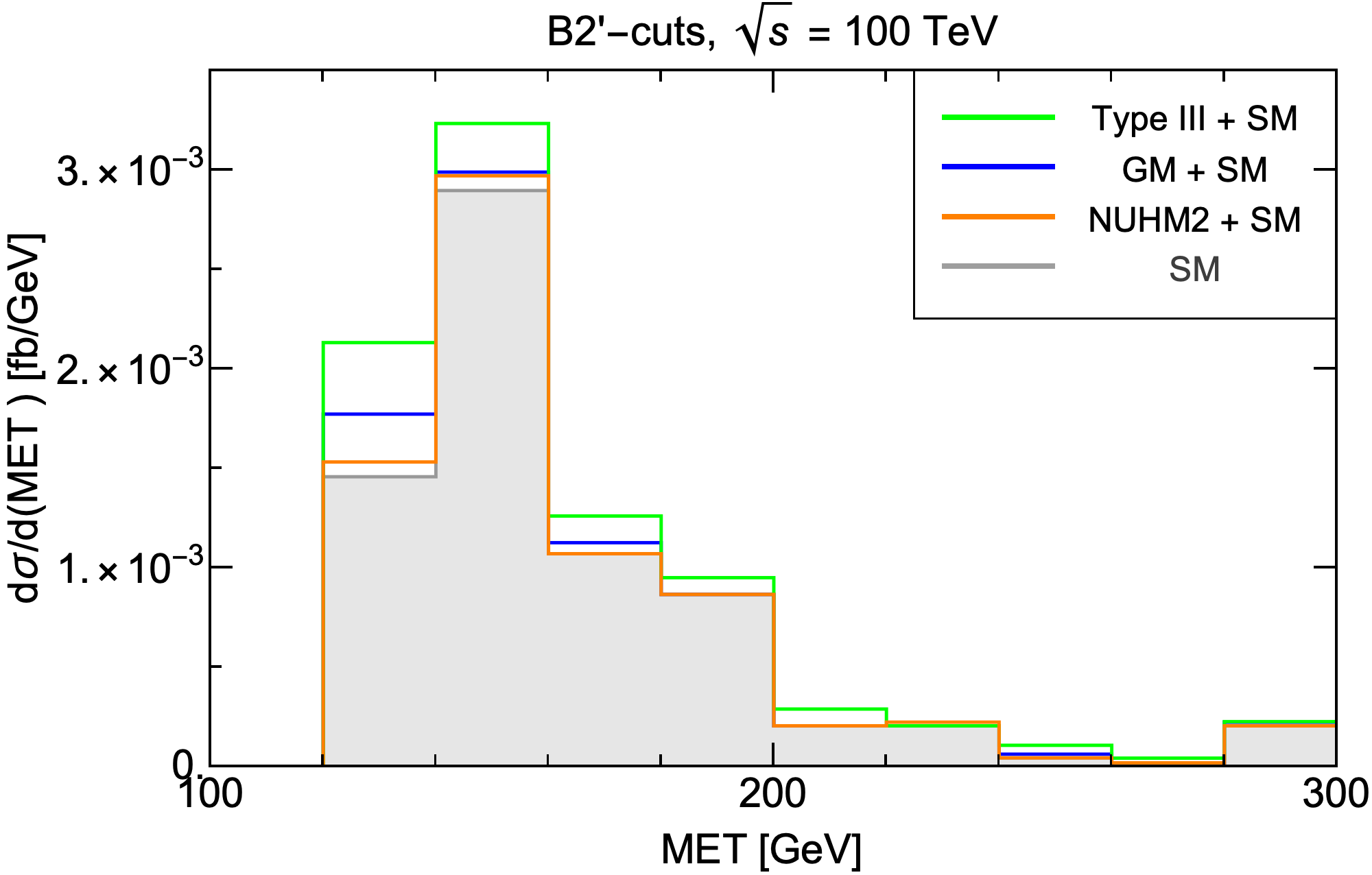}
  \caption{}
\end{subfigure}
\vspace*{-0.1in}
\caption{MCT distribution after (a) B2-cuts at  $\sqrt{s}=27$~TeV and (b) B2$^{\prime}$-cuts at $\sqrt{s}=100$~TeV; and $\slashed{E}_{T}$ distribution after (c) B2-cuts at  $\sqrt{s}=27$~TeV and (d) B2$^{\prime}$-cuts at $\sqrt{s}=100$~TeV.}
\label{fig:b2}
\end{figure}

After the B2 and B2$^{\prime}$-cuts at $\sqrt{s}=27$~TeV and $100$~TeV, respectively, we list the significance $S/\sqrt{S+B}$ for the type-III seesaw signal, the NUHM2 signal and the GM model signal for $\mathcal{L}$ = 3~ab$^{-1}$ and 15~ab$^{-1}$. Besides, we also consider the impact of 3$\%$ systematic uncertainties and calculate the significance which we show in parenthesis in Table~\ref{type3 27tev}. We see that on considering 3$\%$ systematic uncertainties the significance for each signal reduces, for example, at $\sqrt{s}=27$~TeV  and  for $\mathcal{L}$ = 15~ab$^{-1}$, the type-III seesaw signal BM point yields a significance of 7.8 which reduces to 6.3 on considering 3$\%$ systematic uncertainties. The table shows that the type-III seesaw signal yields a significance higher than 5$\sigma$ at both energies for $\mathcal{L}$ = 15~ab$^{-1}$ while the other two BSM scenarios don't.  We have plotted the MCT distribution and the $\slashed{E}_{T}$ distribution for the total SM background and various signals on top of it, after imposing the B2 and B2$^{\prime}$-cuts at the respective energies in Fig.~\ref{fig:b2}.  

\clearpage

%%%%%%%%%%%%%%%%%%%%%%%%%%%%%%%%%%%%%%%%%%%%%%%
\subsection{Type-II Seesaw/Georgi-Machacek Model Analysis}
\label{ssec:type2analysis}
%%%%%%%%%%%%%%%%%%%%%%%%%%%%%%%%%%%%%%%%%%%%%%%

In the GM model, since the SSdL and $\slashed{E}_{T}$ originate from $\Delta^{\pm\pm}$ of mass 300~GeV, the MCT distribution should peak and then sharply fall around 300~GeV.
Therefore, an efficient cut after S1-cuts to scoop out the GM model signal would be to require the MCT to be $\le$ 300~GeV.

Since we explicitly have two forward jets in the GM model signal, hence requiring the number of jets $\ge$ 2 would be a suitable cut to retain most of the GM model signal. Hence after S1- cuts, we apply two additional cuts, namely, MCT $\le 300$~GeV and njet $\ge$ 2 and this entire set of cut is called the C1-cuts:
\begin{itemize}
    \item S1-cuts + MCT $\le 300$~GeV + njet $\ge$ 2.
\end{itemize}

\begin{figure}[h]
\centering
\begin{subfigure}[h]{0.5\textwidth}
  \centering
  \includegraphics[width=1\linewidth]{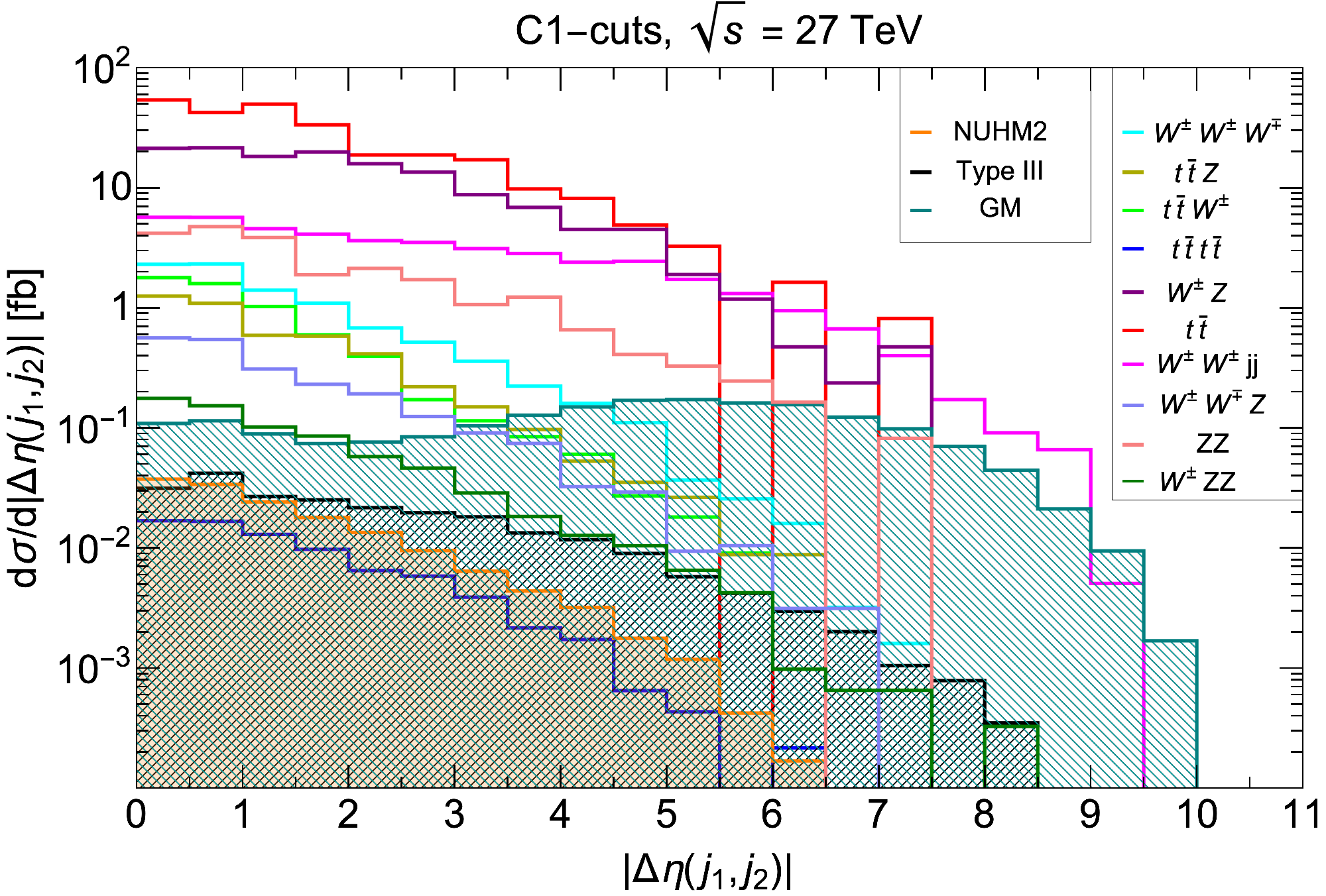}
  \caption{}
  \label{fig:deletac1type2}
\end{subfigure}%
\begin{subfigure}[h]{0.5\textwidth}
  \centering
  \includegraphics[width=1\linewidth]{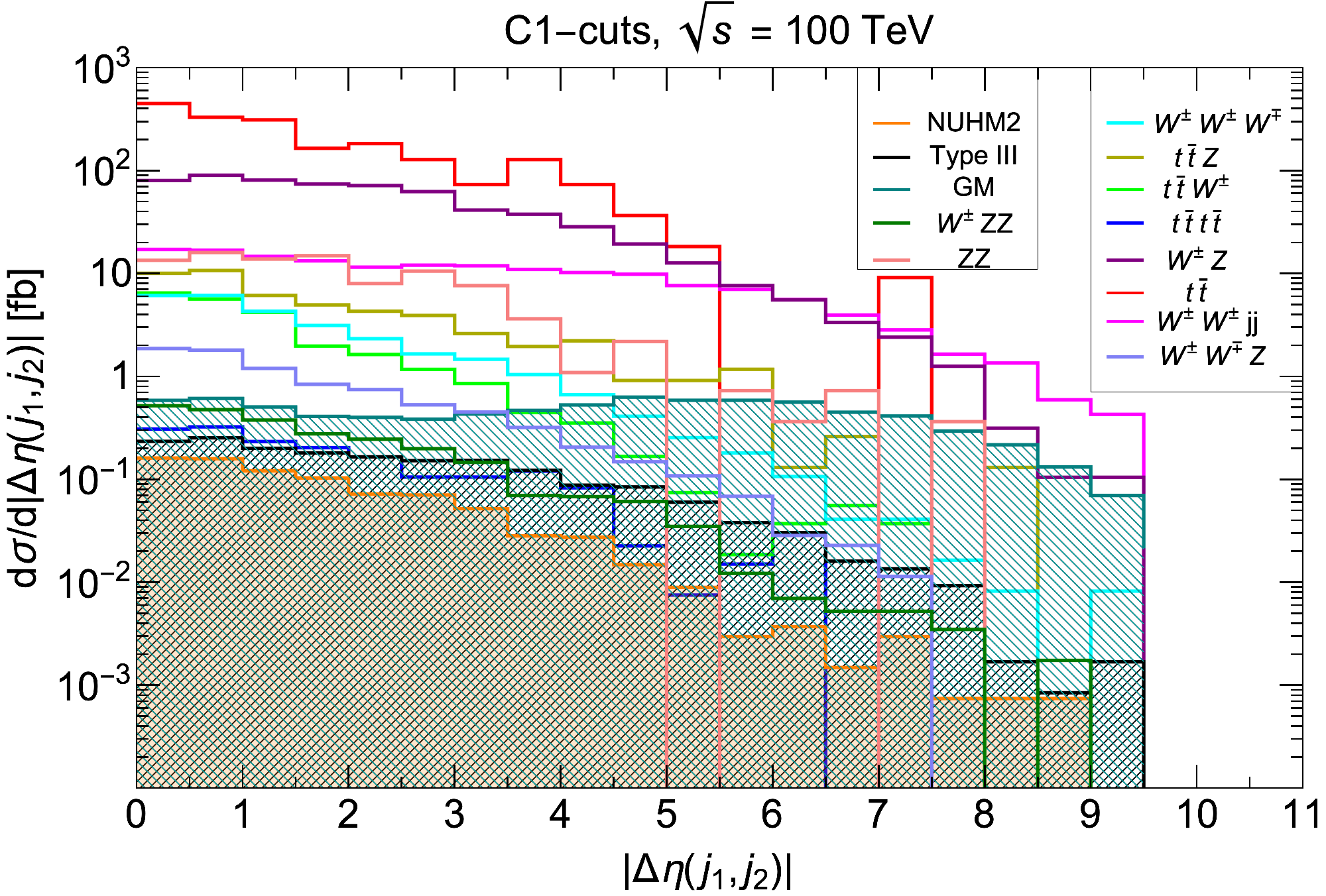}
  \caption{}
  \label{fig:deletac1type2100}
\end{subfigure}
\vspace*{-0.1in}
\caption{$\Delta \eta (j_1, j_2)$ distribution after C1-cuts at (a) $\sqrt{s}=27$~TeV (b) and $\sqrt{s}=100$~TeV .}
\label{fig:deletac1}
\end{figure}

After applying the C1-cuts, we plot the distribution of the pseudorapidity ($\eta$) difference between the two leading jets, $\Delta \eta (j_1, j_2)$, in Fig.~\ref{fig:deletac1type2} and \ref{fig:deletac1type2100} at $\sqrt{s}=27$~TeV and $100$~TeV, respectively.  Due to the presence of two explicit forward jets in the GM model signal, we expect $\Delta \eta (j_1, j_2)$ should peak towards higher values and it indeed peaks around 5 in Fig.~\ref{fig:deletac1}. 
One naively expects that the $W^{\pm}W^{\pm}jj$ process in the SM should also peak towards higher value of $\Delta \eta (j_1, j_2)$.  The reason that Fig.~\ref{fig:deletac1} does not feature this is the following.  The $W^{\pm}W^{\pm}jj$ process includes two types of processes: (a) those of ${\cal O}(\alpha^2 \alpha_S^0)$, and (b) those of ${\cal O}(\alpha_S \alpha)$, both at the amplitude level.  The processes of type (a) do peak towards higher value of $\Delta \eta (j_1, j_2)$ after certain cuts as depicted in Ref.~\cite{Ballestrero:2018anz}.  But here we have included both types of diagrams and our choice of cuts are very different from those used in Ref.~\cite{Ballestrero:2018anz}.  Hence, the peak due to type (a) processes is overshadowed once type (b) processes are included and becomes less prominent.

Therefore, requiring $\Delta \eta (j_1, j_2) > 5$ is an extremely efficient cut to not only reduce the SM background but also to almost eliminate the signals from the other two BSM models.  We now call this full set of cuts to be the C2-cuts:
\begin{itemize}
    \item C1-cuts + $\Delta \eta (j_1, j_2) > 5$.
\end{itemize}

\begin{figure}[h]
\centering
\begin{subfigure}[h]{0.5\textwidth}
  \centering
  \includegraphics[width=1\linewidth]{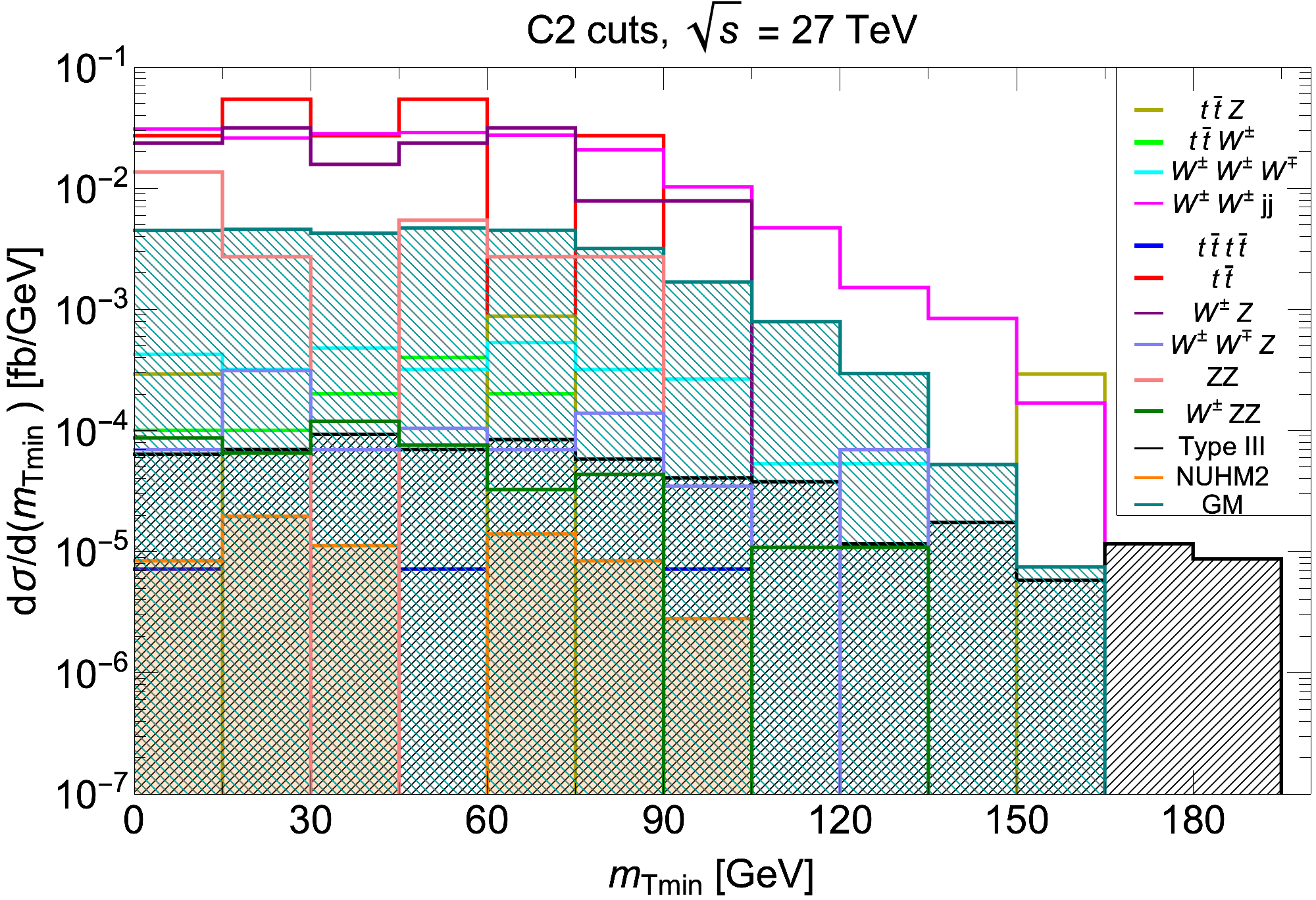}
  \caption{}
  \label{fig:mtminc2type2}
\end{subfigure}%
\begin{subfigure}[h]{0.5\textwidth}
  \centering
  \includegraphics[width=1\linewidth]{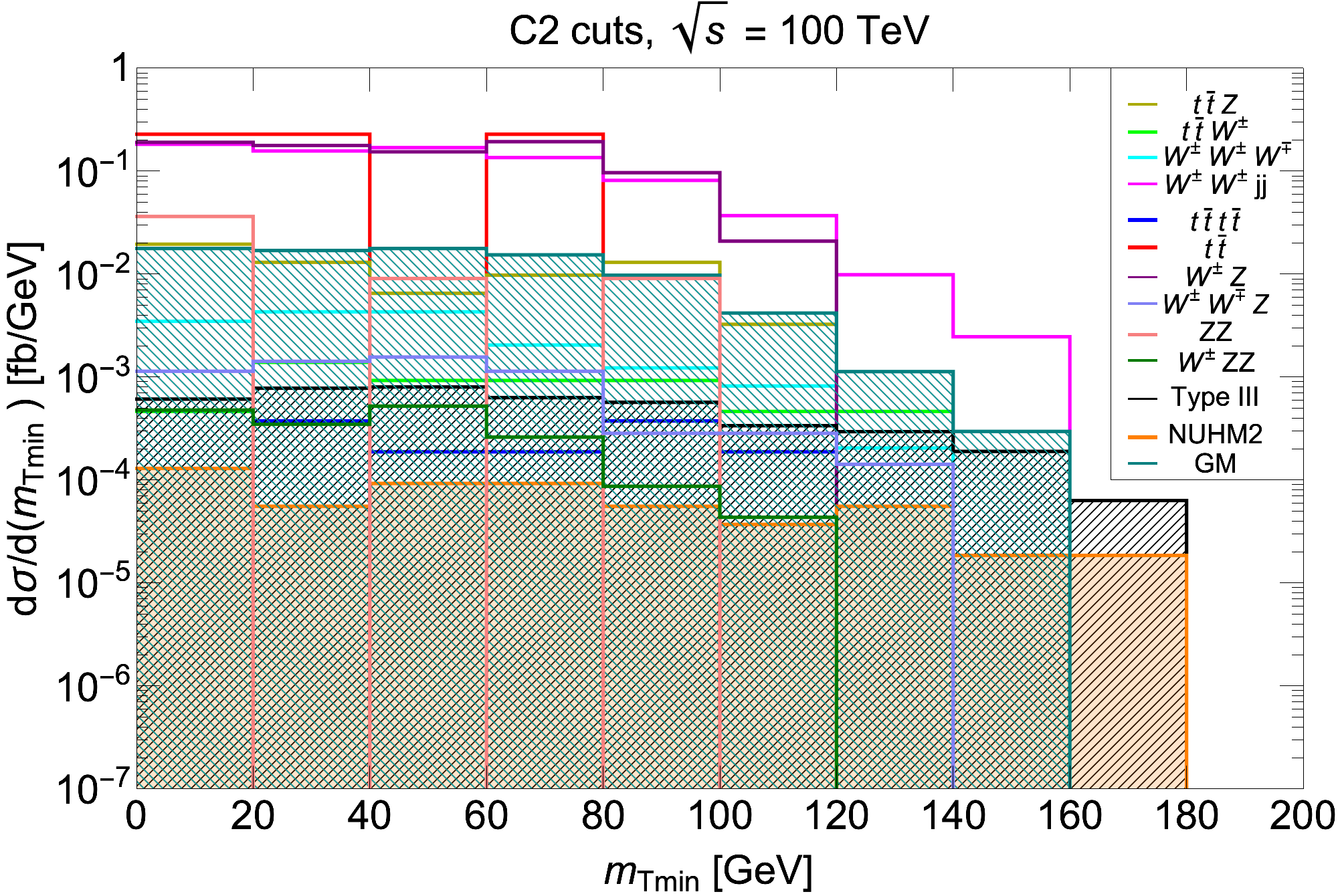}
  \caption{}
  \label{fig:mtminc2type2100}
\end{subfigure}
\vspace*{-0.1in}
\caption{$m_{T_{\rm min}}$ distribution after C2-cuts for (a) $\sqrt{s}=27$~TeV (b) and $\sqrt{s}=100$~TeV .}
\end{figure}

After the C2-cuts, we plot the $m_{T_{\rm min}}$ distribution at $\sqrt{s}=27$~TeV and $100$~TeV in Fig.~\ref{fig:mtminc2type2} and \ref{fig:mtminc2type2100}, respectively.

As can be inferred from Fig.~\ref{fig:mtminc2type2}, a further cut of $m_{T_{\rm min}} > 105$~GeV will make the largest SM backgrounds $t\bar{t}$ and $W^{\pm}Z$ vanish. We also impose a small $\slashed{E}_{T}$ cut of $\slashed{E}_{T} > 50$~GeV just to ensure we pick up those events with some non-zero $\slashed{E}_{T}$ in the final state since we are looking for SSdB + $\slashed{E}_{T}$ signature.  We therefore propose the C3-cuts:
\begin{itemize}
    \item C2-cuts + $m_{T_{\rm min}}$ > 105~GeV + $\slashed{E}_{T}$ > 50~GeV.
\end{itemize}

Similarly, Fig.~\ref{fig:mtminc2type2100} shows that a further cut of $m_{T_{\rm min}} > 120$~GeV would be needed to make $t\bar{t}$ and $W^{\pm}Z$ backgrounds vanish. We therefore propose the C3$^{\prime}$-cuts:
\begin{itemize}
    \item C2-cuts + $m_{T_{\rm min}}$ > 120~GeV + $\slashed{E}_{T}$ > 50~GeV.
\end{itemize}

The cut flow for this scenario is summarized in Table~\ref{type II 27tev}.

After the final sets of cuts, namely the C3 and C3$^\prime$-cuts for $\sqrt{s}=$ 27 and 100 TeV, respectively, we calculate and list the significance $S/\sqrt{S+B}$ for the GM model signal, the type-III seesaw signal, and the NUHM2 signal for $\mathcal{L}$ = 3~ab$^{-1}$ and 15~ab$^{-1}$.  Additionally, we also show the impact of 3$\%$ systematic uncertainties on the significance of the signal in parenthesis in Table~\ref{type II 27tev}. 

\begin{table}[]
\centering
\resizebox{\columnwidth}{!}{%
\begin{tabular}{|l|l|llllll|}
\hline
\hline
                        &          & \multicolumn{6}{c|}{$\sqrt{s}=27$~TeV}                                             \\ \hline 
\multirow{2}{*}{Process}                 & \multirow{2}{*}{K-factor} & \multicolumn{1}{l|}{\multirow{2}{*}{$\sigma$ (NLO) {[}ab{]}}} & \multicolumn{1}{l|}{\multirow{2}{*}{C1 {[}ab{]}}} & \multicolumn{1}{l|}{\multirow{2}{*}{C2 {[}ab{]}}} & \multicolumn{1}{l|}{\multirow{2}{*}{C3 {[}ab{]}}} & \multicolumn{2}{c|}{Significance} \\ \cline{7-8} 
\multicolumn{1}{|l|}{} & \multicolumn{1}{|l|}{} & \multicolumn{1}{|l|}{} & \multicolumn{1}{|l|}{} & \multicolumn{1}{|l|}{} & \multicolumn{1}{|l|}{} &  \multicolumn{1}{c|}{$\mathcal{L} = $ 3 ab$^{-1}$} & \multicolumn{1}{c|}{$\mathcal{L} = $ 15 ab$^{-1}$} \\ \hline
NUHM2                   & 1.17     & \multicolumn{1}{l|}{$4.2 \cdot 10^4$}        & \multicolumn{1}{l|}{77.0}        & \multicolumn{1}{l|}{0.97}        & \multicolumn{1}{l|}{0}        & \multicolumn{1}{l|}{0}                                        & 0                                       \\ 
type-III                & 1.16     & \multicolumn{1}{l|}{$4.36 \cdot 10^4$}       & \multicolumn{1}{l|}{117.86}        & \multicolumn{1}{l|}{8.6}         & \multicolumn{1}{l|}{1.4}         & \multicolumn{1}{l|}{$0.224$ ($0.2$)}                                         & $0.5$ ($0.3$)                                        \\ 
GM                      & 1.26     & \multicolumn{1}{l|}{$5.6 \cdot 10^4$}        & \multicolumn{1}{l|}{976.52}        & \multicolumn{1}{l|}{428.4}        & \multicolumn{1}{l|}{16.35}      & \multicolumn{1}{l|}{$2.5$ ($2.2$)}                                       & $5.5$ ($3.6$)                                       \\ \hline
$t\bar{t}$              & 1.72     & \multicolumn{1}{l|}{$4.1 \cdot 10^9$}        & \multicolumn{1}{l|}{131044.0}           & \multicolumn{1}{l|}{2850.0}           & \multicolumn{1}{l|}{0}           & \multicolumn{1}{l|}{-}                                           & -                                             \\ 
$t\bar{t}t\bar{t}$      & 1.27     & \multicolumn{1}{l|}{$1.1 \cdot 10^5$}        & \multicolumn{1}{l|}{40.0}         & \multicolumn{1}{l|}{0.32}         & \multicolumn{1}{l|}{0}         & \multicolumn{1}{l|}{-}                                           & -                                             \\ 
$t\bar{t}W^{\pm}$       & 1.24     & \multicolumn{1}{l|}{$1.5 \cdot 10^6$}        & \multicolumn{1}{l|}{2938.5}        & \multicolumn{1}{l|}{15.1}        & \multicolumn{1}{l|}{0}        & \multicolumn{1}{l|}{-}                                           & -                                             \\ 
$t\bar{t}Z$             & 1.39     & \multicolumn{1}{l|}{$4.4 \cdot 10^6$}        & \multicolumn{1}{l|}{2260.2}        & \multicolumn{1}{l|}{22.0}         & \multicolumn{1}{l|}{4.4}           & \multicolumn{1}{l|}{-}                                           & -                                            \\ 
$W^{\pm}W^{\pm}jj$      & 1.04     & \multicolumn{1}{l|}{$2.5 \cdot 10^6$}        & \multicolumn{1}{l|}{21638.0}        & \multicolumn{1}{l|}{2693.0}       & \multicolumn{1}{l|}{108.7}        & \multicolumn{1}{l|}{-}                                           & -                                            \\ 
$W^{\pm}W^{\pm}W^{\mp}$ & 2.45     & \multicolumn{1}{l|}{$8.0 \cdot 10^5$}        & \multicolumn{1}{l|}{4622.6}        & \multicolumn{1}{l|}{41.61}        & \multicolumn{1}{l|}{1.6}         & \multicolumn{1}{l|}{-}                                           & -   \\ 
$W^{\pm}Z$              & 1.88     & \multicolumn{1}{l|}{$1.2 \cdot 10^8$}        & \multicolumn{1}{l|}{69464.0}           & \multicolumn{1}{l|}{2126.4}           & \multicolumn{1}{l|}{0}           & \multicolumn{1}{l|}{-}                                           & -  
\\ $ZZ$              & 1.7     & \multicolumn{1}{l|}{$4.1 \cdot 10^7$}        & \multicolumn{1}{l|}{$1.1 \cdot 10^4$}           & \multicolumn{1}{l|}{408.8}           & \multicolumn{1}{l|}{0}           & \multicolumn{1}{l|}{-}                                           & -  \\ 
$W^{\pm}W^{\mp}Z$              & 2.0    & \multicolumn{1}{l|}{$5.2 \cdot 10^5$}        & \multicolumn{1}{l|}{$1.1 \cdot 10^3$}           & \multicolumn{1}{l|}{13.04}           & \multicolumn{1}{l|}{1.0}           & \multicolumn{1}{l|}{-}                                           & -  \\
$W^{\pm}ZZ$              & 2.0     & \multicolumn{1}{l|}{$1.6 \cdot 10^5$}        & \multicolumn{1}{l|}{351.5}           & \multicolumn{1}{l|}{6.7}           & \multicolumn{1}{l|}{0.3}           & \multicolumn{1}{l|}{-}                                           & -  \\\hline
Total BG                & $-$      & \multicolumn{1}{l|}{$4.3 \cdot 10^9$}        & \multicolumn{1}{l|}{$2.4 \cdot 10^5$}       & \multicolumn{1}{l|}{$8.2 \cdot 10^3$}       & \multicolumn{1}{l|}{116.0}        & \multicolumn{1}{l|}{-}                                           & -                                             \\ \hline \hline
                        &          & \multicolumn{6}{c|}{$\sqrt{s}=100$~TeV}                                             \\ \hline 
\multirow{2}{*}{Process}                 & \multirow{2}{*}{K-factor} & \multicolumn{1}{l|}{\multirow{2}{*}{$\sigma$ (NLO) {[}ab{]}}} & \multicolumn{1}{l|}{\multirow{2}{*}{C1 {[}ab{]}}} & \multicolumn{1}{l|}{\multirow{2}{*}{C2 {[}ab{]}}} & \multicolumn{1}{l|}{\multirow{2}{*}{C3$^{\prime}$ {[}ab{]}}} & \multicolumn{2}{c|}{Significance} \\ \cline{7-8} 
\multicolumn{1}{|l|}{} & \multicolumn{1}{|l|}{} & \multicolumn{1}{|l|}{} & \multicolumn{1}{|l|}{} & \multicolumn{1}{|l|}{} & \multicolumn{1}{|l|}{} &  \multicolumn{1}{c|}{$\mathcal{L} = $ 3 ab$^{-1}$} & \multicolumn{1}{c|}{$\mathcal{L} = $ 15 ab$^{-1}$} \\ \hline
NUHM2                   & 1.17     & \multicolumn{1}{l|}{$3.71\cdot 10^5$}        & \multicolumn{1}{l|}{415.13}       & \multicolumn{1}{l|}{11.1}                  & \multicolumn{1}{l|}{1.9}                  & \multicolumn{1}{l|}{$0.22$ ($0.17$)}                                        & $0.48$ ($0.24$)                                       \\ 
type-III                & 1.16      & \multicolumn{1}{l|}{$4.2\cdot 10^5$}         & \multicolumn{1}{l|}{900.8}      & \multicolumn{1}{l|}{85.6}                   & \multicolumn{1}{l|}{11.0}                    & \multicolumn{1}{l|}{$1.23$ ($1.0$)}                                        & $2.7$ ($1.4$)                                     \\ 
GM                      & 1.26      & \multicolumn{1}{l|}{$3.7\cdot 10^5$}         & \multicolumn{1}{l|}{4135.6}         & \multicolumn{1}{l|}{1663.6}                    & \multicolumn{1}{l|}{28.0}                    & \multicolumn{1}{l|}{$3.02$ ($2.4$)}                                        & $6.75$ ($3.5$)                                     \\ \hline
$t\bar{t}$              & 1.72      & \multicolumn{1}{l|}{$4.6\cdot 10^{10}$} & \multicolumn{1}{l|}{$9.5\cdot 10^5$}      & \multicolumn{1}{l|}{$1.4\cdot 10^4$} &    \multicolumn{1}{l|}{0}                      & \multicolumn{1}{l|}{-}                                           & -                                            \\ 
$t\bar{t}t\bar{t}$      & 1.27     & \multicolumn{1}{l|}{$3.75\cdot 10^6$}        & \multicolumn{1}{l|}{860}        & \multicolumn{1}{l|}{26.3}                    & \multicolumn{1}{l|}{0}                    & \multicolumn{1}{l|}{-}                                           & -                                            \\ 
$t\bar{t}W^{\pm}$       & 1.24    & \multicolumn{1}{l|}{$9.3\cdot 10^6$}         & \multicolumn{1}{l|}{$1.1\cdot 10^4$}        & \multicolumn{1}{l|}{111.3}                   & \multicolumn{1}{l|}{9.3}                   & \multicolumn{1}{l|}{-}                                           & -                                            \\ 
$t\bar{t}Z$             & 1.39      & \multicolumn{1}{l|}{$6.51\cdot 10^7$}        & \multicolumn{1}{l|}{$2.5\cdot 10^4$}       & \multicolumn{1}{l|}{$1.3\cdot 10^3$}                   & \multicolumn{1}{l|}{0}                      & \multicolumn{1}{l|}{-}                                           & -                                            \\ 
$W^{\pm}W^{\pm}jj$      & 1.04      & \multicolumn{1}{l|}{$1.6\cdot 10^7$}         & \multicolumn{1}{l|}{$8.0\cdot 10^4$}      & \multicolumn{1}{l|}{$1.5\cdot 10^4$}                   & \multicolumn{1}{l|}{214.0}                   & \multicolumn{1}{l|}{-}                                           & -                                            \\ 
$W^{\pm}W^{\pm}W^{\mp}$ & 2.45      & \multicolumn{1}{l|}{$4.1\cdot 10^6$}         & \multicolumn{1}{l|}{$1.4\cdot 10^4$}       & \multicolumn{1}{l|}{327.5}                   & \multicolumn{1}{l|}{4.1}                   & \multicolumn{1}{l|}{-}                                           & -                                            \\ 
$W^{\pm}Z$              & 1.88      & \multicolumn{1}{l|}{$5.2\cdot 10^8$}         & \multicolumn{1}{l|}{$3.1\cdot 10^5$}       & \multicolumn{1}{l|}{$1.7\cdot 10^4$}                      & \multicolumn{1}{l|}{0}                      & \multicolumn{1}{l|}{-}                                           & -                                    \\ $ZZ$              & 1.7      & \multicolumn{1}{l|}{$1.8 \cdot 10^8$}         & \multicolumn{1}{l|}{$4.6 \cdot 10^4$}       & \multicolumn{1}{l|}{$1.1 \cdot 10^3$}                      & \multicolumn{1}{l|}{0}                      & \multicolumn{1}{l|}{-}                                           & -                                            \\ 
$W^{\pm}W^{\mp}Z$              & 2.0     & \multicolumn{1}{l|}{$2.8 \cdot 10^6$}         & \multicolumn{1}{l|}{$4.2 \cdot 10^3$}       & \multicolumn{1}{l|}{119.5}                      & \multicolumn{1}{l|}{2.8}                      & \multicolumn{1}{l|}{-}                                           & -                                            \\
$W^{\pm}ZZ$              & 2.0     & \multicolumn{1}{l|}{$8.7 \cdot 10^5$}         & \multicolumn{1}{l|}{$1.2 \cdot 10^3$}       & \multicolumn{1}{l|}{34.8}                      & \multicolumn{1}{l|}{0}                      & \multicolumn{1}{l|}{-}                                           & -                                            \\\hline
Total BG                & $-$       & \multicolumn{1}{l|}{$4.72\cdot 10^{10}$}        & \multicolumn{1}{l|}{$1.4 \cdot 10^6$}      & \multicolumn{1}{l|}{$4.9 \cdot 10^4$}                  & \multicolumn{1}{l|}{230.2}                   & \multicolumn{1}{l|}{-}                                           & -                                            \\ \hline \hline
\end{tabular}%
}
\caption{Cut flow table for cleaner GM model signal.}
\label{type II 27tev}
\end{table}

\begin{figure}[t]
\centering
\begin{subfigure}[t]{0.5\textwidth}
  \centering
  \includegraphics[width=1\linewidth]{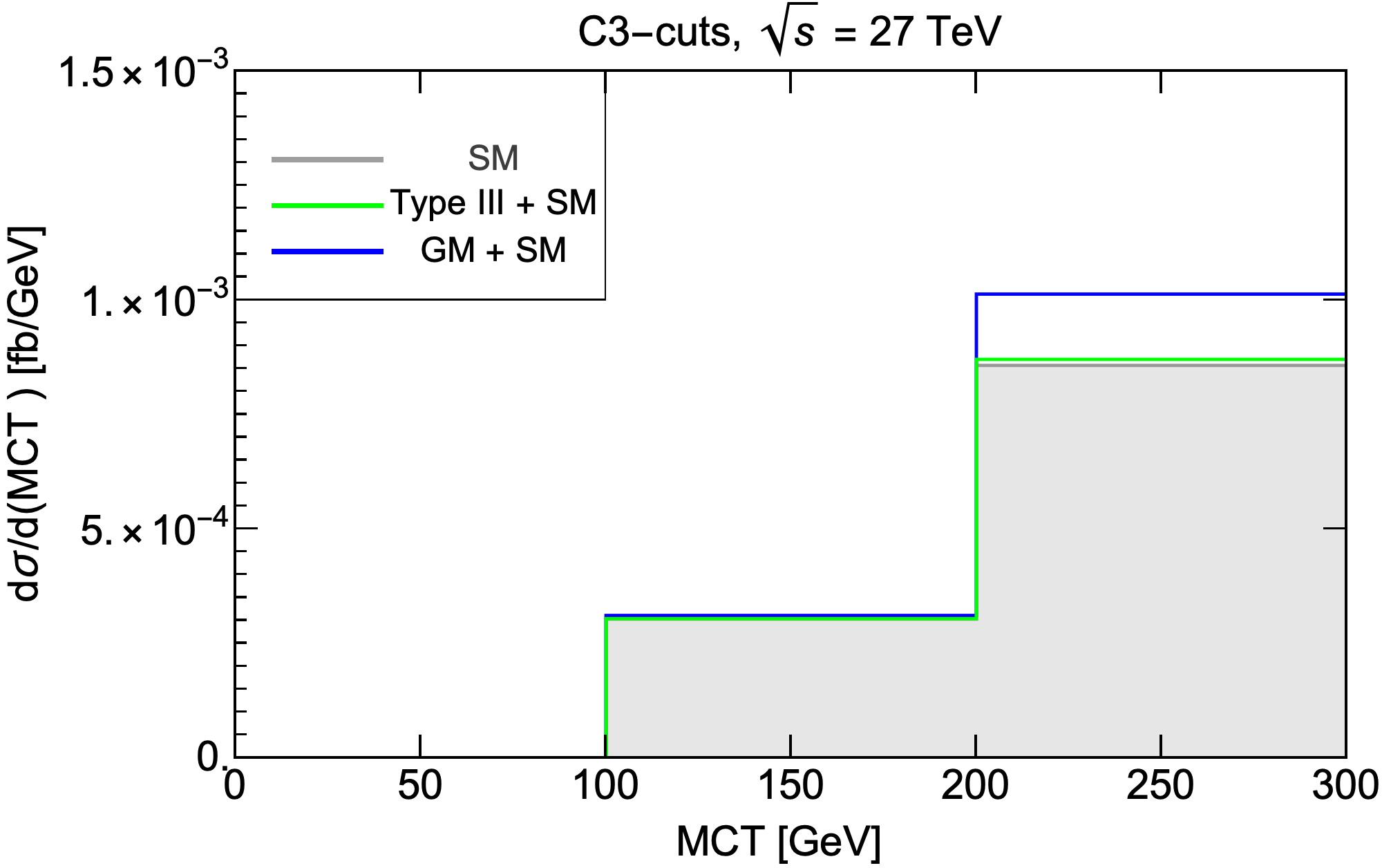}
  \caption{}
  \label{fig:mctc3type2}
\end{subfigure}%
\begin{subfigure}[t]{0.5\textwidth}
  \centering
  \includegraphics[width=1\linewidth]{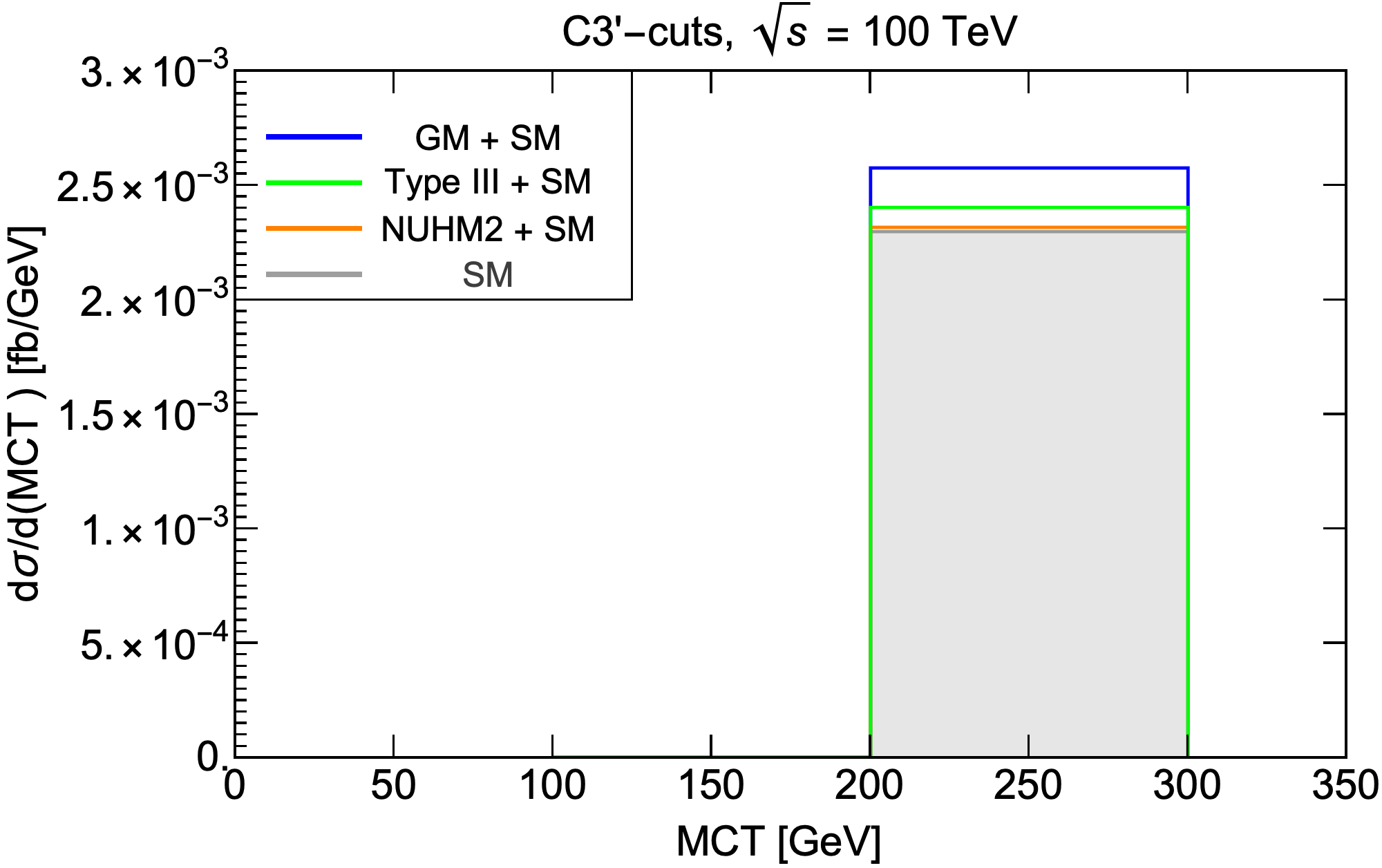}
  \caption{}
  \label{fig:mctc3type3100}
\end{subfigure}
\begin{subfigure}[t]{0.5\textwidth}
  \centering
  \includegraphics[width=1\linewidth]{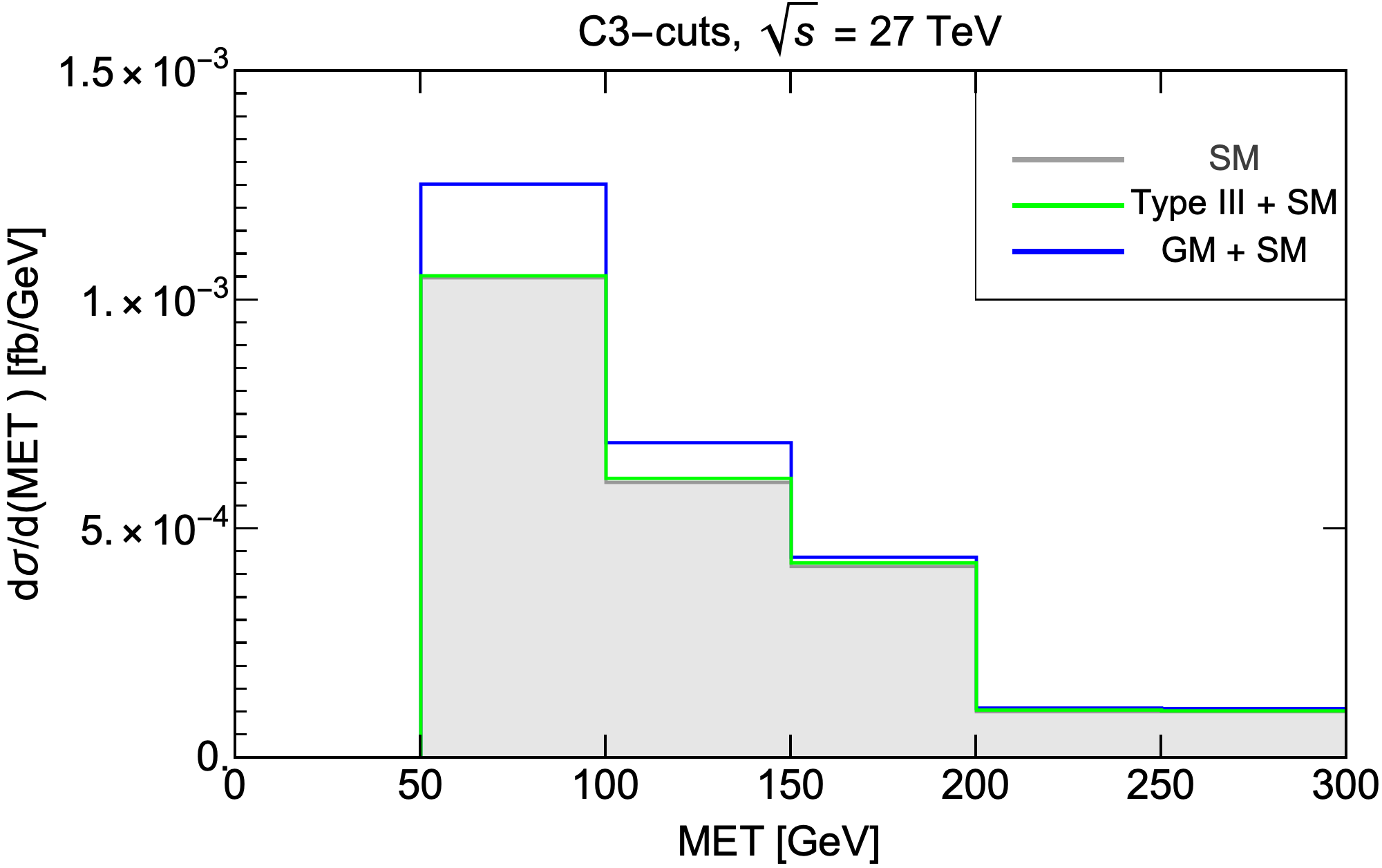}
  \caption{}
  \label{fig:metc3type2}
\end{subfigure}%
\begin{subfigure}[t]{0.5\textwidth}
  \centering
  \includegraphics[width=1\linewidth]{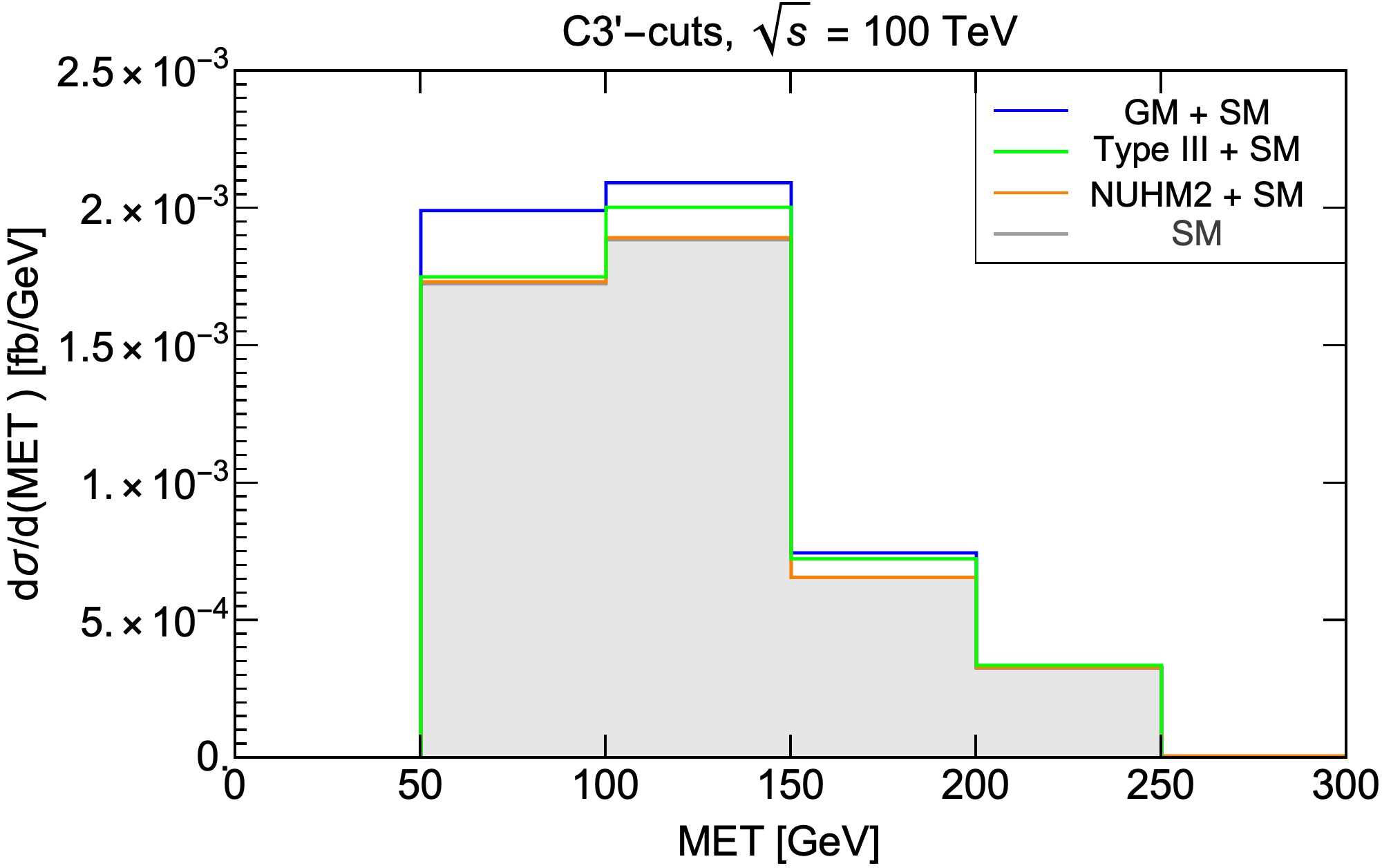}
  \caption{}
  \label{fig:metc3type3100}
\end{subfigure}
\vspace*{-0.1in}
\caption{MCT distribution after (a) C3-cuts at  $\sqrt{s}=27$~TeV and (b) C3$^{\prime}$-cuts at $\sqrt{s}=100$~TeV; and $\slashed{E}_{T}$ distribution after (c) C3-cuts at  $\sqrt{s}=27$~TeV and (d) C3$^{\prime}$-cuts at $\sqrt{s}=100$~TeV.}
\label{fig:c3}
\end{figure}

As can be inferred from Table~\ref{type II 27tev}, with the C3 (C3$^\prime$)-cuts at $\sqrt{s}=27$ ($100$)~TeV, the GM model signal can be observed above the 5$\sigma$ level with IL = 15~ab$^{-1}$ while the other two BSM scenarios yield a significance below 5$\sigma$.
In Fig.~\ref{fig:c3}, we show the MCT distribution and the $\slashed{E}_{T}$ distribution for the total SM background and the various signals on top of it, after the C3 and  C3$^\prime$-cuts at the respective energies.

\clearpage
As seen in the above background and signal evaluations, the dominant SM backgrounds, $t\bar{t}$ and $W^{\pm}Z$ vanish completely after all the final cuts.  This happens because after the S1-cuts, the surviving events from the $t\bar{t}$ background will have one lepton originating from real $W$, another lepton coming from semi-leptonic decay of one of the $b$-jets and the $\slashed{E}_{T}$ would also dominantly come from leptonic decays of $W$. Some of these surviving events will survive the C2-cuts also as these surviving events are accompanied by two or more jets originating from either semi-leptonic decay of $b$-jets or misidentification of the $b$-jet as some lighter jet. In that case, $m_{T_{\rm min}}$ would be mostly bounded by $m_W$. Therefore, a cut of $m_{T_{\rm min}} > m_W$ would reduce the $t\bar{t}$ background as well as the $W^{\pm}Z$ background to a great extent which indeed is the case as seen in the above analyses.

Note that, we have taken $m(\Delta^{\pm\pm}) = 300$~GeV ({\em not} at par with the mass of intermediate states in the other two new physics model signature) in order to highlight the fact that experimental limits on $m(\Delta^{\pm\pm})$ do allow us to take such low mass of $\Delta^{\pm\pm}$. However, even with such small $m(\Delta^{\pm\pm})$, we still do not obtain $5\sigma$ significance unless for the scheme of IL = 15~ab$^{-1}$ for both 27 and 100 TeV because the SM background of $W^{\pm}W^{\pm}jj$ is very much similar to the GM model signal. Thus, the cuts employed to extract the GM model do not reduce the $W^{\pm}W^{\pm}jj$ background as efficiently as they reduce the other SM backgrounds.  Since the GM model signal is produced via the VBF process, taking $m(\Delta^{\pm\pm})$ $\sim$ $m(\Sigma_i^{\pm})$ $\sim$ $770 \rm~ GeV$, would yield a much lower cross section and one might need to resort to a higher-energy collider to obtain $5\sigma$ significance for the GM model signal.
%Note that, it has been possible to observe the GM model signal with 5$\sigma$ significance at HL-LHC while for NUHM2 signal one needs HE-LHC and for type-III seesaw signal one needs 100 TeV collider to obtain 5$\sigma$ significance because $m(\Delta^{\pm\pm})$ in GM model is much less than $m(\Sigma_i^{\pm})$ in type-III seesaw and $m(\tilde{Z}_4, \tilde{W}_2^{\pm})$ in NUHM2 model. Since the GM model signal is produced via VBF process, hence taking $m(\Delta^{\pm\pm})$ $\sim$ $m(\Sigma_i^{\pm})$ $\sim$ $770 \rm~ GeV$, would yield much lower cross section and one might need to upgrade to HE-LHC to obtain 5$\sigma$ significance even for GM model signal. However, we have taken $m(\Delta^{\pm\pm}) = 300 \rm~ GeV$ ($not$ at par with the mass of intermediate states in the other two new physics model signature) in order to highlight the fact that experimental limits on $m(\Delta^{\pm\pm})$ do allow us to obtain a 5$\sigma$ significance at HL-LHC for GM model signal while experimental mass constraints compel us to upgrade to higher energies to obtain 5$\sigma$ significance for NUHM2 and type-III seesaw signal.
%%%%%%%%%%%%%%%%%%%%%%%%%%%%%%%%%%%%%%%%%%%%%%%
%%%%%%%%%%%%%%%%%%%%%%%%%%%%%%%%%%%%%%%%%%%%%%%

%%%%%%%%%%%%%%%%%%%%%%%%%%%%%%%%%%%%%%%%%%%%%%%
%%%%%%%%%%%%%%%%%%%%%%%%%%%%%%%%%%%%%%%%%%%%%%%
\section{Conclusions}
\label{sec:conclude}

%In this paper, we focus on using the signature of SSdB + $\slashed{E}_{T}$ to search for new physics and study how various models with such a signature can be distinguished by imposing suitable cuts.  To reduce the SM background, we consider leptonic decays of the diboson, yielding SSdL + $\slashed{E}_{T}$ in the final state.  
In this paper, our goal is to catalogue various BSM scenarios that can give rise to the SSdB+$\slashed{E}_{T}$ (precisely SSdL+$\slashed{E}_{T}$, as considering leptonic decays of the diboson gets rid of the large SM QCD backgrounds) signature in experiments and extract these signals from SM background by imposing suitable cuts. Since more than one BSM scenario qualify, we also focus on devising suitable cuts to distinguish these BSM models from one another. We have analyzed three new physics models: the NUHM2 scenario of natural SUSY models, the type-III seesaw model, and the GM model. We carefully select the imposed cuts for each model to obtain a sufficiently large significance for its signal. Assuming $\mathcal{L}$ = 15~ab$^{-1}$ and $\sqrt{s}=27$ ($100$)~TeV, the C3 (C3$^{\prime}$)-cuts and the B2 (B2$^{\prime}$)-cuts are needed to observe clean GM model and type-III seesaw model signals, respectively, at a level above $5\sigma$ significance.  For the NUHM2 model, a clean signal at a level above $5\sigma$ significance can be seen with the A3 (A3$^{\prime}$)-cuts for data collected from $\mathcal{L}$ = 3~ab$^{-1}$ and $\sqrt{s}=27$ ($100$)~TeV.

%Though $\sqrt{s}=27$~TeV is sufficient to observe these three BSM models at a level above $5\sigma$ significance, we extend our analysis for FCC-hh at $\sqrt{s}=100$~TeV for completeness.}
%%%%%%%%%%%%%%%%%%%%%%%%%%%%%%%%%%%%%%%%%%%%%%%
%%%%%%%%%%%%%%%%%%%%%%%%%%%%%%%%%%%%%%%%%%%%%%%

%%%%%%%%%%%%%%%%%%%%%%%%%%%%%%%%%%%%%%%%%%%%%
%%%%%%%%%%%%%%%%%%%%%%%%%%%%%%%%%%%%%%%%%%%%%
\section*{Acknowledgments} 

We thank Howard Baer for useful discussions.  The work of CWC and DS was supported in part by the Ministry of Science and Technology (MOST) of Taiwan under Grant Nos.~108-2112-M-002-005-MY3 and 109-2811-M-002-570.

%%%%%%%%%%%%%%%%%%%%%%%%%%%%%%%%%%%%%%%%%%%%%
%%%%%%%%%%%%%%%%%%%%%%%%%%%%%%%%%%%%%%%%%%%%%
\bibliographystyle{utphys}

\bibliography{reference}

\providecommand{\href}[2]{#2}\begingroup\raggedright\begin{thebibliography}{100}

\bibitem{Aad:2012tfa}
{\bfseries ATLAS} Collaboration, G.~Aad {\em et~al.}, ``{Observation of a new
  particle in the search for the Standard Model Higgs boson with the ATLAS
  detector at the LHC},''
  \href{http://dx.doi.org/10.1016/j.physletb.2012.08.020}{{\em Phys. Lett. B}
  {\bfseries 716} (2012) 1--29},
  \href{http://arxiv.org/abs/1207.7214}{{\ttfamily arXiv:1207.7214 [hep-ex]}}.

\bibitem{Chatrchyan:2012ufa}
{\bfseries CMS} Collaboration, S.~Chatrchyan {\em et~al.}, ``{Observation of a
  New Boson at a Mass of 125 GeV with the CMS Experiment at the LHC},''
  \href{http://dx.doi.org/10.1016/j.physletb.2012.08.021}{{\em Phys. Lett. B}
  {\bfseries 716} (2012) 30--61},
  \href{http://arxiv.org/abs/1207.7235}{{\ttfamily arXiv:1207.7235 [hep-ex]}}.

\bibitem{Matalliotakis:1994ft}
D.~Matalliotakis and H.~P. Nilles, ``{Implications of nonuniversality of soft
  terms in supersymmetric grand unified theories},''
  \href{http://dx.doi.org/10.1016/0550-3213(94)00487-Y}{{\em Nucl. Phys. B}
  {\bfseries 435} (1995) 115--128},
  \href{http://arxiv.org/abs/hep-ph/9407251}{{\ttfamily arXiv:hep-ph/9407251}}.

\bibitem{Baer:2005bu}
H.~Baer, A.~Mustafayev, S.~Profumo, A.~Belyaev, and X.~Tata, ``{Direct,
  indirect and collider detection of neutralino dark matter in SUSY models with
  non-universal Higgs masses},''
  \href{http://dx.doi.org/10.1088/1126-6708/2005/07/065}{{\em JHEP} {\bfseries
  07} (2005) 065}, \href{http://arxiv.org/abs/hep-ph/0504001}{{\ttfamily
  arXiv:hep-ph/0504001}}.

\bibitem{Baer:2016hfa}
H.~Baer, V.~Barger, H.~Serce, and X.~Tata, ``{Natural generalized mirage
  mediation},'' \href{http://dx.doi.org/10.1103/PhysRevD.94.115017}{{\em Phys.
  Rev. D} {\bfseries 94} no.~11, (2016) 115017},
  \href{http://arxiv.org/abs/1610.06205}{{\ttfamily arXiv:1610.06205
  [hep-ph]}}.

\bibitem{Randall:1998uk}
L.~Randall and R.~Sundrum, ``{Out of this world supersymmetry breaking},''
  \href{http://dx.doi.org/10.1016/S0550-3213(99)00359-4}{{\em Nucl. Phys. B}
  {\bfseries 557} (1999) 79--118},
  \href{http://arxiv.org/abs/hep-th/9810155}{{\ttfamily arXiv:hep-th/9810155}}.

\bibitem{Baer:2018hwa}
H.~Baer, V.~Barger, and D.~Sengupta, ``{Anomaly mediated SUSY breaking model
  retrofitted for naturalness},''
  \href{http://dx.doi.org/10.1103/PhysRevD.98.015039}{{\em Phys. Rev. D}
  {\bfseries 98} no.~1, (2018) 015039},
  \href{http://arxiv.org/abs/1801.09730}{{\ttfamily arXiv:1801.09730
  [hep-ph]}}.

\bibitem{Baer:2020kwz}
H.~Baer, V.~Barger, S.~Salam, D.~Sengupta, and K.~Sinha, ``{Status of weak
  scale supersymmetry after LHC Run 2 and ton-scale noble liquid WIMP
  searches},'' \href{http://dx.doi.org/10.1140/epjst/e2020-000020-x}{{\em Eur.
  Phys. J. ST} {\bfseries 229} no.~21, (2020) 3085--3141},
  \href{http://arxiv.org/abs/2002.03013}{{\ttfamily arXiv:2002.03013
  [hep-ph]}}.

\bibitem{Foot:1988aq}
R.~Foot, H.~Lew, X.~G. He, and G.~C. Joshi, ``{Seesaw Neutrino Masses Induced
  by a Triplet of Leptons},'' \href{http://dx.doi.org/10.1007/BF01415558}{{\em
  Z. Phys. C} {\bfseries 44} (1989) 441}.

\bibitem{Magg:1980ut}
M.~Magg and C.~Wetterich, ``{Neutrino Mass Problem and Gauge Hierarchy},''
  \href{http://dx.doi.org/10.1016/0370-2693(80)90825-4}{{\em Phys. Lett. B}
  {\bfseries 94} (1980) 61--64}.

\bibitem{Schechter:1980gr}
J.~Schechter and J.~W.~F. Valle, ``{Neutrino Masses in SU(2) x U(1)
  Theories},'' \href{http://dx.doi.org/10.1103/PhysRevD.22.2227}{{\em Phys.
  Rev. D} {\bfseries 22} (1980) 2227}.

\bibitem{Mohapatra:1979ia}
R.~N. Mohapatra and G.~Senjanovic, ``{Neutrino Mass and Spontaneous Parity
  Nonconservation},'' \href{http://dx.doi.org/10.1103/PhysRevLett.44.912}{{\em
  Phys. Rev. Lett.} {\bfseries 44} (1980) 912}.

\bibitem{Lazarides:1980nt}
G.~Lazarides, Q.~Shafi, and C.~Wetterich, ``{Proton Lifetime and Fermion Masses
  in an SO(10) Model},''
  \href{http://dx.doi.org/10.1016/0550-3213(81)90354-0}{{\em Nucl. Phys. B}
  {\bfseries 181} (1981) 287--300}.

\bibitem{Georgi:1985nv}
H.~Georgi and M.~Machacek, ``{DOUBLY CHARGED HIGGS BOSONS},''
  \href{http://dx.doi.org/10.1016/0550-3213(85)90325-6}{{\em Nucl. Phys. B}
  {\bfseries 262} (1985) 463--477}.

\bibitem{Baer:2006rs}
H.~Baer and X.~Tata, {\em {Weak scale supersymmetry: From superfields to
  scattering events}}.
\newblock Cambridge University Press, 5, 2006.

\bibitem{Baer:2018hpb}
H.~Baer, V.~Barger, J.~S. Gainer, D.~Sengupta, H.~Serce, and X.~Tata, ``{LHC
  luminosity and energy upgrades confront natural supersymmetry models},''
  \href{http://dx.doi.org/10.1103/PhysRevD.98.075010}{{\em Phys. Rev. D}
  {\bfseries 98} no.~7, (2018) 075010},
  \href{http://arxiv.org/abs/1808.04844}{{\ttfamily arXiv:1808.04844
  [hep-ph]}}.

\bibitem{Nilles:1982dy}
H.~P. Nilles, M.~Srednicki, and D.~Wyler, ``{Weak Interaction Breakdown Induced
  by Supergravity},''
  \href{http://dx.doi.org/10.1016/0370-2693(83)90460-4}{{\em Phys. Lett. B}
  {\bfseries 120} (1983) 346}.

\bibitem{Frere:1983ag}
J.~M. Frere, D.~R.~T. Jones, and S.~Raby, ``{Fermion Masses and Induction of
  the Weak Scale by Supergravity},''
  \href{http://dx.doi.org/10.1016/0550-3213(83)90606-5}{{\em Nucl. Phys. B}
  {\bfseries 222} (1983) 11--19}.

\bibitem{Derendinger:1983bz}
J.~P. Derendinger and C.~A. Savoy, ``{Quantum Effects and SU(2) x U(1) Breaking
  in Supergravity Gauge Theories},''
  \href{http://dx.doi.org/10.1016/0550-3213(84)90162-7}{{\em Nucl. Phys. B}
  {\bfseries 237} (1984) 307--328}.

\bibitem{Maniatis:2009re}
M.~Maniatis, ``{The Next-to-Minimal Supersymmetric extension of the Standard
  Model reviewed},'' \href{http://dx.doi.org/10.1142/S0217751X10049827}{{\em
  Int. J. Mod. Phys. A} {\bfseries 25} (2010) 3505--3602},
  \href{http://arxiv.org/abs/0906.0777}{{\ttfamily arXiv:0906.0777 [hep-ph]}}.

\bibitem{Ellwanger:2009dp}
U.~Ellwanger, C.~Hugonie, and A.~M. Teixeira, ``{The Next-to-Minimal
  Supersymmetric Standard Model},''
  \href{http://dx.doi.org/10.1016/j.physrep.2010.07.001}{{\em Phys. Rept.}
  {\bfseries 496} (2010) 1--77},
  \href{http://arxiv.org/abs/0910.1785}{{\ttfamily arXiv:0910.1785 [hep-ph]}}.

\bibitem{ATLAS:2019xhj}
{\bfseries ATLAS} Collaboration, M.~Aaboud {\em et~al.}, ``{Measurement of $ZZ$
  production in the $\ell\ell\nu\nu$ final state with the ATLAS detector in
  $pp$ collisions at $\sqrt{s} = 13$ TeV},''
  \href{http://dx.doi.org/10.1007/JHEP10(2019)127}{{\em JHEP} {\bfseries 10}
  (2019) 127}, \href{http://arxiv.org/abs/1905.07163}{{\ttfamily
  arXiv:1905.07163 [hep-ex]}}.

\bibitem{Barger:2010aq}
V.~Barger, G.~Shaughnessy, and B.~Yencho, ``{Many Leptons at the LHC from the
  NMSSM},'' \href{http://dx.doi.org/10.1103/PhysRevD.83.055006}{{\em Phys. Rev.
  D} {\bfseries 83} (2011) 055006},
  \href{http://arxiv.org/abs/1011.3526}{{\ttfamily arXiv:1011.3526 [hep-ph]}}.

\bibitem{Baer:2015tva}
H.~Baer, V.~Barger, P.~Huang, D.~Mickelson, M.~Padeffke-Kirkland, and X.~Tata,
  ``{Natural SUSY with a bino- or wino-like LSP},''
  \href{http://dx.doi.org/10.1103/PhysRevD.91.075005}{{\em Phys. Rev. D}
  {\bfseries 91} no.~7, (2015) 075005},
  \href{http://arxiv.org/abs/1501.06357}{{\ttfamily arXiv:1501.06357
  [hep-ph]}}.

\bibitem{Barman:2017swy}
R.~K. Barman, G.~Belanger, B.~Bhattacherjee, R.~Godbole, G.~Mendiratta, and
  D.~Sengupta, ``{Invisible decay of the Higgs boson in the context of a
  thermal and nonthermal relic in MSSM},''
  \href{http://dx.doi.org/10.1103/PhysRevD.95.095018}{{\em Phys. Rev. D}
  {\bfseries 95} no.~9, (2017) 095018},
  \href{http://arxiv.org/abs/1703.03838}{{\ttfamily arXiv:1703.03838
  [hep-ph]}}.

\bibitem{Barman:2020vzm}
R.~K. Barman, G.~B\'elanger, B.~Bhattacherjee, R.~Godbole, D.~Sengupta, and
  X.~Tata, ``{Current bounds and future prospects of light neutralino dark
  matter in NMSSM},'' \href{http://dx.doi.org/10.1103/PhysRevD.103.015029}{{\em
  Phys. Rev. D} {\bfseries 103} no.~1, (2021) 015029},
  \href{http://arxiv.org/abs/2006.07854}{{\ttfamily arXiv:2006.07854
  [hep-ph]}}.

\bibitem{Baer:2013yha}
H.~Baer, V.~Barger, P.~Huang, D.~Mickelson, A.~Mustafayev, W.~Sreethawong, and
  X.~Tata, ``{Same sign diboson signature from supersymmetry models with light
  higgsinos at the LHC},''
  \href{http://dx.doi.org/10.1103/PhysRevLett.110.151801}{{\em Phys. Rev.
  Lett.} {\bfseries 110} no.~15, (2013) 151801},
  \href{http://arxiv.org/abs/1302.5816}{{\ttfamily arXiv:1302.5816 [hep-ph]}}.

\bibitem{Baer:2017gzf}
H.~Baer, V.~Barger, J.~S. Gainer, M.~Savoy, D.~Sengupta, and X.~Tata,
  ``{Aspects of the same-sign diboson signature from wino pair production with
  light higgsinos at the high luminosity LHC},''
  \href{http://dx.doi.org/10.1103/PhysRevD.97.035012}{{\em Phys. Rev. D}
  {\bfseries 97} no.~3, (2018) 035012},
  \href{http://arxiv.org/abs/1710.09103}{{\ttfamily arXiv:1710.09103
  [hep-ph]}}.

\bibitem{Jana:2019tdm}
S.~Jana, N.~Okada, and D.~Raut, ``{Displaced Vertex and Disappearing Track
  Signatures in type-III Seesaw},''
  \href{http://arxiv.org/abs/1911.09037}{{\ttfamily arXiv:1911.09037
  [hep-ph]}}.

\bibitem{Aaboud:2017vwy}
{\bfseries ATLAS} Collaboration, M.~Aaboud {\em et~al.}, ``{Search for squarks
  and gluinos in final states with jets and missing transverse momentum using
  36 fb$^{-1}$ of $\sqrt{s}=13$ TeV pp collision data with the ATLAS
  detector},'' \href{http://dx.doi.org/10.1103/PhysRevD.97.112001}{{\em Phys.
  Rev. D} {\bfseries 97} no.~11, (2018) 112001},
  \href{http://arxiv.org/abs/1712.02332}{{\ttfamily arXiv:1712.02332
  [hep-ex]}}.

\bibitem{Vami:2019slp}
{\bfseries ATLAS, CMS} Collaboration, T.~A. Vami, ``{Searches for gluinos and
  squarks},'' \href{http://dx.doi.org/10.22323/1.350.0168}{{\em PoS} {\bfseries
  LHCP2019} (2019) 168}, \href{http://arxiv.org/abs/1909.11753}{{\ttfamily
  arXiv:1909.11753 [hep-ex]}}.

\bibitem{ATLAS:2019oho}
{\bfseries ATLAS} Collaboration, ``{Search for direct top squark pair
  production in the 3-body decay mode with a final state containing one lepton,
  jets, and missing transverse momentum in $\sqrt{s}=13$TeV $pp$ collision data
  with the ATLAS detector},''.

\bibitem{CMS:2019ysk}
{\bfseries CMS} Collaboration, A.~M. Sirunyan {\em et~al.}, ``{Search for
  direct top squark pair production in events with one lepton, jets, and
  missing transverse momentum at 13 TeV with the CMS experiment},''
  \href{http://dx.doi.org/10.1007/JHEP05(2020)032}{{\em JHEP} {\bfseries 05}
  (2020) 032}, \href{http://arxiv.org/abs/1912.08887}{{\ttfamily
  arXiv:1912.08887 [hep-ex]}}.

\bibitem{Craig:2013cxa}
N.~Craig, ``{The State of Supersymmetry after Run I of the LHC},'' in {\em
  {Beyond the Standard Model after the first run of the LHC}}.
\newblock 9, 2013.
\newblock \href{http://arxiv.org/abs/1309.0528}{{\ttfamily arXiv:1309.0528
  [hep-ph]}}.

\bibitem{Barbieri:1987fn}
R.~Barbieri and G.~F. Giudice, ``{Upper Bounds on Supersymmetric Particle
  Masses},'' \href{http://dx.doi.org/10.1016/0550-3213(88)90171-X}{{\em Nucl.
  Phys. B} {\bfseries 306} (1988) 63--76}.

\bibitem{Papucci:2011wy}
M.~Papucci, J.~T. Ruderman, and A.~Weiler, ``{Natural SUSY Endures},''
  \href{http://dx.doi.org/10.1007/JHEP09(2012)035}{{\em JHEP} {\bfseries 09}
  (2012) 035}, \href{http://arxiv.org/abs/1110.6926}{{\ttfamily arXiv:1110.6926
  [hep-ph]}}.

\bibitem{Kitano:2006gv}
R.~Kitano and Y.~Nomura, ``{Supersymmetry, naturalness, and signatures at the
  LHC},'' \href{http://dx.doi.org/10.1103/PhysRevD.73.095004}{{\em Phys. Rev.
  D} {\bfseries 73} (2006) 095004},
  \href{http://arxiv.org/abs/hep-ph/0602096}{{\ttfamily arXiv:hep-ph/0602096}}.

\bibitem{Baer:2013gva}
H.~Baer, V.~Barger, and D.~Mickelson, ``{How conventional measures overestimate
  electroweak fine-tuning in supersymmetric theory},''
  \href{http://dx.doi.org/10.1103/PhysRevD.88.095013}{{\em Phys. Rev. D}
  {\bfseries 88} no.~9, (2013) 095013},
  \href{http://arxiv.org/abs/1309.2984}{{\ttfamily arXiv:1309.2984 [hep-ph]}}.

\bibitem{Mustafayev:2014lqa}
A.~Mustafayev and X.~Tata, ``{Supersymmetry, Naturalness, and Light
  Higgsinos},'' \href{http://dx.doi.org/10.1007/s12648-014-0504-8}{{\em Indian
  J. Phys.} {\bfseries 88} (2014) 991--1004},
  \href{http://arxiv.org/abs/1404.1386}{{\ttfamily arXiv:1404.1386 [hep-ph]}}.

\bibitem{Baer:2014ica}
H.~Baer, V.~Barger, D.~Mickelson, and M.~Padeffke-Kirkland, ``{SUSY models
  under siege: LHC constraints and electroweak fine-tuning},''
  \href{http://dx.doi.org/10.1103/PhysRevD.89.115019}{{\em Phys. Rev. D}
  {\bfseries 89} no.~11, (2014) 115019},
  \href{http://arxiv.org/abs/1404.2277}{{\ttfamily arXiv:1404.2277 [hep-ph]}}.

\bibitem{Baer:2012cf}
H.~Baer, V.~Barger, P.~Huang, D.~Mickelson, A.~Mustafayev, and X.~Tata,
  ``{Radiative natural supersymmetry: Reconciling electroweak fine-tuning and
  the Higgs boson mass},''
  \href{http://dx.doi.org/10.1103/PhysRevD.87.115028}{{\em Phys. Rev. D}
  {\bfseries 87} no.~11, (2013) 115028},
  \href{http://arxiv.org/abs/1212.2655}{{\ttfamily arXiv:1212.2655 [hep-ph]}}.

\bibitem{LEP:2003aa}
{\bfseries LEP, ALEPH, DELPHI, L3, OPAL, LEP Electroweak Working Group, SLD
  Electroweak Group, SLD Heavy Flavor Group} Collaboration, t.~S. Electroweak,
  ``{A Combination of preliminary electroweak measurements and constraints on
  the standard model},'' \href{http://arxiv.org/abs/hep-ex/0312023}{{\ttfamily
  arXiv:hep-ex/0312023}}.

\bibitem{Baer:2016lpj}
H.~Baer, V.~Barger, M.~Savoy, and H.~Serce, ``{The Higgs mass and natural
  supersymmetric spectrum from the landscape},''
  \href{http://dx.doi.org/10.1016/j.physletb.2016.05.010}{{\em Phys. Lett. B}
  {\bfseries 758} (2016) 113--117},
  \href{http://arxiv.org/abs/1602.07697}{{\ttfamily arXiv:1602.07697
  [hep-ph]}}.

\bibitem{Baer:2018rhs}
H.~Baer, V.~Barger, D.~Sengupta, and X.~Tata, ``{Is natural higgsino-only dark
  matter excluded?},''
  \href{http://dx.doi.org/10.1140/epjc/s10052-018-6306-y}{{\em Eur. Phys. J. C}
  {\bfseries 78} no.~10, (2018) 838},
  \href{http://arxiv.org/abs/1803.11210}{{\ttfamily arXiv:1803.11210
  [hep-ph]}}.

\bibitem{Peccei:1977hh}
R.~D. Peccei and H.~R. Quinn, ``{CP Conservation in the Presence of
  Instantons},'' \href{http://dx.doi.org/10.1103/PhysRevLett.38.1440}{{\em
  Phys. Rev. Lett.} {\bfseries 38} (1977) 1440--1443}.

\bibitem{Peccei:1977ur}
R.~D. Peccei and H.~R. Quinn, ``{Constraints Imposed by CP Conservation in the
  Presence of Instantons},''
  \href{http://dx.doi.org/10.1103/PhysRevD.16.1791}{{\em Phys. Rev. D}
  {\bfseries 16} (1977) 1791--1797}.

\bibitem{Weinberg:1977ma}
S.~Weinberg, ``{A New Light Boson?},''
  \href{http://dx.doi.org/10.1103/PhysRevLett.40.223}{{\em Phys. Rev. Lett.}
  {\bfseries 40} (1978) 223--226}.

\bibitem{Wilczek:1977pj}
F.~Wilczek, ``{Problem of Strong $P$ and $T$ Invariance in the Presence of
  Instantons},'' \href{http://dx.doi.org/10.1103/PhysRevLett.40.279}{{\em Phys.
  Rev. Lett.} {\bfseries 40} (1978) 279--282}.

\bibitem{Kim:1979if}
J.~E. Kim, ``{Weak Interaction Singlet and Strong CP Invariance},''
  \href{http://dx.doi.org/10.1103/PhysRevLett.43.103}{{\em Phys. Rev. Lett.}
  {\bfseries 43} (1979) 103}.

\bibitem{Shifman:1979if}
M.~A. Shifman, A.~I. Vainshtein, and V.~I. Zakharov, ``{Can Confinement Ensure
  Natural CP Invariance of Strong Interactions?},''
  \href{http://dx.doi.org/10.1016/0550-3213(80)90209-6}{{\em Nucl. Phys. B}
  {\bfseries 166} (1980) 493--506}.

\bibitem{Dine:1981rt}
M.~Dine, W.~Fischler, and M.~Srednicki, ``{A Simple Solution to the Strong CP
  Problem with a Harmless Axion},''
  \href{http://dx.doi.org/10.1016/0370-2693(81)90590-6}{{\em Phys. Lett. B}
  {\bfseries 104} (1981) 199--202}.

\bibitem{Zhitnitsky:1980tq}
A.~R. Zhitnitsky, ``{On Possible Suppression of the Axion Hadron Interactions.
  (In Russian)},'' {\em Sov. J. Nucl. Phys.} {\bfseries 31} (1980) 260.

\bibitem{Aprile:2017iyp}
{\bfseries XENON} Collaboration, E.~Aprile {\em et~al.}, ``{First Dark Matter
  Search Results from the XENON1T Experiment},''
  \href{http://dx.doi.org/10.1103/PhysRevLett.119.181301}{{\em Phys. Rev.
  Lett.} {\bfseries 119} no.~18, (2017) 181301},
  \href{http://arxiv.org/abs/1705.06655}{{\ttfamily arXiv:1705.06655
  [astro-ph.CO]}}.

\bibitem{Paige:2003mg}
F.~E. Paige, S.~D. Protopopescu, H.~Baer, and X.~Tata, ``{ISAJET 7.69: A Monte
  Carlo event generator for pp, anti-p p, and e+e- reactions},''
  \href{http://arxiv.org/abs/hep-ph/0312045}{{\ttfamily arXiv:hep-ph/0312045}}.

\bibitem{Franceschini:2008pz}
R.~Franceschini, T.~Hambye, and A.~Strumia, ``{Type-III see-saw at LHC},''
  \href{http://dx.doi.org/10.1103/PhysRevD.78.033002}{{\em Phys. Rev. D}
  {\bfseries 78} (2008) 033002},
  \href{http://arxiv.org/abs/0805.1613}{{\ttfamily arXiv:0805.1613 [hep-ph]}}.

\bibitem{Arhrib:2009mz}
A.~Arhrib, B.~Bajc, D.~K. Ghosh, T.~Han, G.-Y. Huang, I.~Puljak, and
  G.~Senjanovic, ``{Collider Signatures for Heavy Lepton Triplet in Type I+III
  Seesaw},'' \href{http://dx.doi.org/10.1103/PhysRevD.82.053004}{{\em Phys.
  Rev. D} {\bfseries 82} (2010) 053004},
  \href{http://arxiv.org/abs/0904.2390}{{\ttfamily arXiv:0904.2390 [hep-ph]}}.

\bibitem{Bandyopadhyay:2011aa}
P.~Bandyopadhyay, S.~Choi, E.~J. Chun, and K.~Min, ``{Probing Higgs bosons via
  the type III seesaw mechanism at the LHC},''
  \href{http://dx.doi.org/10.1103/PhysRevD.85.073013}{{\em Phys. Rev. D}
  {\bfseries 85} (2012) 073013},
  \href{http://arxiv.org/abs/1112.3080}{{\ttfamily arXiv:1112.3080 [hep-ph]}}.

\bibitem{Goswami:2017jqs}
D.~Goswami and P.~Poulose, ``{Direct searches of Type III seesaw triplet
  fermions at high energy $e^+e^-$ collider},''
  \href{http://dx.doi.org/10.1140/epjc/s10052-017-5478-1}{{\em Eur. Phys. J. C}
  {\bfseries 78} no.~1, (2018) 42},
  \href{http://arxiv.org/abs/1702.07215}{{\ttfamily arXiv:1702.07215
  [hep-ph]}}.

\bibitem{Das:2020gnt}
A.~Das, S.~Mandal, and T.~Modak, ``{Testing triplet fermions at the
  electron-positron and electron-proton colliders using fat jet signatures},''
  \href{http://dx.doi.org/10.1103/PhysRevD.102.033001}{{\em Phys. Rev. D}
  {\bfseries 102} no.~3, (2020) 033001},
  \href{http://arxiv.org/abs/2005.02267}{{\ttfamily arXiv:2005.02267
  [hep-ph]}}.

\bibitem{Ashanujjaman:2021jhi}
S.~Ashanujjaman and K.~Ghosh, ``{Type-III Seesaw: Phenomenological Implications
  of the Information Lost in Decoupling from High-Energy to Low-Energy},''
  \href{http://arxiv.org/abs/2102.09536}{{\ttfamily arXiv:2102.09536
  [hep-ph]}}.

\bibitem{Das:2020uer}
A.~Das and S.~Mandal, ``{Bounds on the triplet fermions in type-III seesaw and
  implications for collider searches},''
  \href{http://dx.doi.org/10.1016/j.nuclphysb.2021.115374}{{\em Nucl. Phys. B}
  {\bfseries 966} (2021) 115374},
  \href{http://arxiv.org/abs/2006.04123}{{\ttfamily arXiv:2006.04123
  [hep-ph]}}.

\bibitem{Biggio:2019eeo}
C.~Biggio, E.~Fernandez-Martinez, M.~Filaci, J.~Hernandez-Garcia, and
  J.~Lopez-Pavon, ``{Global Bounds on the Type-III Seesaw},''
  \href{http://dx.doi.org/10.1007/JHEP05(2020)022}{{\em JHEP} {\bfseries 05}
  (2020) 022}, \href{http://arxiv.org/abs/1911.11790}{{\ttfamily
  arXiv:1911.11790 [hep-ph]}}.

\bibitem{delAguila:2008cj}
F.~del Aguila and J.~A. Aguilar-Saavedra, ``{Distinguishing seesaw models at
  LHC with multi-lepton signals},''
  \href{http://dx.doi.org/10.1016/j.nuclphysb.2008.12.029}{{\em Nucl. Phys. B}
  {\bfseries 813} (2009) 22--90},
  \href{http://arxiv.org/abs/0808.2468}{{\ttfamily arXiv:0808.2468 [hep-ph]}}.

\bibitem{Sirunyan:2020pjd}
{\bfseries CMS} Collaboration, A.~M. Sirunyan {\em et~al.}, ``{Search for
  disappearing tracks in proton-proton collisions at $\sqrt{s} =$ 13 TeV},''
  \href{http://dx.doi.org/10.1016/j.physletb.2020.135502}{{\em Phys. Lett. B}
  {\bfseries 806} (2020) 135502},
  \href{http://arxiv.org/abs/2004.05153}{{\ttfamily arXiv:2004.05153
  [hep-ex]}}.

\bibitem{Mohapatra:1980yp}
R.~N. Mohapatra and G.~Senjanovic, ``{Neutrino Masses and Mixings in Gauge
  Models with Spontaneous Parity Violation},''
  \href{http://dx.doi.org/10.1103/PhysRevD.23.165}{{\em Phys. Rev. D}
  {\bfseries 23} (1981) 165}.

\bibitem{Pati:1974yy}
J.~C. Pati and A.~Salam, ``{Lepton Number as the Fourth Color},''
  \href{http://dx.doi.org/10.1103/PhysRevD.10.275}{{\em Phys. Rev. D}
  {\bfseries 10} (1974) 275--289}. [Erratum: Phys.Rev.D 11, 703--703 (1975)].

\bibitem{Mohapatra:1974hk}
R.~N. Mohapatra and J.~C. Pati, ``{Left-Right Gauge Symmetry and an
  Isoconjugate Model of CP Violation},''
  \href{http://dx.doi.org/10.1103/PhysRevD.11.566}{{\em Phys. Rev. D}
  {\bfseries 11} (1975) 566--571}.

\bibitem{Senjanovic:1975rk}
G.~Senjanovic and R.~N. Mohapatra, ``{Exact Left-Right Symmetry and Spontaneous
  Violation of Parity},''
  \href{http://dx.doi.org/10.1103/PhysRevD.12.1502}{{\em Phys. Rev. D}
  {\bfseries 12} (1975) 1502}.

\bibitem{Kuchimanchi:1993jg}
R.~Kuchimanchi and R.~N. Mohapatra, ``{No parity violation without R-parity
  violation},'' \href{http://dx.doi.org/10.1103/PhysRevD.48.4352}{{\em Phys.
  Rev. D} {\bfseries 48} (1993) 4352--4360},
  \href{http://arxiv.org/abs/hep-ph/9306290}{{\ttfamily arXiv:hep-ph/9306290}}.

\bibitem{Babu:2008ep}
K.~S. Babu and R.~N. Mohapatra, ``{Minimal Supersymmetric Left-Right Model},''
  \href{http://dx.doi.org/10.1016/j.physletb.2008.09.018}{{\em Phys. Lett. B}
  {\bfseries 668} (2008) 404--409},
  \href{http://arxiv.org/abs/0807.0481}{{\ttfamily arXiv:0807.0481 [hep-ph]}}.

\bibitem{Babu:2014vba}
K.~S. Babu and A.~Patra, ``{Higgs Boson Spectra in Supersymmetric Left-Right
  Models},'' \href{http://dx.doi.org/10.1103/PhysRevD.93.055030}{{\em Phys.
  Rev. D} {\bfseries 93} no.~5, (2016) 055030},
  \href{http://arxiv.org/abs/1412.8714}{{\ttfamily arXiv:1412.8714 [hep-ph]}}.

\bibitem{Basso:2015pka}
L.~Basso, B.~Fuks, M.~E. Krauss, and W.~Porod, ``{Doubly-charged Higgs and
  vacuum stability in left-right supersymmetry},''
  \href{http://dx.doi.org/10.1007/JHEP07(2015)147}{{\em JHEP} {\bfseries 07}
  (2015) 147}, \href{http://arxiv.org/abs/1503.08211}{{\ttfamily
  arXiv:1503.08211 [hep-ph]}}.

\bibitem{Zee:1985id}
A.~Zee, ``{Quantum Numbers of Majorana Neutrino Masses},''
  \href{http://dx.doi.org/10.1016/0550-3213(86)90475-X}{{\em Nucl. Phys. B}
  {\bfseries 264} (1986) 99--110}.

\bibitem{Babu:1988ki}
K.~S. Babu, ``{Model of 'Calculable' Majorana Neutrino Masses},''
  \href{http://dx.doi.org/10.1016/0370-2693(88)91584-5}{{\em Phys. Lett. B}
  {\bfseries 203} (1988) 132--136}.

\bibitem{ArkaniHamed:2002qx}
N.~Arkani-Hamed, A.~G. Cohen, E.~Katz, A.~E. Nelson, T.~Gregoire, and J.~G.
  Wacker, ``{The Minimal moose for a little Higgs},''
  \href{http://dx.doi.org/10.1088/1126-6708/2002/08/021}{{\em JHEP} {\bfseries
  08} (2002) 021}, \href{http://arxiv.org/abs/hep-ph/0206020}{{\ttfamily
  arXiv:hep-ph/0206020}}.

\bibitem{Babu:2020hun}
K.~S. Babu, P.~S.~B. Dev, S.~Jana, and A.~Thapa, ``{Unified framework for
  $B$-anomalies, muon $g − 2$ and neutrino masses},''
  \href{http://dx.doi.org/10.1007/JHEP03(2021)179}{{\em JHEP} {\bfseries 03}
  (2021) 179}, \href{http://arxiv.org/abs/2009.01771}{{\ttfamily
  arXiv:2009.01771 [hep-ph]}}.

\bibitem{Gunion:1989ci}
J.~F. Gunion, R.~Vega, and J.~Wudka, ``{Higgs triplets in the standard
  model},'' \href{http://dx.doi.org/10.1103/PhysRevD.42.1673}{{\em Phys. Rev.
  D} {\bfseries 42} (1990) 1673--1691}.

\bibitem{Babu:2009aq}
K.~S. Babu, S.~Nandi, and Z.~Tavartkiladze, ``{New Mechanism for Neutrino Mass
  Generation and Triply Charged Higgs Bosons at the LHC},''
  \href{http://dx.doi.org/10.1103/PhysRevD.80.071702}{{\em Phys. Rev. D}
  {\bfseries 80} (2009) 071702},
  \href{http://arxiv.org/abs/0905.2710}{{\ttfamily arXiv:0905.2710 [hep-ph]}}.

\bibitem{Bonnet:2009ej}
F.~Bonnet, D.~Hernandez, T.~Ota, and W.~Winter, ``{Neutrino masses from higher
  than d=5 effective operators},''
  \href{http://dx.doi.org/10.1088/1126-6708/2009/10/076}{{\em JHEP} {\bfseries
  10} (2009) 076}, \href{http://arxiv.org/abs/0907.3143}{{\ttfamily
  arXiv:0907.3143 [hep-ph]}}.

\bibitem{Bhattacharya:2016qsg}
S.~Bhattacharya, S.~Jana, and S.~Nandi, ``{Neutrino Masses and Scalar Singlet
  Dark Matter},'' \href{http://dx.doi.org/10.1103/PhysRevD.95.055003}{{\em
  Phys. Rev. D} {\bfseries 95} no.~5, (2017) 055003},
  \href{http://arxiv.org/abs/1609.03274}{{\ttfamily arXiv:1609.03274
  [hep-ph]}}.

\bibitem{Kumericki:2012bh}
K.~Kumericki, I.~Picek, and B.~Radovcic, ``{TeV-scale Seesaw with Quintuplet
  Fermions},'' \href{http://dx.doi.org/10.1103/PhysRevD.86.013006}{{\em Phys.
  Rev. D} {\bfseries 86} (2012) 013006},
  \href{http://arxiv.org/abs/1204.6599}{{\ttfamily arXiv:1204.6599 [hep-ph]}}.

\bibitem{Cai:2017mow}
Y.~Cai, T.~Han, T.~Li, and R.~Ruiz, ``{Lepton Number Violation: Seesaw Models
  and Their Collider Tests},''
  \href{http://dx.doi.org/10.3389/fphy.2018.00040}{{\em Front. in Phys.}
  {\bfseries 6} (2018) 40}, \href{http://arxiv.org/abs/1711.02180}{{\ttfamily
  arXiv:1711.02180 [hep-ph]}}.

\bibitem{Chiang:2012dk}
C.-W. Chiang, T.~Nomura, and K.~Tsumura, ``{Search for doubly charged Higgs
  bosons using the same-sign diboson mode at the LHC},''
  \href{http://dx.doi.org/10.1103/PhysRevD.85.095023}{{\em Phys. Rev. D}
  {\bfseries 85} (2012) 095023},
  \href{http://arxiv.org/abs/1202.2014}{{\ttfamily arXiv:1202.2014 [hep-ph]}}.

\bibitem{Chiang:2015amq}
C.-W. Chiang, A.-L. Kuo, and T.~Yamada, ``{Searches of exotic Higgs bosons in
  general mass spectra of the Georgi-Machacek model at the LHC},''
  \href{http://dx.doi.org/10.1007/JHEP01(2016)120}{{\em JHEP} {\bfseries 01}
  (2016) 120}, \href{http://arxiv.org/abs/1511.00865}{{\ttfamily
  arXiv:1511.00865 [hep-ph]}}.

\bibitem{Zyla:2020zbs}
{\bfseries Particle Data Group} Collaboration, P.~A. Zyla {\em et~al.},
  ``{Review of Particle Physics},''
  \href{http://dx.doi.org/10.1093/ptep/ptaa104}{{\em PTEP} {\bfseries 2020}
  no.~8, (2020) 083C01}.

\bibitem{Blasi:2017xmc}
S.~Blasi, S.~De~Curtis, and K.~Yagyu, ``{Effects of custodial symmetry breaking
  in the Georgi-Machacek model at high energies},''
  \href{http://dx.doi.org/10.1103/PhysRevD.96.015001}{{\em Phys. Rev. D}
  {\bfseries 96} no.~1, (2017) 015001},
  \href{http://arxiv.org/abs/1704.08512}{{\ttfamily arXiv:1704.08512
  [hep-ph]}}.

\bibitem{Chiang:2018xpl}
C.-W. Chiang, A.-L. Kuo, and K.~Yagyu, ``{One-loop renormalized Higgs boson
  vertices in the Georgi-Machacek model},''
  \href{http://dx.doi.org/10.1103/PhysRevD.98.013008}{{\em Phys. Rev. D}
  {\bfseries 98} no.~1, (2018) 013008},
  \href{http://arxiv.org/abs/1804.02633}{{\ttfamily arXiv:1804.02633
  [hep-ph]}}.

\bibitem{Melfo:2011nx}
A.~Melfo, M.~Nemevsek, F.~Nesti, G.~Senjanovic, and Y.~Zhang, ``{Type II Seesaw
  at LHC: The Roadmap},''
  \href{http://dx.doi.org/10.1103/PhysRevD.85.055018}{{\em Phys. Rev. D}
  {\bfseries 85} (2012) 055018},
  \href{http://arxiv.org/abs/1108.4416}{{\ttfamily arXiv:1108.4416 [hep-ph]}}.

\bibitem{Aoki:2011pz}
M.~Aoki, S.~Kanemura, and K.~Yagyu, ``{Testing the Higgs triplet model with the
  mass difference at the LHC},''
  \href{http://dx.doi.org/10.1103/PhysRevD.85.055007}{{\em Phys. Rev. D}
  {\bfseries 85} (2012) 055007},
  \href{http://arxiv.org/abs/1110.4625}{{\ttfamily arXiv:1110.4625 [hep-ph]}}.

\bibitem{Chiang:2012cn}
C.-W. Chiang and K.~Yagyu, ``{Testing the custodial symmetry in the Higgs
  sector of the Georgi-Machacek model},''
  \href{http://dx.doi.org/10.1007/JHEP01(2013)026}{{\em JHEP} {\bfseries 01}
  (2013) 026}, \href{http://arxiv.org/abs/1211.2658}{{\ttfamily arXiv:1211.2658
  [hep-ph]}}.

\bibitem{Barnett:1988mx}
R.~M. Barnett, J.~F. Gunion, and H.~E. Haber, ``{LIKE SIGN DILEPTONS AS A
  SIGNAL FOR GLUINO PRODUCTION},'' in {\em {1988 DPF Summer Study on
  High-energy Physics in the 1990s (Snowmass 88)}}.
\newblock 10, 1988.

\bibitem{Baer:1989hr}
H.~Baer, X.~Tata, and J.~Woodside, ``{Gluino Cascade Decay Signatures at the
  Tevatron Collider},'' \href{http://dx.doi.org/10.1103/PhysRevD.41.906}{{\em
  Phys. Rev. D} {\bfseries 41} (1990) 906--915}.

\bibitem{Baer:1991xs}
H.~Baer, X.~Tata, and J.~Woodside, ``{Multi - lepton signals from supersymmetry
  at hadron super colliders},''
  \href{http://dx.doi.org/10.1103/PhysRevD.45.142}{{\em Phys. Rev. D}
  {\bfseries 45} (1992) 142--160}.

\bibitem{Barnett:1993ea}
R.~M. Barnett, J.~F. Gunion, and H.~E. Haber, ``{Discovering supersymmetry with
  like sign dileptons},''
  \href{http://dx.doi.org/10.1016/0370-2693(93)91623-U}{{\em Phys. Lett. B}
  {\bfseries 315} (1993) 349--354},
  \href{http://arxiv.org/abs/hep-ph/9306204}{{\ttfamily arXiv:hep-ph/9306204}}.

\bibitem{Alwall:2011uj}
J.~Alwall, M.~Herquet, F.~Maltoni, O.~Mattelaer, and T.~Stelzer, ``{MadGraph 5
  : Going Beyond},'' \href{http://dx.doi.org/10.1007/JHEP06(2011)128}{{\em
  JHEP} {\bfseries 06} (2011) 128},
  \href{http://arxiv.org/abs/1106.0522}{{\ttfamily arXiv:1106.0522 [hep-ph]}}.

\bibitem{Alwall:2014hca}
J.~Alwall, R.~Frederix, S.~Frixione, V.~Hirschi, F.~Maltoni, O.~Mattelaer,
  H.~S. Shao, T.~Stelzer, P.~Torrielli, and M.~Zaro, ``{The automated
  computation of tree-level and next-to-leading order differential cross
  sections, and their matching to parton shower simulations},''
  \href{http://dx.doi.org/10.1007/JHEP07(2014)079}{{\em JHEP} {\bfseries 07}
  (2014) 079}, \href{http://arxiv.org/abs/1405.0301}{{\ttfamily arXiv:1405.0301
  [hep-ph]}}.

\bibitem{Sjostrand:2014zea}
T.~Sj\"ostrand, S.~Ask, J.~R. Christiansen, R.~Corke, N.~Desai, P.~Ilten,
  S.~Mrenna, S.~Prestel, C.~O. Rasmussen, and P.~Z. Skands, ``{An introduction
  to PYTHIA 8.2},'' \href{http://dx.doi.org/10.1016/j.cpc.2015.01.024}{{\em
  Comput. Phys. Commun.} {\bfseries 191} (2015) 159--177},
  \href{http://arxiv.org/abs/1410.3012}{{\ttfamily arXiv:1410.3012 [hep-ph]}}.

\bibitem{deFavereau:2013fsa}
{\bfseries DELPHES 3} Collaboration, J.~de~Favereau, C.~Delaere, P.~Demin,
  A.~Giammanco, V.~Lema\^\i{}tre, A.~Mertens, and M.~Selvaggi, ``{DELPHES 3, A
  modular framework for fast simulation of a generic collider experiment},''
  \href{http://dx.doi.org/10.1007/JHEP02(2014)057}{{\em JHEP} {\bfseries 02}
  (2014) 057}, \href{http://arxiv.org/abs/1307.6346}{{\ttfamily arXiv:1307.6346
  [hep-ex]}}.

\bibitem{Cacciari:2008gp}
M.~Cacciari, G.~P. Salam, and G.~Soyez, ``{The anti-$k_t$ jet clustering
  algorithm},'' \href{http://dx.doi.org/10.1088/1126-6708/2008/04/063}{{\em
  JHEP} {\bfseries 04} (2008) 063},
  \href{http://arxiv.org/abs/0802.1189}{{\ttfamily arXiv:0802.1189 [hep-ph]}}.

\bibitem{CMS:2012feb}
{\bfseries CMS} Collaboration, S.~Chatrchyan {\em et~al.}, ``{Identification of
  b-Quark Jets with the CMS Experiment},''
  \href{http://dx.doi.org/10.1088/1748-0221/8/04/P04013}{{\em JINST} {\bfseries
  8} (2013) P04013}, \href{http://arxiv.org/abs/1211.4462}{{\ttfamily
  arXiv:1211.4462 [hep-ex]}}.

\bibitem{CMS:2017wtu}
{\bfseries CMS} Collaboration, A.~M. Sirunyan {\em et~al.}, ``{Identification
  of heavy-flavour jets with the CMS detector in pp collisions at 13 TeV},''
  \href{http://dx.doi.org/10.1088/1748-0221/13/05/P05011}{{\em JINST}
  {\bfseries 13} no.~05, (2018) P05011},
  \href{http://arxiv.org/abs/1712.07158}{{\ttfamily arXiv:1712.07158
  [physics.ins-det]}}.

\bibitem{CMS-DP-2018-026}
{\bfseries CMS Collaboration} Collaboration, ``{Tau Identification Performance
  in 2017 Data at $\sqrt{s}=13$ TeV},''.
  \url{https://cds.cern.ch/record/2622155}.

\bibitem{CMS:2015pac}
{\bfseries CMS} Collaboration, V.~Khachatryan {\em et~al.}, ``{Reconstruction
  and identification of \ensuremath{\tau} lepton decays to hadrons and
  \ensuremath{\nu}$_τ$ at CMS},''
  \href{http://dx.doi.org/10.1088/1748-0221/11/01/P01019}{{\em JINST}
  {\bfseries 11} no.~01, (2016) P01019},
  \href{http://arxiv.org/abs/1510.07488}{{\ttfamily arXiv:1510.07488
  [physics.ins-det]}}.

\bibitem{Beenakker:1996ed}
W.~Beenakker, R.~Hopker, and M.~Spira, ``{PROSPINO: A Program for the
  production of supersymmetric particles in next-to-leading order QCD},''
  \href{http://arxiv.org/abs/hep-ph/9611232}{{\ttfamily arXiv:hep-ph/9611232}}.

\bibitem{Gherghetta:1999sw}
T.~Gherghetta, G.~F. Giudice, and J.~D. Wells, ``{Phenomenological consequences
  of supersymmetry with anomaly induced masses},''
  \href{http://dx.doi.org/10.1016/S0550-3213(99)00429-0}{{\em Nucl. Phys. B}
  {\bfseries 559} (1999) 27--47},
  \href{http://arxiv.org/abs/hep-ph/9904378}{{\ttfamily arXiv:hep-ph/9904378}}.

\bibitem{Feng:1999hg}
J.~L. Feng and T.~Moroi, ``{Supernatural supersymmetry: Phenomenological
  implications of anomaly mediated supersymmetry breaking},''
  \href{http://dx.doi.org/10.1103/PhysRevD.61.095004}{{\em Phys. Rev. D}
  {\bfseries 61} (2000) 095004},
  \href{http://arxiv.org/abs/hep-ph/9907319}{{\ttfamily arXiv:hep-ph/9907319}}.

\bibitem{ATLAS:2016pbt}
{\bfseries ATLAS} Collaboration, ``{Search for doubly-charged Higgs bosons in
  same-charge electron pair final states using proton-proton collisions at
  $\sqrt{s}=13\,\mathrm{TeV}$ with the ATLAS detector},''.

\bibitem{CMS:2017dzg}
{\bfseries CMS} Collaboration, A.~M. Sirunyan {\em et~al.}, ``{Measurements of
  the $\mathrm {p}\mathrm {p}\rightarrow \mathrm{Z}\mathrm{Z}$ production cross
  section and the $\mathrm{Z}\rightarrow 4\ell $ branching fraction, and
  constraints on anomalous triple gauge couplings at $\sqrt{s} = 13\,\text
  {TeV} $},'' \href{http://dx.doi.org/10.1140/epjc/s10052-018-5567-9}{{\em Eur.
  Phys. J. C} {\bfseries 78} (2018) 165},
  \href{http://arxiv.org/abs/1709.08601}{{\ttfamily arXiv:1709.08601
  [hep-ex]}}. [Erratum: Eur.Phys.J.C 78, 515 (2018)].

\bibitem{Nhung:2013tfu}
D.~T. Nhung, L.~D. Ninh, and M.~M. Weber, ``{NLO $W W Z$ production at the
  LHC},'' in {\em {9th Rencontres du Vietnam}: {Windows on the Universe}},
  pp.~219--222.
\newblock 2013.
\newblock \href{http://arxiv.org/abs/1310.6159}{{\ttfamily arXiv:1310.6159
  [hep-ph]}}.

\bibitem{Nhung:2013jta}
D.~T. Nhung, L.~D. Ninh, and M.~M. Weber, ``{NLO corrections to WWZ production
  at the LHC},'' \href{http://dx.doi.org/10.1007/JHEP12(2013)096}{{\em JHEP}
  {\bfseries 12} (2013) 096}, \href{http://arxiv.org/abs/1307.7403}{{\ttfamily
  arXiv:1307.7403 [hep-ph]}}.

\bibitem{Hankele:2007sb}
V.~Hankele and D.~Zeppenfeld, ``{QCD corrections to hadronic WWZ production
  with leptonic decays},''
  \href{http://dx.doi.org/10.1016/j.physletb.2008.02.014}{{\em Phys. Lett. B}
  {\bfseries 661} (2008) 103--108},
  \href{http://arxiv.org/abs/0712.3544}{{\ttfamily arXiv:0712.3544 [hep-ph]}}.

\bibitem{NLOMultilegWorkingGroup:2008bxd}
{\bfseries NLO Multileg Working Group} Collaboration, Z.~Bern {\em et~al.},
  ``{The NLO multileg working group: Summary report},'' in {\em {5th Les
  Houches Workshop on Physics at TeV Colliders}}, pp.~1--120.
\newblock 3, 2008.
\newblock \href{http://arxiv.org/abs/0803.0494}{{\ttfamily arXiv:0803.0494
  [hep-ph]}}.

\bibitem{Ballestrero:2018anz}
A.~Ballestrero {\em et~al.}, ``{Precise predictions for same-sign W-boson
  scattering at the LHC},''
  \href{http://dx.doi.org/10.1140/epjc/s10052-018-6136-y}{{\em Eur. Phys. J. C}
  {\bfseries 78} no.~8, (2018) 671},
  \href{http://arxiv.org/abs/1803.07943}{{\ttfamily arXiv:1803.07943
  [hep-ph]}}.

\end{thebibliography}\endgroup

\end{document}